\documentclass[%
    final,      % fertiges Dokument
   11pt,
   bigheadings,       % gro�e �berschriften
   a4paper,
   BCOR5mm,          % Zusaetzlicher Rand auf der Innenseite
   DIV11,            % Seitengroesse (siehe Koma Skript Dokumentation !)
   1.1headlines,     % Zeilenanzahl der Kopfzeilen
   pagesize,         % Schreibt die Papiergroesse in die Datei.
   twoside,          % Seitenraender f�r zweiseitiges Layout
   openright,        % Kapitel beginnen immer auf der rechten Seite
   titlepage,        % Titel als einzelne Seite ('titlepage' Umgebung)
   parindent,        % Einger�ckt (Standard)
   headsepline,      % Linie unter Kolumnentitel
   nochapterprefix,  % keine Ausgabe von 'Kapitel:'
   bibtotoc,         % Bibliographie ins TOC
   tocindent,        % eingereuckte Gliederung
   listsindent,      % eingereuckte LOT, LOF
   pointlessnumbers, % �berschriftnummerierung ohne Punkt, siehe DUDEN !
   cleardoubleplain,
   fleqn,            % Formeln werden linksbuendig angezeigt
]{scrbook}%     Klassen: scrartcl, scrreprt, scrbook
\usepackage[latin1]{inputenc}
\usepackage{xspace} % Define commands that don't eat spaces.
\usepackage{ifpdf} %\ifpdf \else \fi
\usepackage{calc} % Calculation with LaTeX
\usepackage[english]{babel} % Languagesetting
\usepackage[table]{xcolor} % Farben
\usepackage[]{graphicx} % Bilder
\usepackage[]{amsmath} % Amsmath - Mathematik Basispaket
\usepackage{ragged2e} % Besserer Flatternsatz (Linksbuendig, statt Blocksatz)

\usepackage[T1]{fontenc} % T1 Schrift Encoding (notwendig f�r die meisten Type 1 Schriften)
\usepackage{textcomp}	 % Zusatzliche Symbole (Text Companion font extension)

\usepackage[fixamsmath,disallowspaces]{mathtools} % Erweitert amsmath und behebt einige Bugs
\usepackage{fixmath}
\usepackage[all,warning]{onlyamsmath} % Warnt bei Benutzung von Befehlen die mit amsmath inkompatibel sind.
\usepackage{icomma} % Erlaubt die Benutzung von Kommas im Mathematikmodus

\usepackage{amssymb}
\usepackage{booktabs}
\usepackage{tabularx} % tabularx nach hyperref laden
\usepackage{multirow}

\usepackage{url} % Setzen von URLs. In Verbindung mit hyperref sind diese auch aktive Links.
\usepackage[stable,perpage, ragged,  multiple]{footmisc} % Fussnoten
\usepackage[english]{varioref} % Intelligente Querverweise
\usepackage{enumitem} % Listen

\usepackage[%
	square,	% for square brackets;
	comma,	% to use commas as separaters;
	sort,		% orders multiple citations into the sequence in which they appear in the list of references;
]{natbib}

\usepackage{float}             % Stellt die Option [H] fuer Floats zur Verfgung
\usepackage{subfigure}
\usepackage{floatflt}
\usepackage{wrapfig}	        % Bilder von Text Umfliessen lassen
\usepackage{units}

\usepackage{setspace}
\typearea[current]{last}
\usepackage[%
   automark,         % automatische Aktualisierung der Kolumnentitel
   nouppercase,      % Grossbuchstaben verhindern
]{scrpage2}

\clearscrheadings
\clearscrplain
\ohead{\pagemark} % Oben aussen: Seitenzahlen
\ihead{\headmark} % Oben innen: Kapitel und Section

\automark[section]{chapter} %[rechts]{links}
\setheadsepline{.4pt}[\color{black}] % Linie zwischen Kopf und Textk�rper

\deffootnote{1.5em}{1em}{\makebox[1.5em][l]{\thefootnotemark}}
\newcommand\SectionFontStyle{\sffamily}
\setkomafont{chapter}{\huge\SectionFontStyle}    % Chapter
\setkomafont{sectioning}{\SectionFontStyle} %  % Titelzeilen % \bfseries
\setkomafont{pagenumber}{\bfseries\SectionFontStyle}             % Seitenzahl
\setkomafont{pagehead}{\small\sffamily}        % Kopfzeile
\setkomafont{descriptionlabel}{\itshape}        % Kopfzeile
\usepackage{caption}
\DeclareCaptionOption{parskip}[]{}
\DeclareCaptionOption{parindent}[]{}

\usepackage[subfigure]{tocloft}

\usepackage[breaklinks=true,dvipdfmx]{hyperref}
\newcommand{\eps}{\varepsilon}
\newcommand{\follows}{\; \Rightarrow \;}

\newcommand{\MeV}{\hbox{ MeV}}
\newcommand{\GeV}{\hbox{ GeV}}
\newcommand{\TeV}{\hbox{ TeV}}
\newcommand{\fm}{\hbox{ fm}} 

\def\abs#1{\mathinner{\lvert#1\rvert}}

\def\bra#1{\ensuremath{\langle{#1}\vert}}
\def\ket#1{\ensuremath{\vert{#1}\rangle}}
\def\braket#1#2{\ensuremath{\langle{#1}\mkern1.2mu\vert\mkern1.2mu{#2}\rangle}}

\renewcommand{\exp}[1]{\mbox{exp}\left[#1\right]}

\DeclareMathOperator{\sign}{sign}

\DeclareMathOperator{\atanh}{atanh}

\DeclareMathOperator{\dn}{d}
\DeclareMathOperator{\Dn}{D}

\begin{document}

\thispagestyle{plain}

\begin{center}
 \includegraphics[scale=0.25]{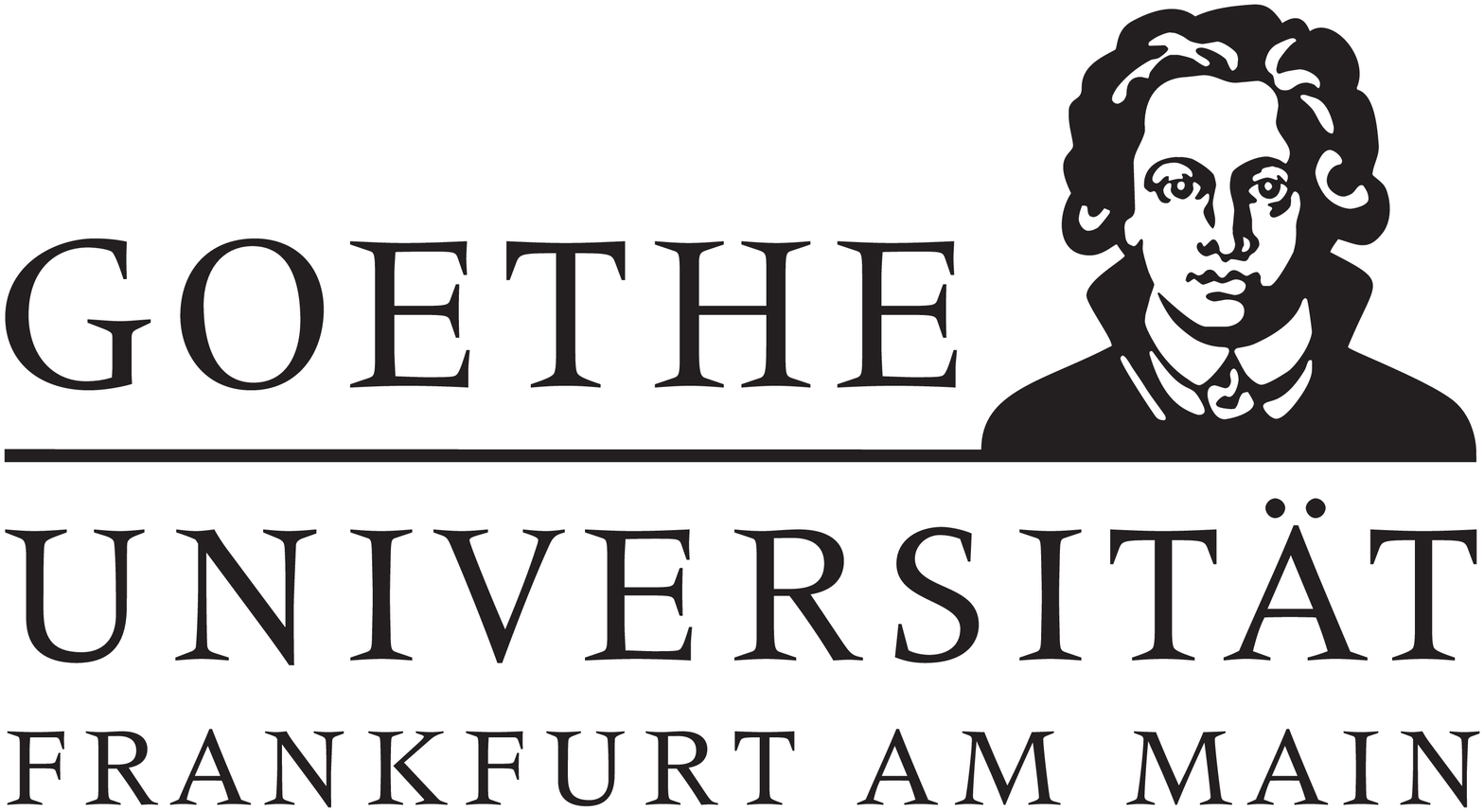}\\\vspace{3cm}
\begin{LARGE}{\bf A parton recombination approach to heavy ion collisions at RHIC and LHC}\\\vspace{1cm}
Diploma thesis
\end{LARGE}
\\\vspace{2cm}
by Daniel Krieg\\
December 2008
\end{center}

\cleardoublepage
\thispagestyle{plain}
\begin{center}
{\Large\textit{\mdseries{F\"ur Daniela}}}
\smallskip
\end{center}
\smallskip
\noindent
\vfill\eject

\frontmatter
\tableofcontents
\clearpage
\listoffigures
\listoftables

\mainmatter

\chapter{Introduction}
\section{Curiosity and doubt}
The knowledge of mankind has grown rapidly in the last century. Especially through the fundamental research in physics. It was and is driven by human curiosity, since typically the fundamental research has no direct application in everyday or technical life. Nevertheless, over the time these insights into natural laws have shown their profitableness in the technical revolutions.

So on the one hand, the growth of the knowledge of mankind has always been driven by curiosity, since
``curiosity is always the starting point for solutions to a problem''\footnote{Galileo Galilei}.
On the other hand, physical models and theories can never be right or true in a rigourous sense, but only falsified. So beside our success in describing nature, ``we absolutely must leave room for doubt or there is no progress and no learning. There is no learning without having to pose a question. And a question requires doubt''\footnote{Richard P. Feynman, Galileo Symposium in Italy, 1964}.
In order to learn about nature, we have to be curious and ask new question, but we also must have the courage to doubt our new found answers.

But curiosity is not a feature of science itself, it is an innateness of mankind. Most of the questions asked by physicists may seem academic and irrelevant to the scientific layman. Maybe still complicated, the questions that are tried to be answered by nuclear phyicist, by colliding atomic particles at the speed of light, are those questions every human reflects: Where are we from and why are we here?

\vspace{1cm}
Where we are from is widely believed to be the so called big bang, during which time, space, matter and energy turned into existence. Why or what exactly happend at time zero no one can tell because at this singluarity in spacetime the law of physics broke down. Asking about before also makes no sense, because without time there is no before.

\vspace{1cm}
Why we are here is a more philosophical question which can be understood in many different ways. While the physicist wonders about the asymmetry between matter and antimatter, which is an essential fact for our existence, others would ask this question more spirtiually. But the creation of the universe via a big bang would not contradict the existence of an omnipotent power that is responsible for ``life, the universe and everything''\footnote{Douglas Adams, The Hitchhiker's Guide to the Galaxy}...

\section{The fundamental building blocks}
But let us go back to a question we believe to have answered to a wide degree: What are we made of?
This question has a long history and mankind has had great success in finding the fundamental building blocks that make up the matter around us. Most all of these findings where mainly made in the last century.

The theory of the greek classical ``elements'' earth, water, air and fire, that go back to the philosopher Empedokles, persisted to the sevententh century until the new definition for an element by the chemicist Robert Boyle marked the beginning of the modern chemistry.

The chemicists classified many new elements, which soon required a new ordering schema. The periodic system of the elements was introduced which sorts the elements by their mass and attributes to groups of eight. The electron, the first subatomic particle, was discovered in 1897 by Joseph John Thomson while its existence was predicted by George Johnstone Stoney 23 years before. Thomsons so-called ``plumpudding''-model of the atom was replaced in 1911 by Ernest Rutherford after his famous scattering experiment. With Rutherfords interpretation, where a positively charged core is surrounded by a cloud of electrons, it was then possible to explain the chemical properties in the periodic system by the atomic electron shells. In 1919 Rutherford discovered the proton, and with the discovery of the neutron in 1932 by his student James Chadwick the description of the atomic nuclei was complete. On the one side, the atomic physics focused on the electron shells, Max Born postulated discrete orbits for the electrons and with the advent of quantum mechanics physicists were able to explain the atomic spectra.

\vspace{1cm}
On the other side the nuclear physicists were studying the structure of the atomic core. Around the 50's more and more new particles like the proton and neutron, called hadrons, were found. Soon there was such an overwhelming richness, that now the nuclear physicists needed a new ordering schema. Murray Gell-Mann proposed his ``eightfold-way'', where the hadrons could be grouped to octets of similiar mass. This was the predecessor of the quark model in which the protrons, neutrons and all other hadrons are no elementary particles, but made up of quarks, where the mesons are quark-antiquark combinations and the baryons consist of three quarks.

Today the standard model of particle physics classifies the fundamental particles into fermions as the particles of matter and the bosons as the particles mediating the interactions. The standard model (SM) has been very successful and marks a milestone in our understanding of the elementary particles and their interaction. But it falls short of being a complete theory since it lacks the description of gravity. Also the many free parameters that have to be determined by experiment are a big shortcoming of the model.

So there are many attempts to extend the SM (e.g. supersymmetry) or to discard it in favor of a new theory (e.g. stringtheory). But up to now it is the best answer to the question ``What we are made of''.

\section{Back to the Big Bang}
When we go back in time, we can see different forces responsible for the evolution. At the beginning the universe was extremly hot and dense and all fundamental particles were unbound and free like in plasma. Due to the expansion the temperature was going down and the quarks were pulled together by the strong force and formed hadrons; this is called hadronization. Most of them are unstable, but protons and neutrons assembled to stable clusters. During the ongoing cooling the electrons were bound to these clusters by the electromagnetic force. The atomic elements and molecules began to develop. Finally, the gravity pulled the neutral atoms together and so larger clusters of matter started to grow forming the stars and planets making up our known universe.

\vspace{1cm}
The gravity and the electromagnetic force are quite good understood, so one main topic of current research is the study of early universe, the regime of the strong force. In fact, this force is so strong that it is not possible to separate two quarks. Therefore, no one has ever seen (or measured) a quark. The hope is that if we could heat up the hadrons to the temperatures directly after the big bang, we could create this plasma state of the quarks where they are free.
With the largest machines in the world, physicists collide nuclei at the speed of light to create such a hot and dense system resembling the universe a few nanoseconds after the big bang to study the phase transition of the strongly interacting matter. The theoretical description of these collisions are very involved and due to the complex dynamics of the fundamental degrees of freedom, this is not yet understood from first principles.

\vspace{1cm}
In this thesis, I will analyze a phenomenologial model called recombination or coalescence which is a relevant mechanism for hadronization in heavy ion collisions (HIC) at the highest energies that are available today. It is structured as follows:
\begin{itemize}
 \item In the next chapter I give a short introduction into the phenomenology HIC, explain different models to describe the collisions and discuss how to probe the QGP.
\item The third chapter adresses the hadronization from a QGP, focusing on the recombination approach and its ingredients.
\item The results are presented in chapter four where I first give an overview of the parameters and parameterizations, followed by a detailed discussion on the different results and predictions of recombination.
\item The last chapter contains final remarks and conclusions.
\end{itemize}

\chapter{Heavy Ion Collisions}
\section{Phase diagram of the strong force}
The main goal of the current and past heavy ion programs is the search for a new state of matter
called the Quark-Gluon-Plasma (QGP) \cite{Bass:1998vz}. The quarks are the constituents of hadrons. They interact via their color charge by the strong force meditated by the gluons. Due to the strong coupling quarks can not be observed as single particles, only as clusters, where the color of all quarks add up to ``white''. This so called confinement is expected to break at sufficiently high temperatures and/or densities. The quarks and gluons are then free unbound particles like the electrons and protons in a plasma. That is why this state of matter is referred to as QGP and describes a new phase different to the normal ground state.

\begin{wrapfigure}{p}{0.5\textwidth}
 \centering
\includegraphics[scale=0.4]{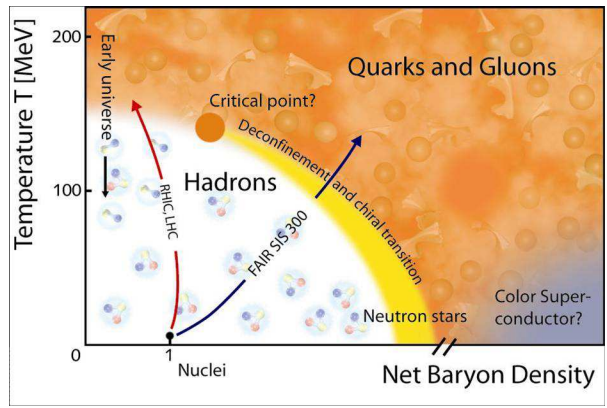}
\caption[The (assumed) phase diagram of the strong force]{The (assumed) phase diagram of the strong force (picture taken from \cite{CBM:2007introduction}).}
\label{fig:phase_diagram}
\end{wrapfigure}
The nuclear matter in the accelerators is highly compressed and heated up during the collision. At suffienctly high beam energies, we expect the system to go through a phase transition from the hadronic phase to the QGP. Due to the large pressure, the systems expands rapdily, cools down and crosses the phase boundary again. The quarks hadronize and form new hadrons. In this still very dense system, the hadrons scatter mostly inelastic and create exited hadronic states until the system is too dilute. At this so called chemical freeze-out, the inelastic cross sections are neglectable. After the kinetic freeze-out, there are also no more elastic interactions between the hadrons.

The livetime of the QGP is very short and only the hadrons coming from the kinetic freeze-out can be measured in the detectors. So there is no way to see if we have created a QGP or not. There are two different ways to examine the early phase of the collision:
\begin{itemize}
 \item We can study different probes that can escape the early system mostly undisturbed and try to extract information from these measures. That would be particles with a low cross-section in the medium like photons, pair-created leptons (dileptons) and heavy-quark mesons ($J/\psi$).
\item The other possibility is to look for characterisitic properties of the hadrons from freeze-out, which can be associated with a QGP formation, e.g. collective flow phenomena like the elliptic flow.
\end{itemize}

\section{Theoretical descriptions of heavy ion collision}
\subsection{Statistical thermal model}
The thermal model is a statistical approach to particle multiplicity in heavy ion collisions. One assumes a globally equilibriated thermal source described by temperature $T$ and chemical potentials $\mu_Q$ for the charges $Q$. The thermodynamical observables are then evaluated as an average over the statistical ensembles \cite{nucl-th/0304013}. Therefore, one has to calculate the grand canonical partition function
\begin{align}
 Z^{\mbox{GC}} (T,V,\mu_Q) = \mbox{Tr}\left[e^{-\beta\left(H-\sum_i \mu_i Q_i\right)}\right],
\end{align}
where the Hamiltonian $H$ depends on the equation of state (EoS), which connects the energy density with the pressure. The EoS depends on the degrees of freedom and so changes throughout the evolution of the system. The EoS of the QGP is subject to current research and can only be determined within QCD. Below the phase boundary the system can be described by a hadron resonance gas with a known EoS.

The applicability of the thermal model is very limited. The assumptions of global equilibrium and a static source are at best questionable. The predicted particle multiplicity, when using the EoS of the hadron gas, are in good agreement, but since this does not require the formation of a QGP, the predicitve powers are very limited.

\subsection{Quantumchromodynamics (QCD)}
The only possibility to really calculate the shape and order of the phase transition would be to solve the equations of motion for QCD, the gauge theory for the strongly interacting colored particles.
The approach from Quantumelectrodynamics (QED), where pertubation theory is very successful because of the small coupling constant $\alpha \approx 1/137$, is barely applicable in QCD due to the large coupling of the strong force and the non-abelian nature of the corresponding symmetry group $SU(3)$. The pertubative QCD (pQCD) works only at large momentum transfers, where the running coupling constant becomes small.

The other way is to discretize the langrangian and solve it numerically on a lattice (LQCD). Since this requires vast computational powers, there is no chance of a dynamical description from LQCD. But it can give insights into the order of the phase transition, the critical energy density and the equation of state \cite{Fodor:2001pe, de_Forcrand:2006hh, Stephanov:2007fk}.
These are needed as input to hydro- and thermodynamical models.

Due to the lack of lattice data at high baryo-chemical potentials, I will later use the simple phenomenological MIT bag model to describe the confinement and the phase transition.
\subsubsection{Fragmentation (pQCD)}
The hadron production at sufficiently large momentum transfer can be described by pQCD. The quark-antiquark pairs created in hard scatterings are connected by a color string and as they depart from each other back to back, the potential energy between them rises until the string breaks and creates a new quark-antiquark pair. This process continues until most of the kinetic energy of the original parton is converted. The hadrons, formed from the created quarks, carry a momentum fraction of $P_a$ and are all emitted in the direction of the original parton. This is called a jet.

The probability that the jet of parton $q$ emits a hadron $h$ with a momentum fraction $z$ is called fragmentation function $D_{q\rightarrow h}(z)$ and can be measured in $e^+e^-$ annihilations.  The invariant cross section for a hadron $h$ with momentum $P$ can then calculated with
\begin{align}
 E \dfrac{\dn\sigma_h}{\dn^3P} = \sum_a \int_0^1 \dfrac{\dn z}{z^2} D_{q\rightarrow h}(z) \dfrac{\dn \sigma_q}{\dn^3 P_q}
\end{align}
For proton-proton collisions the situation is comparable to the $e^+e^-$ annihilations, but in heavy ion collisions (A-A collision) with a large fireball volume, the jets lose energy as they travel through the medium and will be modified. This modification is prominently observed in the so-called away-side jet supression: When two partons are created near the edge of the fireball, one jet can be emitted directly into vacuum while the other one has to travel through the whole dense medium and will therefore be suppressed.

But for a dense medium, the description in terms of a jet from a single quark is questionable and one has to account for multiple parton fragmentation (higher twist). In the extreme case of a very dense phase space with abundant quarks, they might simply recombine. That means e.g. if a $u$ and $\bar{d}$ quark are near in phase space, they are confined to form a $\pi^+$. This recombination mechanism plays an important role in the mid-$p_T$ range while it is dominated by fragmentation at high-$p_T$. I will dicuss the recombination (also called coalescence) approach in the next chapter.

\subsubsection{MIT bag model}
\label{sec:hic-mit_bag}
In the MIT bag model \cite{Chodos:1974je,Chodos:1975ix, Myhrer:1983wx, Theberge:1980ye} the quarks and gluons are confined in volume with radius $R$ equal to the radius of a nuclei. Outside of this volume, the virtuell quark-antiquark-pairs excert a pressure $B$ on this bag, which prevents the partons inside to escape. They can be described by thermodynamics as ideal gas.

The pressure of this gas depends on the temperature and the density which is expressed by the chemical potential $\mu_B$. If this pressure becomes greater than the bag pressure, the partons inside can escape their confinement.

So the phase boundary \footnote{The derivations of this formula can be found in the appendix~\ref{app:bag_model}.} within the bag model is at 
\begin{align}
T = \sqrt{\dfrac{1}{C}\left[-(\dfrac{1}{3}\mu_B)^2g_q+\sqrt{(\dfrac{1}{3}\mu_B)^4 g_q \left(g_q-\dfrac{C}{\pi^2}\right) + T_C^4 C^2}\right]}
\label{eqn:mit_bag_model}
\end{align}
with $C=\dfrac{\pi^2}{15}\left(7g_q+4g_g\right)$ and the critical temperature $T_C=175\MeV$ at zero chemical potential.

\subsection{Transport theory}
The transport model \cite{nucl-th/9803035,hep-ph/9909407} is a microscopic approach which models the trajectory of every single particle. The evolution of the system is calulated with the Boltzmann equation
\begin{align}
 p^\mu \partial_\mu f(x,p) = \dfrac{\partial f}{\partial t} = S_{\mbox{coll}}
\label{eqn:boltzmann_equation}
\end{align}
where the collision term depends on the cross-sections and determines the time-dependence of the distribution function. Since only two-body collisions are considered in the gain and loss terms, one has to assume a dilute gas of particle with large mean free path so that three- or more-body collisions are neglectable. But the advantage is that since it is a microscopic theory, it can describe non-equilibrium systems.

\subsection{Hydrodynamics}
The complement to transport theory is hydrodynamics, where one assumes a local equilibrium within a dense system, e.g. a very small mean free path of the particles like in a fluid. It is a very common model in HIC \cite{nucl-th/0305084,hep-ph/0006129}, especially, since the matter created at the Relativistic Heavy Ion Collider (RHIC) seems to behave like a perfect fluid (minimal viscosity) \cite{Romatschke:2007mq, Tannenbaum:2006ch}.

With the equilibrium assumption, we can set the collision term in eq.~\eqref{eqn:boltzmann_equation} to zero, and with the baryon number current $N^\mu$ and the energy-momentum tensor $T^{\mu\nu}$, the basic formulas read
\begin{align}
 \partial_\mu N^\mu =& 0,\\
\partial_\mu T^{\mu\nu} =& 0.
\end{align}
State-of-the-art calculations are 3-dimensional with three distinct fluid, e.g. two for the colliding nuclei and a third for the evolving collision zone. For simpler systems the equations can be solved analytically which has been done by Bjorken \cite{Bjorken:1982qr} and Landau \cite{Landau:1953gs}.

The hydro results depend on the needed input, namely the initial conditions (Glauber model or CGC), the equation of state (LQCD) and the freeze-out prescription.

\subsubsection{Collision geometry}
The interpretation of the time evolution of the system in the transverse plane is motivated by hydrodynamical ideas. The high compression of the nuclear matter generates a high pressure in the fireball. The difference of the inner medium and the surrounding vaccum results in a pressure gradient. This leads to an expansion in the transverse direction. The initial transverse area, determined by the overlap zone of the nuclei, increases rapidly until the freeze-out. So the spectrum of the particles can be described by the emission from a boosted thermal source, where the transverse velocity of the particles comes from the thermal motion with a superimposed transverse boost.

Based on these ideas, the particle spectra are fitted by the so-called blast-wave model that describes a thermal source with a transverse and a longitudinal boost. Because the longitudinal rapidity is generally taken to be equal to the space-time rapidity by invoking Bjorkens longitudinal boost invariance, the model is characterized by three parameters namely temperature $T$, baryo-chemical potential $\mu_B$ and transverse expansion rapidity $\eta_T$.

\begin{figure}[ht]
\centering
\subfigure[\label{fig:impactparameter}]{\includegraphics[scale=0.7]{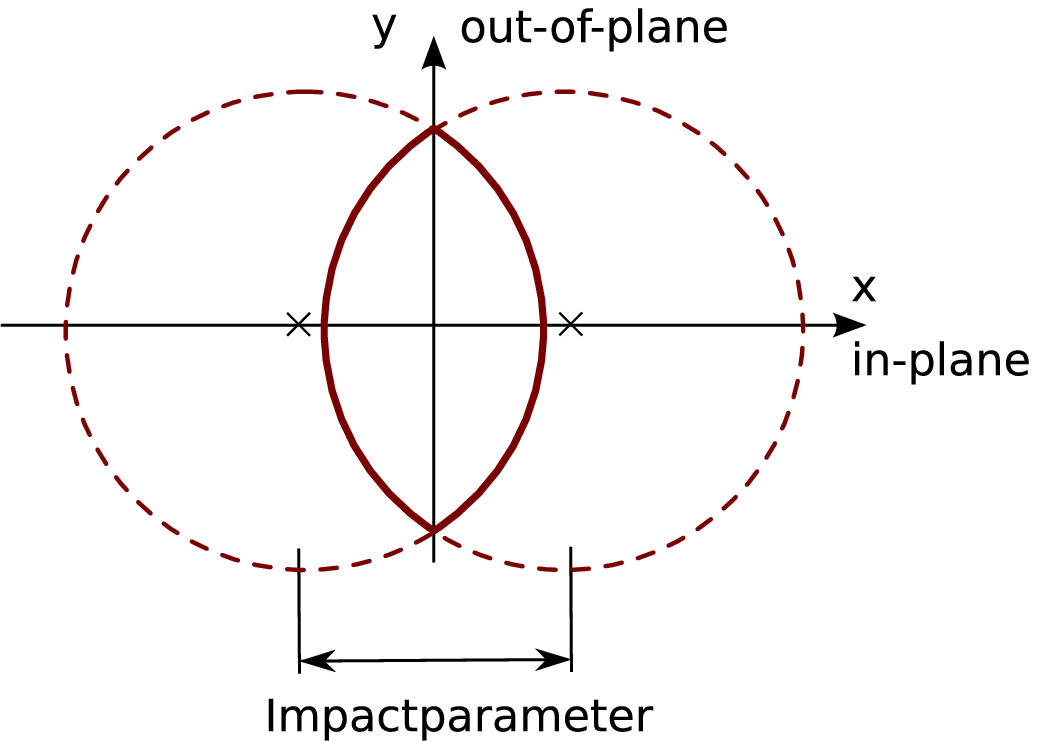}}
\subfigure[\label{fig:inital_asymmetry}]{\includegraphics[scale=0.7]{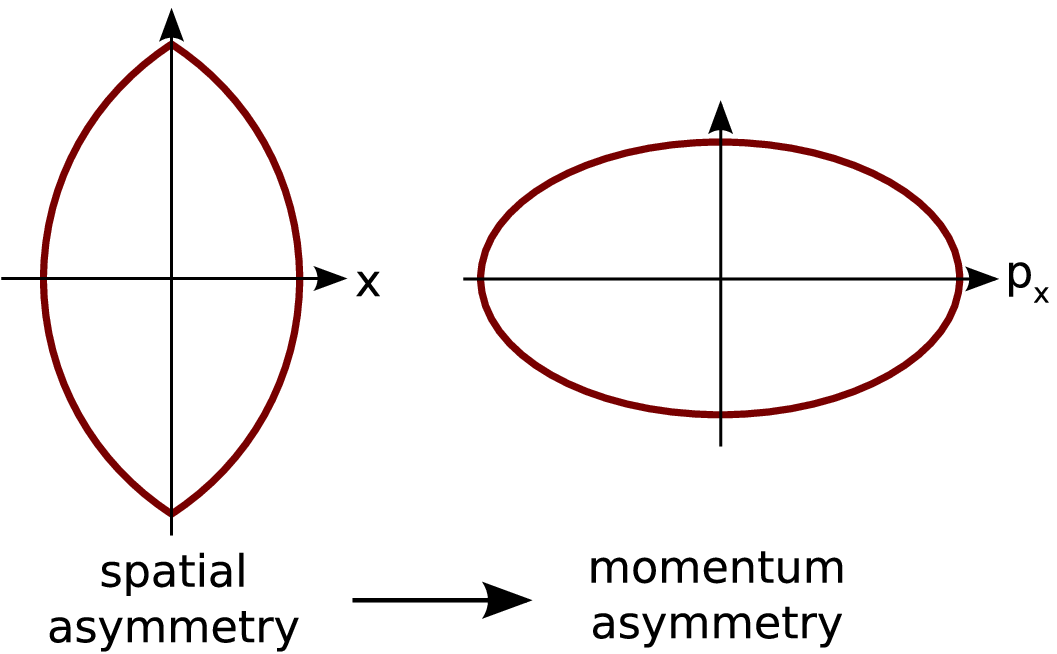}}
\subfigure[\label{fig:expansion}]{\includegraphics[scale=0.7]{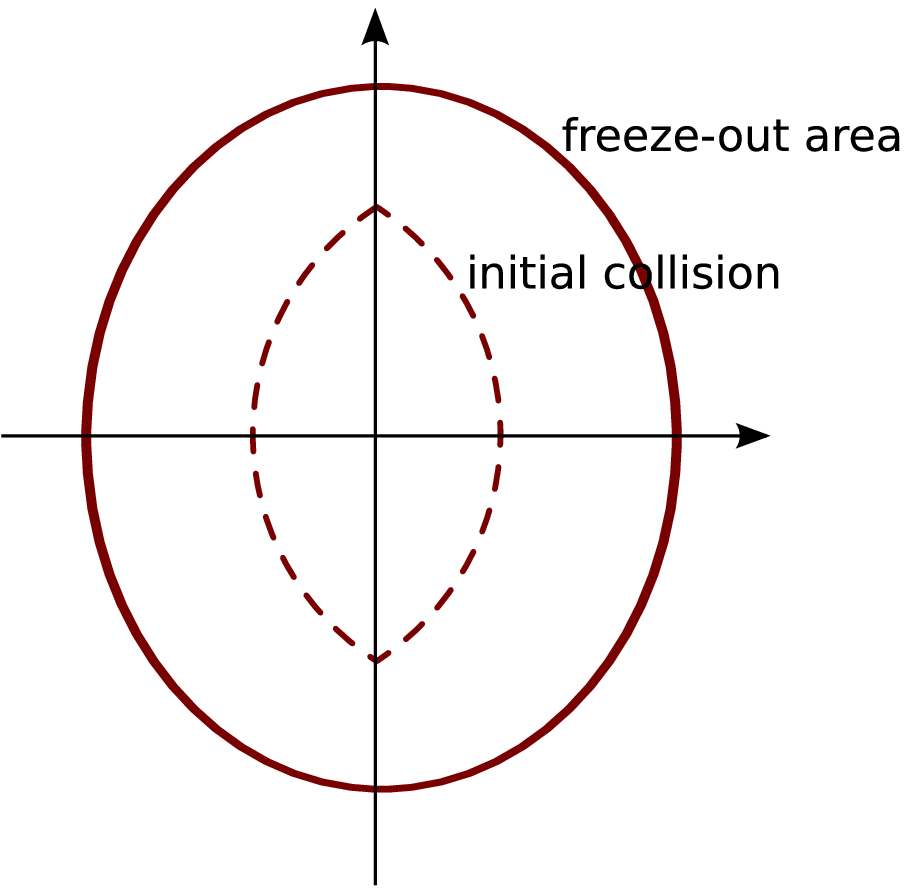}}
 \caption[Collision geometry in the transverse plane]{Collision geometry in the transverse plane. a) Impactparameter.
b) The spatial asymmetry is converted into a perpendicular momentum asymmetry due to the different pressure gradients.
c) The initial spatial asymmetry weakens during the expansion, so the eccentricity at freeze-out is smaller than the initial one.}
\label{fig:collision_geometry}
\end{figure}

For collisions with non-zero impact parameter the radial size and therefore the pressure gradient varies with the azimuthal angle $\varphi$. It is defined with respect to the plane in which the nuclei collide. The direction with $\varphi=0$ or $\varphi=\pi$ is called ``in-plane'' and perpendicular to that ``out-of-plane'' (Fig.~\ref{fig:impactparameter}). The protons and neutrons in the collision area that will take part in the scattering are called ``participants'', while the other nucleons, which just pass by, are called ``spectators''.

The radial size in-plane is smaller than out-of-plane which leads to a larger in-plane pressure gradient. The spatial azimuthal asymmetry therefore generates a momentum asymmetry that is perpendicular to the spatial one (Fig.~\ref{fig:inital_asymmetry}).

During the expansion, the increasing momentum asymmetry weakens the spatial asymmetry and so destroys its own origin; it is self-quenching. So the spatial eccentricity at freeze-out will differ from the initial one (Fig.~\ref{fig:expansion}). Depending on the hydrodynamical evolution and the freeze-out time it can be smaller, zero or even negative. A negative freeze-out eccentricity would mean a change of direction from an elongation out-of-plane to in-plane.

\chapter{Hadronization from a QGP}

In the previous chapter, I shortly discussed different theories for the description of heavy ion collisions and the phase diagram of the strong force. Currently, the collision dynamics are studied extensively within microscopical transport theory and hydrodynamics while the location and nature of the phase transition is subject to research in lattice QCD.

The main question of my thesis is, assuming we have crossed the phase border in a HIC and created a QGP, how do the quarks form hadrons again when the plasma cools down. Therefore I will analyze an analytic model for the colinear recombination approach \cite{Fries:2003kq}. Beside studying probes from the early phase of the collision, this offers the possibility to study the influence of a QGP creation on different observables at the freeze-out.

This chapter has the following structure:
\begin{itemize}
 \item The first section contains the recombination formalism,
 \item the second one addresses the form of the hypersurface where the phase transition and freeze-out take place,
\item  the third one introduces the blast-wave model which models the quark densities depending on the collision geometry and
\item the fourth section will contain the formulas for the observables from the recombination approach with the quark density and freeze-out hypersurface as dicussed.
\end{itemize}

\section{Colinear recombination of quarks}
\subsection{Introduction}
The results from the Relativisitic Heavy Ion Collider (RHIC) are a confirmative sign for a very dense phase space so that recombination seems to play a major role in the hadronization process. While the strong nuclear suppression (ratio of A-A to p-p collisions) of pions is widely seen as the experimental confirmation of jet quenching predicted by pQCD, there are several key observations that cannot be explained by fragmentation. These are the very different results for mesons and baryons namely
\begin{itemize}
 \item the nuclear suppression factor,
\item the elliptic flow and
\item the large baryon / meson ratios.
\end{itemize}
For collisions with a higher center-of-mass energy, the influence of recombination can be expected to be even higher. The predictions for the Large Hadron Collider (LHC) are therefore of great relevance.

I will now discuss the implications of an analytical coalescence formalism that constrains the recombining quarks to have colinear momenta. For a detailed discussions of the derivations I refer to \cite{Fries:2003kq}. While the onset of fragmentation sets the upper $P_T$ bound of applicability, the colinearity constraint leads to an energy violation at low $P_T$. Therefore, the results for the transverse momentum spectra of this recombination model will be valid only for $P_T \gtrsim 1 \GeV$. But the most prominent and promising observable of recombination, namely the elliptic flow $v_2$, can be expected to give valid results down to several hundred $\MeV$ \cite{Fries:2003kq}, since the azimuthal angle dependence in these flow coefficients is only expressed relatively to the absolute yield. And also the very good description of the data (see sec.~\ref{sec:elliptic_flow_results}) justifies the use down to low $p_T$. So such an analytical model can still serve as a good guidance in studying the general features and point the way for dynamical studies.

\subsection{Non-relativistic model}
Let us start with a volume V with a homogen distribution $w_a(p)$ of various quarks $a$ which will therefore only depend on the momentum. We assume that only constituent quarks with colinear momentum $p_a$ will recombine to form a hadron with momentum $p_h = \sum_a p_a$. This simple picture then leads us to a first non-relativistic equation for the hadron multiplicity which is the momentum integral over the product of the quark densities times the probability for these quarks to form a hadron (namely the wavefunctions overlap squared). For simplicity, I will stick to mesons (quark-antiquark pairs) and will give the generalization to baryons later. The spatial wavefunctions for the quarks and the meson $M$ are:
\begin{equation}
\braket{x}{q_1,p_1;q_2, p_2} = V^{-1} e^{i(p_1 x_1 + p_2 x_2)}
\end{equation}
\begin{equation}
\braket{x}{M,P_M} = V^{-1/2} e^{i(\frac{x_1+x_2}{2} \cdot P_M)} \phi_M (x_1-x_2)
\end{equation}
Defining the center of mass and relative coordinates resp. momentum as $R = (x_1+x_2)/2$ and $y=x_1-x_2$ resp. $P = (p_1+p_2)/2$ and $q=p_1-p_2$, the overlap is
\begin{align}
&\braket{q_1,p_1;q_2, p_2}{M,P_M}\nonumber\\
= &V^{-3/2} \int \dn^3 x_1 \dn^3 x_2 \exp{i\left(P_M\cdot R - p_1 x_1 - p_2 x_2\right)} \phi_M (y)\nonumber\\
= &V^{-3/2} \int \dn^3 R \dn^3 y \exp{i\left(\left(P_M-P\right)R - q y\right )} \phi_M (y)\nonumber\\
= &(2\pi)^3 V^{-3/2} \delta^3\left(P_M-P\right) \phi_M (q)
\end{align}
$\phi_M(q)$ is then the fourier transform in the relative momentum. Due to the absolute value squared, the deltafunction has to be rewritten as
\begin{equation}
\delta^3(P_M-P) = \dfrac{1}{(2\pi)^3} \int \dn^3 x \exp{i\left(P_M-P\right)x} = \dfrac{V}{(2\pi)^3}
\end{equation}
Using this in the proposed formula for the hadron multiplicity
\begin{equation}
N_M = C_M V^3 \int \frac{\dn^3 P}{(2\pi)^3} \frac{\dn^3 p_1}{(2\pi)^3} \frac{\dn^3 p_2}{(2\pi)^3} w_{q_1}(p_1)  w_{q_2}(p_2)
\abs{\braket{q_1,p_1;q_2, p_2}{M,P}}^2
\end{equation}
with a spin degeneracy factor $C_M$ (mesons) or $C_B$ (baryons) respectively, one finds the momentum distribution
\begin{equation}
\dfrac{\dn N_M}{\dn^3P} = C_M \dfrac{V}{(2\pi)^3} \int \frac{\dn^3 q}{(2\pi)^3} w_{q_1}(\frac{P}{2}+q)  w_{q_2}(\frac{P}{2}-q)
\abs{\phi_M(q)}^2
\end{equation}
Assuming an exponential parton distribution $w_a(p)=\gamma_a \exp{-\dfrac{p}{T}}$, the $q$-dependence will drop from the density product. Due to the normalization of the wavefunction, the integral vanishes and one is left with
\begin{equation}
\dfrac{\dn N_M}{\dn^3P} = C_M \dfrac{V}{(2\pi)^3} \gamma_a \gamma_b \exp{-\dfrac{P}{T}}
\end{equation}
The spectrum does not depend on the form of the wavefunction, it is just exponential. The only dependence on the specific hadron is in the degeneracy factor and the quark fugacities $\gamma$.

Even in this simple approach one can explain the large baryon/meson ratio at RHIC, which pQCD fails to predict. It simple comes down to the degeneracies
\begin{equation}
\dfrac{\dn N_B}{\dn N_M} \approx \dfrac{C_B}{C_M}
\end{equation}
which then results a proton-to-pion ratio of about 2, in contrast to about 0.2 from pQCD fragmentation.

\subsection{Relativistic}
To make the approach more general and consistent, one starts from a density matrix $\rho_{ab}$ for the quarks $a$ and $b$. The number of mesons is then given by
\begin{equation}
N_M = \sum_{a.b} \int \dfrac{\dn^3 P}{(2\pi)^3}\bra{M,P} \rho_{ab} \ket{M,P}
\end{equation}
Inserting complete sets of coordinates
\begin{equation}
N_M = \sum_{a,b} \int \dfrac{\dn^3 P}{(2\pi)^3} \dn^3 \hat r_1 \dn^3\hat r_2 \dn^3\hat r_1' \dn^3\hat r_2'
\braket{M;P}{\hat r_1,\hat r_2} \bra{\hat r_1,\hat r_2} \rho_{ab} \ket{\hat r_1',\hat r_2'} \braket{\hat r_1',\hat r_2'}{M,P}
\end{equation}
and using the definition of the wigner function
\begin{equation}
\bra{\hat r_1,\hat r_2} \rho_{ab} \ket{\hat r_1',\hat r_2'} = \int \dfrac{\dn^3 p_1}{(2\pi)^3}\dfrac{\dn^3 p_2}{(2\pi)^3} \; \exp{-i \left(p_1 r_1' + p_2 r_2'\right)} W_{ab}(r_1,r_2;p_1,p_2).
\end{equation}
With the center of mass (c.m.) coordinate $r_{1,2} = \left(\hat r_{1,2} + \hat r_{1,2}'\right) /2$ and the relative coordinate $r_{1,2}' = \hat r_{1,2} - \hat r_{1,2}'$ of the correlation coordinates, one can define the relevant quantities for the hadron. By assuming small spatial variations in the wigner function, after some steps one arrives at
\begin{align}
 E\dfrac{\dn N_M}{\dn^3 P} =& C_M \int_{\Sigma} \dfrac{\dn^3 R \, P_\mu u^\mu(R)}{(2\pi)^3} \int \dfrac{\dn^3 q}{(2\pi)^3} \nonumber\\
&\times w_a(R; P/2-q) \Phi_M(q) w_b(R;P/2+q)
\end{align}
with the relative momentum $q$ between the quarks, the spatially integrated wigner function of the meson $\Phi_M(q)$ and the future orientated unit vector $u^\mu(R)$ on the freeze-out hypersurface $\Sigma$. When going to local light cone (LLC) coordinates, one can restrict the quark momentum to colinearity by writing it as a fraction $x_a=\dfrac{p_a}{P_H}$ of the hadron momentum. By introducing the ansatz for the light cone wavefunction
\begin{align}
\Phi(q) \overset {\mbox{LLC}}{\rightarrow} \psi(x) = N\prod_a x_a =
\begin{cases}
 x &\mbox{for quarks}\\
\sqrt{30}\, x_1 x_2 &\mbox{for mesons}\\
12\,\sqrt{35}\,x_1 x_2 x_3 \quad&\mbox{for baryons}\\
\end{cases},
\label{eqn:realistic_wavefunctions}
\end{align}
the formula for the invariant yield becomes
\begin{align}
 E\dfrac{\dn N}{\dn^3 P} =& C \int_{\Sigma} \dfrac{P_\mu \dn\sigma^\mu(R)}{(2\pi)^3} \int \Dn \hat x
\abs{\phi(\hat x)}^2\times \prod_a w_a(R; x_a P)
\label{eqn:invariant_yield_general}
\end{align}
with $\hat x = (x_1, x_2, .., x_{n_q})$, the number of constituent quarks $n_q$ and
\begin{align}
 \int \Dn \hat x = \int_0^1 \delta\left(1-\sum_a x_a\right)\, \prod_a \dn x_a
\end{align}
In the case of quarks this is simplified to
\begin{align}
 E\dfrac{\dn N_q}{\dn^3 p} =& g \int_{\Sigma} \dfrac{p_\mu \dn\sigma^\mu(R)}{(2\pi)^3}  w(R; p)
\label{eqn:invariant_yield_quark}
\end{align}

As already shown in the previous section, the recombination formalism alone is capable of explaining the large baryon/meson ratio, but the results will generally depend strongly on the parameterizations of freeze-out hypersurface $\Sigma$ and the used quark density $w_a(R;p)$.

\clearpage
\section{Freeze-out}
\label{sec:freeze-out}
Now I will model the freeze-out hypersurface. The spatial four-vector
\begin{align}
 x_\nu &= \left(t,x,y,z\right)
\end{align}
will be translated into the new coordinates
\begin{align}
x'_\nu &= \left(\tau,\rho,\varphi,\eta\right)
\end{align}
with the transverse radial variable $\rho$, its angle $\varphi$, the time $\tau = \sqrt{t^2-z^2}$ and the space-time rapidity
\begin{align}
 \eta = \dfrac{1}{2} \ln \dfrac{t-z}{t+z}.
\end{align}
For the study of collisions with a nonzero impact parameter, I will directly generalize to an ellipsoidal cylinder. The elliptic transverse freeze-out area
\begin{align}
 r = \sqrt{\dfrac{x^2}{R_x^2}+\dfrac{y^2}{R_y^2}}
\end{align}
has the numerical eccentricity
\begin{align}
 \eps_{\mbox{numerical}} := \dfrac{\sqrt{R_y^2-R_x^2}}{R_y} \equiv e_f
\end{align}
if $R_y > R_x$. To avoid ambiguities with other eccentricity definitions, the numerical eccentricity of an ellipse with the above definition will be called $e$ while the $f$ denotes the freeze-out value.

So the surface radius depends on the angle $\varphi$, which is measured with respect to the x-axis:
\begin{align}
 r(\varphi) &= \rho f(\varphi)\quad \mbox{with}\\
f(\varphi)&=\dfrac{N}{\sqrt{1-e_f^2\sin^2(\varphi)}}\nonumber\\
&= \dfrac{N\alpha^{-1}}{\sqrt{\alpha^2\sin^2(\varphi)+\alpha^{-2}\cos^2(\varphi)}} \label{eqn:fo_eccent_f1}\\
&= \dfrac{N\alpha^{-1}}{\sqrt{(\alpha^{-2}+\alpha^2)+\cos(2\varphi)(\alpha^{-2}-\alpha^2)}}\label{eqn:fo_eccent_f2}\\
\mbox{with}\quad \alpha &= \sqrt[4]{1-e_f^2}
\end{align}
The radial coordinate $\rho$ is constant for a given ellipsoid surface. The length of the short axis $x$ (in-plane) and respectively the long axis $y$ (out-of-plane) are
\begin{align}
 R_x = r(0)=N \rho\\
 R_y=r(\pi/2)=\dfrac{N\rho}{\sqrt{1-e_f^2}} = \dfrac{N\rho}{\alpha^2}
\end{align}
The constant $N$ is chosen such that the transverse area $A_T^f$ is independent of the eccentricity and only depends on the parameter $\rho$. From
\begin{align}
 A_T^f = \pi R_x R_y = N^2\dfrac{\pi \rho^2}{\alpha^2}
\label{eqn:A_T^f}
\end{align}
if follows
\begin{align}
 N &= \alpha,\\
R_x &= \alpha \rho \quad \mbox{and} \quad R_y = \dfrac{\rho}{\alpha}
\end{align}
Looking at eq.~\eqref{eqn:fo_eccent_f1}, one directly sees that the axes are changed under the transformation $a \rightarrow a^{-1}$ and the ellipsoid is flipped by $90^\circ$. This will be used in the next paragraph.

\subsection{Integration measure}
\label{sec:integration_measure}
To obtain a 3-dimensional freeze-out hypersurface, one needs to apply a constraint to the four-vector $x_\nu$ or respectively $x'_\nu$. This is done by constraining the freeze-out time $\tau$ to be a function of the other three variables, so generally $\tau=\tau(\rho,\varphi,\eta)$.

The simplest choice would be to use a constant freeze-out time with $\tau = \tau_0$, but to be a bit more general and to connect the transverse with the longitudinal dynamics, I will perfom the derivations with an explicit $\rho$ dependence only.

According to the Cooper-Frye prescription \cite{Cooper:1974mv} the integration measure is given by
\begin{align}
p_\nu \dn\sigma'^\nu
\end{align}
where $\dn \sigma'_\nu$ is the normalvector on the hypersurface in the new coordinates which is calculated by a generalized cross-product of the tangent vectors. Therefore, I will calculate the Jacobian
\begin{align}
 J_{\mu i} = \dfrac{\partial x_\mu}{\partial x'^i}
\end{align}
With the dependencies
\begin{subequations}
\label{eqn:fo_coordinates}
\begin{align}
 t &= \tau(\rho) \cosh(\eta),\\
x &= \rho f(\varphi) \cos(\varphi),\\
y &= \rho f(\varphi) \sin(\varphi)\quad\mbox{and}\\
z &= \tau(\rho) \sinh(\eta)
\end{align}
\end{subequations}
the tangent vectors read
\begin{subequations}
\begin{align}
\rho_\mu :=& J_{\mu 1} = \left(\dfrac{\partial \tau}{\partial \rho} \cosh\eta, f \cos\varphi, f \sin\varphi ,\dfrac{\partial \tau}{\partial \rho} \sinh\eta\right)\\
\varphi_\mu :=& J_{\mu 2} = \left(0, \rho (-f\sin\varphi+f'\cos\varphi), \rho (f\cos\varphi+f'\sin\varphi),0\right)\\
\eta_\mu :=& J_{\mu 3} = \left(\tau\sinh \eta, 0, 0 , \tau\cosh\eta\right)
\end{align}
\end{subequations}
where
\begin{align}
 f' = \partial_\varphi f = f^3(\varphi) \dfrac{\alpha^{-2}-\alpha^2}{2} \sin(2\varphi).
\end{align}

With the fully anti-symmetric rank-4 tensor $\eps_{\nu\mu\lambda\sigma}$, the infinitesimal normalvector on the hypersurface is calculated as
\begin{align}
 \dn\sigma'_\nu &= \eps_{\nu\mu\lambda\sigma}\; \rho^\mu\dn\rho\; \varphi^\lambda \dn\varphi \; \eta^\sigma \dn\eta\nonumber\\
&= \tau \rho \left(f^2 \cosh\eta, \partial_r \tau \left(f\cos\varphi+f'\sin\varphi\right), \partial_r \tau \left(f\sin\varphi-f'\cos\varphi\right), f^2 \sinh\eta\right) \; \dn\rho\dn\varphi\dn\eta
\end{align}
Similiar to eq.~\eqref{eqn:fo_coordinates}, the particle momentum is parameterized as
\begin{align}
 p_\nu = \left(m_T \cosh y, p_T \cos(\phi), p_T \sin(\phi),m_T \sinh y\right)
\end{align}
which then leads to the integration measure
\begin{align}
 p_\nu \dn \sigma'^\nu =& \dn\rho\dn\varphi\dn\eta \tau(\rho) \rho \nonumber\\
& \left[ f^2(\varphi) m_T \cosh(y-\eta) - \dfrac{\partial \tau}{\partial \rho} p_T 
	\left(f \cos(\varphi-\phi)+f' \sin(\varphi-\phi)\right) \right].
\label{eqn:integration_measure}
\end{align}
If I had assumed a constant $\tau$ independent of $\rho$, the second term would vanish and the integration measure would simplify to
\begin{align}
 p_\nu \dn \sigma'^\nu = \dn\rho\dn\varphi\dn\eta \tau_0 \rho 
f^2(\varphi) m_T \cosh(y-\eta),
\label{eqn:integration_measure_tau_const}
\end{align}
By additionally setting $e_f=0$, one recovers the measure $\tau_0 \rho \dn\rho \; \dn\varphi \dn\eta\; m_T \cosh(y-\eta)$ for a cylindrical freeze-out, since
\begin{align}
e_f = 0 \; \follows \alpha = 1 \; \follows f(\varphi)=1 \; \follows r(\varphi)=\rho=R_x=R_y
\end{align}

\subsection{Surface flow}
\label{sec:surface_flow}
For an ellipsoid the radial direction $\varphi$ is generally not perpendicular to the surface. It is related to the orthogonal direction $\beta$ by
\begin{align}
 \tan \beta = \left(1-e_f^2\right) \tan \varphi = \alpha^4 \tan \varphi.
\end{align}
So for an elliptical freeze-out area I consider the two extreme cases:
\begin{itemize}
 \item Model 1: the system expands perpendicular to the surface
\item Model 2: the system expands radially outwards independent of the surface orientation
\end{itemize}
\begin{figure}
 \centering
 \includegraphics{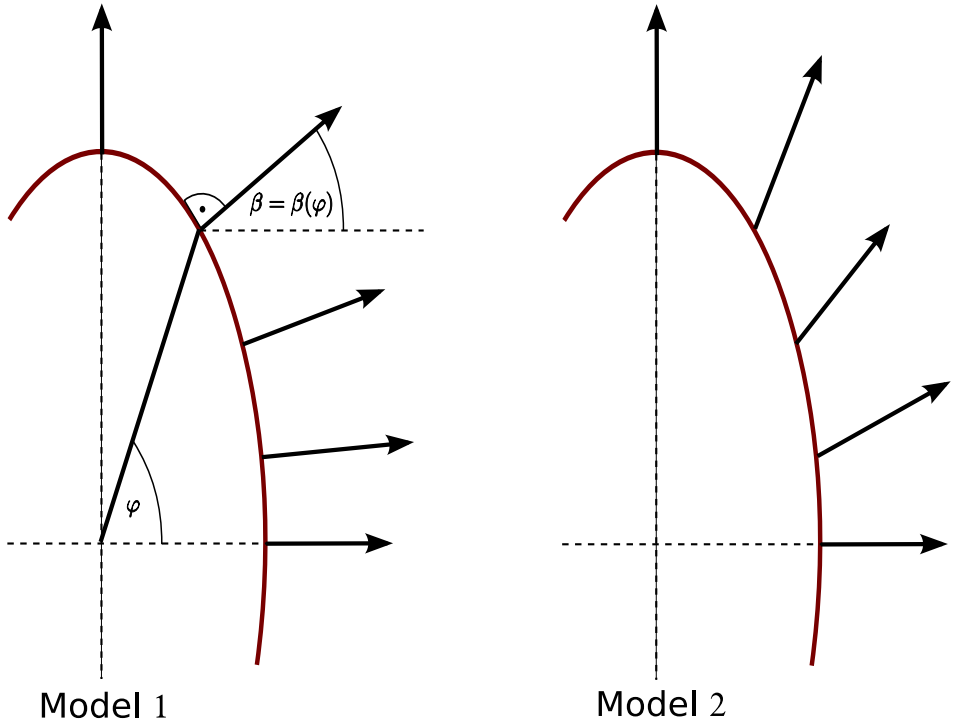}
 \caption[The two extreme cases for the expansion of fireball.]{The two extreme cases for the expansion of fireball. Left: The system expands perpendicular to the surface. Right: The system expands radially outwards.}
 \label{fig:surface_flow_cases}
\end{figure}
Fig.~\ref{fig:surface_flow_cases} depicts the possibilities. When I drop the $\rho$ dependence of $\tau$ (eq.~\eqref{eqn:integration_measure_tau_const}) both models can be related via the integration measure \cite{nucl-th/0409074} as can be shown by rewritting the $x$ and $y$ coordinate in eq.~\eqref{eqn:fo_coordinates} using the angle $\beta$:
\begin{align}
 x'' =& \rho f(\varphi) \cos\varphi\nonumber\\
 =& \rho \dfrac{\cos\varphi}{\sqrt{\alpha^2\sin^2(\varphi)+\alpha^{-2}\cos^2(\varphi)}}\nonumber\\
 =& \rho \dfrac{\sign\left(\cos\varphi\right)}{\sqrt{\alpha^2\tan^2(\varphi)+\alpha^{-2}}}\nonumber\\
 =& \rho \dfrac{\sign\left(\cos\beta\right)}{\sqrt{\alpha^{-6}\tan^2(\beta)+\alpha^{-2}}}\nonumber\\
 =& \rho \dfrac{\alpha^2\cos\beta}{\sqrt{\alpha^{-2}\sin^2(\beta)+\alpha^2\cos^2(\beta)}}\nonumber\\
 =& \rho \left(f_{\alpha \rightarrow \alpha^{-1}} (\beta)\right) \alpha^2\cos\beta
\end{align}
and similar
\begin{align}
 y'' =& \rho f (\varphi) \sin\varphi\nonumber\\
 =& \rho \left[f_{\alpha \rightarrow \alpha^{-1}} (\beta)\right] \alpha^{-2}\sin\beta
\end{align}
The integration measure becomes
\begin{align}
p_\nu \dn \sigma''^\nu &= \dn\rho\dn\beta\dn\eta \tau_0 \rho 
\left[f^2_{\alpha \rightarrow \alpha^{-1}} (\beta)\right] m_T \cosh(y-\eta),
\end{align}
which is similiar to the one from eq.~\eqref{eqn:integration_measure_tau_const}. When renaming $\beta \rightarrow \varphi$, the only difference is the transformed $\alpha$. As already explained above, this transformation is a rotation of the ellipse by $90^\circ$. So the integration over an ellipse elongated out-of-plane with the system expanding perpendicular to the surface is equal to the integration of an in-plane ellipse with a radial expansion. But this exact correspondence is only true if $\partial_\rho \tau = 0$. The consequences of this relation are discussed in section~\ref{sec:geometrical_contrib}.

\clearpage
\section{Blast Wave Model}
\label{sec:blast_wave}
For the parameterisation of the partonic density I take inspiration by hydrodynamical ideas \cite{nucl-th/9307020, nucl-th/0410081} and therefore assume a locally thermalised fireball. With the four-momentum $p_\nu$ of the quark $a$ and the flow four-velocity of the system $u_\nu$, the thermal density for fermions with the degeneracy factor $C$ reads
\begin{align}
 w_a(R) &= C \dfrac{1}{\exp{\left(p_\nu u^\nu-\mu\right)/T}-1}\nonumber\\
&= \exp{-\left(p_\nu u^\nu-\mu\right)/T} \dfrac{1}{1-\exp{-\left(p_\nu u^\nu-\mu\right)/T}}\nonumber\\
&= \sum_{k=1}^\infty (-1)^{k+1}\exp{-k\left(p_\nu u^\nu-\mu\right)/T}
\end{align}
To simplify the derivations, I only use the lowest-order term which is just Maxwell-Boltzmann statistic. This is sufficient, since the higher order terms are neglectable as I have verified.

\subsection{Azimuthal momentum asymmetry}
\label{sec:momentum_asymmetry}
In hydrodynamics the transverse flow is generated by a pressure gradient
\begin{align}
\delta p(\rho,\varphi)=\dfrac{\Delta p}{\Delta r}= \dfrac{\Delta p(\rho)}{f(\varphi)\Delta \rho},
\label{eqn:pressure_gradient}
\end{align}
which will depend on the angle for non-central collisions. The spatial asymmetry (fig.~\ref{fig:inital_asymmetry}) has the initial numerical eccentricity
\begin{align}
e_i(b) = \dfrac{\sqrt{h(b)^2-w(b)^2}}{h(b)} \quad\mbox{with}\\
w(b) = R_A-\dfrac{b}{2}\quad\mbox{and}\\
h(b) = \sqrt{R_A^2-\left(\dfrac{b}{2}\right)^2}
\end{align}
of an ellipse with the long axis $h(b)$ (y-direction) and the short axis $w(b)$ (x-direction) which describes the collision in the transvese plane and only depends on the impact parameter $b$.
So in the x-direction there is a stronger pressure gradient as in the y-direction. This yields a larger expansion velocity of the system and this momentum asymmetry will be perpendicular to the spatial asymmetry. This expansion leads to a decrease in the spatial asymmetry and therefore a decelerated growth of the momentum asymmetry which is called a "self-quenched" behaviour, since the momentum asymmetry destroys its own origin.

The size of the initial transverse area $A_T^i$ is determined by the impact parameter:
\begin{align}
 A_T^i = \pi w(b) h(b) = \dfrac{\pi w(b)^2}{\sqrt{1-e_i^2}} = \dfrac{\pi \left(R_A-b/2\right)^2}{\alpha^2}
\label{eqn:A_T^i}
\end{align}

It is not a piori clear how the spatial eccentricity will evolve until the freeze-out, but it generally depends on the initial spatial eccentricity, so $e_f = e_f(e_i)$. Depending on the dynamics, the system could
\begin{itemize}
 \item freeze-out early and leave a remaining spatial freeze-out eccentricity $e_f < e_i$, where it is still elongated out-of-plane,
\item freeze-out so that the initial spatial eccentricity has been compensated by the expansion and it is now circular with $e_f = 0$
\item or freeze-out late so that the initial spatial eccentricity is overcompensated and reversed so it is elongated in-plane.
\end{itemize}
The second point is just a special case of first one. So one has to distinguish between the elongation of the spatial area at freeze-out. In section~\ref{sec:geometrical_contrib_strength} I will give evidence for the first option and also relate $e_i$ to $e_f$ to study the effects of a non-circular freeze-out area.

\subsection{Comparing asymmetry parameters}
\label{sec:asymmetry_parameters}
\begin{figure}
 \centering
\includegraphics{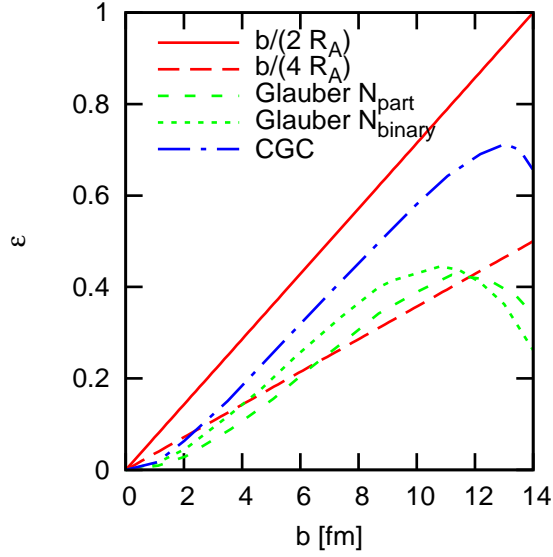}
\caption[Comparison of the impact parameter dependent eccentricity]{Comparison of the impact parameter dependent eccentricity $\eps$ from Glauber model and CGC \cite{nucl-th/0511046} calculations to an eccentricity from constant density (eq.~\ref{eqn:const_density_eccent})}
\label{fig:eccent_compare}
\end{figure}

I would like to state again that the used eccentricity $e$ is the numerical eccentricity of an ellipse while the eccentricity definition generally used in the heavy-ion community is \cite{nucl-ex/0008015}

\begin{align}
\eps = \dfrac{\left\langle y^2\right\rangle -\left\langle x^2\right\rangle }
{\left\langle y^2\right\rangle +\left\langle x^2\right\rangle }
\end{align}
where the average is over the transverse plane weigthed by the nucleon density. For a simple constant density this is simplified to

\begin{align}
 \eps_{\mbox{const}} = \dfrac{R_y^2-R_x^2}{R_y^2+R_x^2} = \dfrac{b}{2R_A}.
\label{eqn:const_density_eccent}
\end{align}

The definition of this (lets call it "geometric") eccentricity $\eps$ differs from the definition of the numerical eccentricity $e$. To compare one with the other, the above equation is written in terms of the numerical eccentricity. 
\begin{align}
\dfrac{b}{2R_A} \equiv \dfrac{e^2}{2-e^2}
\end{align}

Realistic calculations of the eccentricity are the ones from the glauber model \cite{Glauber:1970jm} which differs from the simple approximation $b/(2R_A)$, because it accounts for the thickness of the nuclei. Especially for very peripheral collisions, where only the sparsely populated parts of the nuclei collide, the glauber eccentricity is much smaller. Figure~\ref{fig:eccent_compare} shows a comparison for different conditions \cite{nucl-th/0511046}, including the Color Glas Condensate (CGC) results. The value of $b/(2R_A)$ is generally to large but $\eps_{\mbox{const}}/2=b/(4R_A)$ is in good agreement with the glauber model. Therefore, I define the eccentricity by the relation
\begin{align}
\eps:=&\dfrac{1}{2} \dfrac{e^2}{2-e^2} \equiv \dfrac{b}{4 R_A} \quad\mbox{with the inverted relation}
\label{eqn:eccent_definition}\\
e^2=&\dfrac{4\eps}{2\eps+1}
\label{eqn:eccent_definition_inverse}
\end{align}
The relevance of this redefined eccentricity becomes visible in the following paragraph.

\subsubsection{Expansion velocity}
Since the transverse flow is generated by a pressure gradient and I will not model the pressure profile, I take the ansatz from eq.~\eqref{eqn:pressure_gradient} and assume $\eta_T \sim \dfrac{\eta_T^0}{f(\varphi)}$ with the mean transverse flow rapidity $\eta_T^0$ as a free parameter. To have a much more handy expression, I rewrite the denominator and expand it to fourth order in $\eps$:
\begin{align}
 \dfrac{1}{f(\varphi)}&=\dfrac{\sqrt{1-e_i^2\sin^2(\varphi)}}{\alpha}\nonumber\\
 &=\dfrac{\sqrt{1-\dfrac{4\eps}{2\eps+1} 0.5\left(1-\cos(2\varphi)\right)}}{\sqrt[4]{1-\dfrac{4\eps}{2\eps+1}}}\nonumber\\
&\approx 1 + \left(\eps+\dfrac{11}{8} \eps^3\right) \cos(2\varphi)
+ \left(\dfrac{1}{4}\eps^2+\dfrac{9}{16} \eps^4\right) \cos(4\varphi)
+ \left(\dfrac{1}{8}\eps^3\right) \cos(6\varphi)\nonumber\\
&=:1+\tilde\eta_T(\varphi)
\label{eqn:asymmetry_expansion}
\end{align}
So the elliptic asymmetry $\cos(2\varphi)$ scales mainly linear with the defined eccentricity
\begin{align}
 \eps = \dfrac{\eps_{\mbox{const}}}{2},
\end{align}
which is comparable to the glauber eccentricity from central to mid-central collisions as can be seen in Figure~\ref{fig:eccent_compare}. This leads to a direct linear relation between the elliptic flow and the eccentricity which is also expect from hydrodynamics (see section~\ref{sec:eccent_scaling}) and therefore justifies the choice in eq.~\ref{eqn:eccent_definition}.

Together with a radial profil in $\rho$, the expression reads
\begin{align}
 \eta_T(\varphi,\rho) &= \eta_T^0 N_d \left(\dfrac{\rho}{\rho_0}\right)^{d_\rho} \left(1+\tilde\eta_T(\varphi)\right)
\label{eqn:transverse_rapidity}
\end{align}
The normalization $N_d$ is chosen to be $\dfrac{d_\rho+2}{2}$ so that
\begin{align}
\langle \eta_T \rangle=\dfrac{\int \dn \rho\, \rho \eta_T}{\int \dn\rho\, \rho} \overset{!}{=} \eta_T^0
\end{align}

Now the four velocity $u_\nu$ of the expanding fireball is created with a longitudinal boost $\eta_L$ and a transversal boost $\eta_T$. Taking Bjorkens boost-invariance argument, I set the longitudinal rapidity equal to the space-time rapidity, i.e. $\eta_L \equiv \eta$:
\begin{align}
 u_\nu =  \left(\cosh \eta_L \cosh\eta_T, \sinh\eta_T \cos\beta', \sinh\eta_T \sin\beta',\sinh\eta_L\cosh\eta_T\right)
\end{align}
The angle $\beta'$ depends on the model chosen for the surface expansion from section~\ref{sec:surface_flow} with $\beta'=\beta(\varphi)$ for model 1 (perpendicular expansion) and $\beta'=\varphi$ for model 2 (radial expansion).

\clearpage
\section{Observables from recombination}
\subsection{Invariant yield}
Now I combine the recombination formalism with the discussed freeze-out hypersurface (sec.~\ref{sec:freeze-out}) and the quark density from the blast-wave model (sec.~\ref{sec:blast_wave}). With eq.~\eqref{eqn:invariant_yield_quark} the invariant yield for the quarks read
\begin{align}
E\dfrac{\dn N_q}{\dn^3 p} =& \dfrac{\dn N_q}{p_T \dn p_T \dn \phi \dn y}= g \int \dfrac{p_\nu \dn \sigma'^\nu}{(2\pi)^3} w(R;p)\nonumber\\
 =& \dfrac{g}{(2\pi)^3} \int_0^{\rho_0} \dn \rho \int_0^{2\pi}\dn \varphi \int_{-\infty}^{+\infty}\dn \eta 
\tau(\rho) \rho\nonumber\\
&\times \left[ f^2(\varphi) m_T \cosh(y-\eta) - \dfrac{\partial \tau}{\partial \rho} p_T 
	\left(f \cos(\beta'-\phi)+f' \sin(\beta'-\phi)\right) \right]\nonumber\\
&\times\exp{-\left(p_\nu u^\nu-\mu\right)/T}
\end{align}
where the exponent expands to
\begin{align}
 p_\nu u^\nu = m_T \cosh(y-\eta) \cosh \eta_T - p_T \cos(\beta'-\phi) \sinh \eta_T
\label{eqn:local_frame_energy}
\end{align}
To study the dependence on the particles momentum angle $\phi$, one uses the fourier expansion
\begin{align}
\dfrac{\dn N}{p_T \dn p_T \dn \phi \dn y}
&= \tilde{v}_0
+ 2\sum_{n=1}^\infty \cos(n\phi) \tilde{v}_n
+\sin(n\phi) \tilde{u}_n\nonumber\\
&= \dfrac{\dn N}{2\pi p_T \dn p_T \dn y} \left(1+2\sum_{n=1}^\infty v_n \cos(n\phi)+u_n \sin(n\phi)\right)
\label{eqn:yield_fourier_expansion}
\end{align}
with 
\begin{align}
\tilde{v}_n &=\dfrac{1}{2\pi}\int\dn\phi \cos(n\phi) \dfrac{\dn N}{p_T \dn p_T \dn\phi \dn y},\\
\tilde{u}_n &=\dfrac{1}{2\pi}\int\dn\phi \sin(n\phi) \dfrac{\dn N}{p_T \dn p_T \dn\phi \dn y}
\label{eqn:def_fourier_coefficients}
\end{align}
The $\phi$-integrated yield $\tilde{v}_0 = \dfrac{\dn N}{2\pi P_T \dn P_T \dn y}$ describes the overall transverse momentum dependence, while the other fourier coefficients $\tilde{v}_n$ and $\tilde{u}_n$ encode the deviations from an azimuthal symmetric yield. To have a better comparison, the normalized fourier coefficients $v_n$ and $u_n$, also called flow coefficients, are introduced:
\begin{align}
v_n &:= \dfrac{\tilde{v}_n}{\tilde{v}_0} \equiv \left\langle \cos(n\phi) \right \rangle \quad\mbox{and}\\
 u_n &:= \dfrac{\tilde{u}_n}{\tilde{u}_0} \equiv \left\langle \sin(n\phi) \right \rangle
\label{eqn:def_flow_coefficients}
\end{align}

The full expression reads \footnote{How to solve the integrals can be found in the appendix~\ref{app:integral_flow_coefficients}.}
\begin{align}
\tilde{v}_n=& g\exp{\mu/T} \dfrac{4\pi \rho_0^2}{(2\pi)^3} \int_0^{2\pi} \dn\varphi \int_0^1 \dn \rho' \tau(\rho') \rho'\nonumber\\
&\times \Biggl[K_1 I_n f^2(\varphi) m_T \cos(n\beta')\nonumber\\
&\qquad - K_0 \dfrac{I_{n-1}+I_{n+1}}{2} f(\varphi) p_T \partial_\rho \tau \cos(n\beta')\nonumber\\
&\qquad - K_0 \dfrac{I_{n-1}-I_{n+1}}{2} f^3(\varphi) p_T \partial_\rho \tau \dfrac{\alpha^{-2}-\alpha^2}{2} \dfrac{\cos((n-2)\beta')-\cos((n+2)\beta')}{2}\Biggr]
% \tilde{u}_n=& \exp{\mu/T} \dfrac{4\pi \rho_0^2}{(2\pi)^3} \int_0^{2\pi} \dn\varphi \int_0^1 \dn \rho' \tau(\rho') \rho'\nonumber\\
% &\times \Biggl[K_1 I_n f^2(\varphi) m_T \sin(n\beta')\nonumber\\
% &\qquad - K_0 \dfrac{I_{n-1}+I_{n+1}}{2} f(\varphi) p_T \partial_\rho \tau \sin(n\beta')\nonumber\\
% &\qquad - K_0 \dfrac{I_{n-1}-I_{n+1}}{2} f^3(\varphi) p_T \partial_\rho \tau \dfrac{\alpha^{-2}-\alpha^2}{2} \dfrac{\sin((n-2)\beta')-\sin((n+2)\beta')}{2}\Biggr]
\end{align}
For the coefficients $\tilde{u}_n$, one has to replace the cosine terms with sinus. From that the $\phi$-integrated yield follows as
\begin{align}
\label{eqn:phi_integrated_yield_quark}
\dfrac{\dn N}{2\pi P_T \dn P_T \dn y}=&\tilde{v}_0\nonumber\\
=& g\exp{\mu/T} \dfrac{4\pi \rho_0^2}{(2\pi)^3} \int_0^{2\pi} \dn\varphi \int_0^1 \dn \rho' \tau(\rho') \rho'\nonumber\\
&\times \Biggl[K_1 I_0 f^2(\varphi) m_T - K_0 I_1 f(\varphi) p_T \partial_\rho \tau \Biggr]
\end{align}
The arguments for the modified bessel functions $K_m=K_m\left(k(\rho,\varphi,p_T)\right)$ and $I_m=I_m\left(i(\rho,\varphi,p_T)\right)$ are
\begin{subequations}
\label{eqn:bessel_arguments}
\begin{align}
\left[k\right]^{\mbox{quark}}(\rho,\varphi,p_T) =& \dfrac{m_T \cosh\eta_T(\varphi,\rho,p_T)}{T} \quad\mbox{and respectively}\\
\left[i\right]^{\mbox{quark}}(\rho,\varphi,p_T) =& \dfrac{p_T \sinh\eta_T(\varphi,\rho,p_T)}{T}
\end{align}
\label{eqn:quark_bessel_arguments}
\end{subequations}
Due to the symmetry of the system, most coefficients vanish. The only $\varphi$ dependence is in the function $f(\varphi)$ which only contains terms proportional to $\cos(2n\varphi)$, as can be seen in eq.~\eqref{eqn:asymmetry_expansion}. Therefore, the expectation values $\left\langle \sin(n\phi) \right\rangle$ and $\left\langle \cos\left((2n+1)\phi\right) \right \rangle$ vanish:
\begin{align}
 \tilde{u}_n = 0 = u_n\\
 \tilde{v}_{2n+1} = 0 = u_{2n+1}
\end{align}
This results in the symmetries
\begin{align}
 \dfrac{\dn N}{p_T \dn p_T \dn \phi \dn y} (\phi) =& \dfrac{\dn N}{p_T \dn p_T \dn \phi \dn y} (-\phi)\quad\mbox{and}\\
 \dfrac{\dn N}{p_T \dn p_T \dn \phi \dn y} (\phi) =& \dfrac{\dn N}{p_T \dn p_T \dn \phi \dn y} (\pi-\phi)
\end{align}

For the hadrons the integration over the product of the quark densities involves an additional integral 
over the momentum fractions $\hat x=(x_1,..,x_{n_q})$ times the wavefunction $\left\vert\psi(\hat x)\right\vert^2 $ (eq.~\eqref{eqn:invariant_yield_general}):
\begin{align}
\dfrac{\dn N}{P_T \dn P_T \dn \phi \dn y} = C \int_\Sigma \dfrac{p_\nu \dn \sigma'^\nu}{(2\pi)^3} \int \Dn\hat x \prod_a w_a(R;x_a P) \left\vert\psi(\hat x)\right\vert^2 
\end{align}
Due to the product $\prod_a w_a(R; x_a P)$, the arguments of the besselfunctions are now $x$-weigthed sum of the above arguments (eq.~\eqref{eqn:bessel_arguments}) for the quarks:
\begin{subequations}
\begin{align}
 \left[k\right]^{\mbox{hadron}}(\rho,\varphi,\hat x) =& \sum_{n=1}^{n_q} \dfrac{\sqrt{m_n^2+(x_n P_T)^2} \cosh\eta_T(\varphi,\rho,x_n P_T)}{T},\\
\left[i\right]^{\mbox{hadron}}(\rho,\varphi,\hat x) =& \sum_{n=1}^{n_q} \dfrac{x_n P_T \sinh\eta_T(\varphi,\rho,x_n P_T)}{T}
\end{align}
\end{subequations}
where $n_q$ is the number of valence quarks.

The degeneracy factor $C$ are not a priori clear from QCD. In principle every quark has 3 color and 2 spin degrees of freedom. But since there are no dynamical gluons in this model it would be consistent to require that the quarks have the right quantumnumbers. Therefore, the degeneracy is only determined by the degrees of freedom of the hadron, e.g. $C_p=2$ and $C_\pi=1$. I will not take into account feeddown from resonance decay, except for the $\Lambda$, where the $\Sigma^0$ is too close in mass to be suppressed. Hence I take $C_\Lambda = 4$ \cite{Fries:2003kq}.

\subsubsection{Central collisions}
For collisions with zero impact parameter, the initial as well as the freeze-out eccentricity will be zero. Then $f(\varphi)\equiv1$ and $\eta_T$ is independent of $\varphi$ and so one arrives at
\begin{align}
\dfrac{\dn N}{2\pi P_T \dn P_T \dn y}=&C\exp{\mu/T} \dfrac{4\pi \rho_0^2}{(2\pi)^3} \int_0^1 \dn \rho' \tau(\rho') \rho' \int \Dn\hat x\, \left\vert\psi(\hat x)\right\vert^2 \nonumber\\
&\times \Biggl[K_1 I_0 m_T - K_0 I_1 p_T \partial_\rho \tau \Biggr]
\end{align}
\subsubsection{Peripheral collisions}
The parameter $\rho_0$ measures the size of the transverse freeze-out area $A^f_T=\pi\rho_0^2$ (see eq.~\eqref{eqn:A_T^f}).

The total multiplicity depends on the initially created quark density, which I assume to scale linear with the size of the initial transverse area $A^i_T(b)=\dfrac{\pi w(b)^2}{\sqrt{1-e_i(b)^2}}$ (eq.~\eqref{eqn:A_T^i}), which only depends on the impact parameter $b$. In order to have only one parameter for all centralities, I let $\rho_0$ scale like
\begin{align}
\rho_0(b)=\rho_c\cdot \dfrac{A^i_T(b)}{\pi R_A^2}
\end{align}
and will fix $\rho_c$ for central collisions.

\subsection{Flow components}
\subsubsection{Elliptic flow}
The first non-vanishing fourier coefficient in the $\phi$ expansion is the so called elliptic flow $v_2$ which will therefore dominate the expansion. It measures the momentum asymmetry between the particles emitted in-plane ($x$-direction) and out-of-plane ($y$-direction):
\begin{align}
v_2 :=& \left\langle\cos(2\phi)\right\rangle = \left\langle\cos^2(\phi)-\sin^2(\phi)\right\rangle\nonumber\\
=&\left\langle \dfrac{ p^2_x -p^2_y}{ p^2_x +p^2_y}\right\rangle.
\end{align}
 For $\partial_\rho \tau=0$, it can be calculated by
\begin{align}
 v_2(P_T) = \dfrac{\int_0^{2\pi} \dn\varphi \int_0^1 \dn \rho' \rho' K_1 I_2 f^2(\varphi) \cos(2\beta')}
{\int_0^{2\pi} \dn\varphi \int_0^1 \dn \rho' \rho' K_1 I_0 f^2(\varphi)}
\label{eqn:elliptic_flow_quarks}
\end{align}
The importance of this observable comes from the fact that it is generated by the self-quenching momentum asymmetry (see sec.~\ref{sec:momentum_asymmetry}) and is therefore mostly sensitive to the initial (partonic) stage of the collision.

\subsubsection{Hexadecupole flow}
To study additional effects, I also take a look at the next order $v_4$. By considering the ratio of the different coefficients, one can disentangle the different contribution to the flow. Since the coefficients scale like $v_2\sim \eps$ and $v_4\sim \eps^2$, one expects the ratio of the flow coefficients to be independent of the initial eccentricity. In the section~\ref{sec:cqns_v4}, I will give a more detailed description of this ratio.

\subsubsection{High-\texorpdfstring{$p_T$}{p\_T} flow}
\label{sec:high-pt_flow}
The flow components calculated with recombination are monotonically rising with $p_T$. Since I will not consider contributions from fragmentation which dominates over recombination at high $p_T$, the experimental observed drop at a finite $p_T$ has to be modeled by a phenomenological factor \cite{Fries:2003kq}
\begin{align}
 \kappa(p_T) = \dfrac{1}{1+(p_T/p_0)^2}.
\end{align}
with the additional parameter $p_0$.
It enters the transverse flow rapdity (eq.~\eqref{eqn:transverse_rapidity}) and makes sure that faster partons do not feel the expansion asymmetry that much:
\begin{align}
 \eta_T(\varphi,\rho,p_T) &= \eta_T^0 N_d \left(\dfrac{\rho}{\rho_0}\right)^d \left(1+\tilde\eta_T(\varphi)\kappa(p_T)\right)
\end{align}

\subsection{Constituent quark number scaling}
\subsubsection{Elliptic flow scaling}
\label{sec:cqns_v2}
One major success of recombination is a predicted simple connection between the quark elliptic flow and the hadron elliptic flow which gives strong evidence for a QGP phase in the collision. This connection is called ``constituent quark number scaling'' (CQNS), because the hadron elliptic flow scales with the number of constituent quarks.

To study the connection of the elliptic flow between quarks and hadrons, let us fourier expand the quark density similar to eq.~\eqref{eqn:yield_fourier_expansion} as
\begin{align}
 w_q(p_T,\varphi,\phi) =& \dfrac{1}{2\pi}\int \dn \phi\, w_q + \dfrac{2}{2\pi} \sum_{n=1}^\infty \cos(n\phi)\int \dn \phi\, \cos(n\phi) w_q + \sin(n\phi)\int \dn \phi\, \sin(n\phi) w_q\nonumber\\
=& \exp{k(\varphi,\rho,p_T)\cosh(\eta-y)+\mu/T}
 I_0\nonumber\\
 & \times \left(1+2 \sum_{n=1}^\infty \dfrac{I_n}{I_0} \cos(n\beta' )\cos(n\phi) + \dfrac{I_n}{I_0} \sin(n\beta') \sin(n\phi)\right)\nonumber\\
=& \exp{k(\varphi,\rho,p_T)\cosh(\eta-y)+\mu/T} I_0 2 \sum_{n=0}^\infty v^w_n \cos(n(\beta'-\phi))
\label{eqn:full_fourier_density}
\end{align}
by defining $v^w_0 := 1/2$ and $v^w_n := I_n/I_0$.

To simplify the derivation, I take $\partial_\rho \tau = 0$ which is only a minor simplification compared to the following ones. Because of the orthogonality of the cosine, only the coefficient $v^w_2$ enters the elliptic flow for quarks:
\begin{align}
\left[v_2\right]^{\mbox{quark}}_{p_T} = \dfrac{\int_0^1 \rho' \dn \rho' \int_0^{2\pi} \dn \varphi\, f^2(\varphi) I_0 K_1 v^w_2 \cos(2\beta')}
{\int_0^1 \rho' \dn \rho' \int_0^{2\pi} \dn \varphi\, f^2(\varphi) I_0 K_1},
\end{align}
To compare it to the hadron elliptic flow, I simplify the hadron wavefunction. Since it is the product of the momentum fractions $x_q$, it is already maximal for quarks with equal momentum. Therefore, I use the delta-function approximation
\begin{align}
 \left\vert\psi(\hat x)\right\vert^2 = \prod_{i=1}^{n_q} \delta\left(x_i - \dfrac{1}{n_q}\right)
\label{eqn:delta-shaped_wavefunction}
\end{align}
Then the $x$-integration will brake down and all quarks have the equal momentum fraction $1/n_q$. For a meson one now has a product of two fourier series which can be expanded to
\begin{align}
\label{eq:meson_quark_product}
 w_{q_1}(p_T/2) w_{q_2}(p_T/2) =& \exp{k^{\mbox{meson}}(\varphi,\rho,1/2)\cosh(\eta-y)+2\mu/T}
 I_0^2\left(i(\varphi,\rho,p_T/2)\right) \nonumber\\
&\times4 \sum_{n,m=0}^\infty v^w_n v^w_m \cos(n(\beta'-\phi)) \cos(m(\beta'-\phi))
\end{align}
As I will show in section~\ref{sec:cqns_breaking}, the higher terms can not be neglected, so analytically there is no simple and exact relation between the quark and the hadron $v_2$.

Those who want to derive an analytical expression for the scaling, must use a strong simplification and drop the spatial correlations between the angle and flow velocity from the quark density. In this case, one assumes just a constant asymmetry in the flow profile which is equal at every point in the fireball, and rewrite the density with the help of the quark elliptic flow as \cite{Fries:2003kq}
\begin{align}
 w_q(p_T,\varphi,\phi) = w_q(p_T,\phi) \left(1+2 \left[v_2\right]^{\mbox{quark}}_{p_T} \cos(2 \phi) \right)
\end{align}
The azimuthal asymmetry is only contained in the $\cos(2\phi)$ term which depends only on the hadron emission angle $\phi$. The previous dependence of $\eta_T$ on the spatial angle $\varphi$ is now gone. When assuming a circular freeze-out area ($e_f = 0 \Rightarrow f(\varphi)=1$), the $\varphi$-dependence can be integrated out completely which leads to a much simpler equation for the hadron elliptic flow. One finds for mesons
\begin{align}
\left[v_2\right]^{\mbox{meson}}_{P_T} =& \dfrac{\int_0^1 \rho' \dn \rho' I_0\left[i(\rho,P_T/2)\right] K_1\left[k(\rho,P_T/2)\right]}
{\int_0^1 \rho' \dn \rho' I_0\left[i(\rho,P_T/2)\right] K_1\left[k(\rho,P_T/2)\right]}\nonumber\\
&\times \dfrac{v^{q1}_2 + v^{q2}_2} {1+2 v^{q1}_2 v^{q2}_2}
\end{align}
and for baryons
\begin{align}
\left[v_2\right]^{\mbox{baryon}}_{P_T} =& \dfrac{\int_0^1 \rho' \dn \rho' I_0\left[i(\rho,P_T/3)\right] K_1\left[k(\rho,P_T/3)\right]}
{\int_0^1 \rho' \dn \rho' I_0\left[i(\rho,P_T/3)\right] K_1\left[k(\rho,P_T/3)\right]}\nonumber\\
&\times \dfrac{ \left(v^{q1}_2 + v^{q2}_2 + v^{q3}_2 + 3 v^{q1}_2v^{q2}_2v^{q3}_2\right) }{\left(1+2\left( v^{q1}_2 v^{q2}_2 + v^{q1}_2 v^{q3}_2 + v^{q2}_2 v^{q3}_2\right)\right)}
\end{align}
where $v^q_2$ is to be evaluated at $P_T/2$ for mesons and at $P_T/3$ for baryons. With the additional simplifications of a constant transverse flow rapidity independent of the radius, one arrives at
\begin{align}
 \left[v_2\right]^{\mbox{meson}}_{P_T} = \left[\dfrac{v^{q1}_2 + v^{q2}_2}{1+2 v^{q1}_2 v^{q2}_2}\right]_{P_T/2}
\end{align}
and
\begin{align}
 \left[v_2\right]^{\mbox{baryon}}_{P_T} = \left[\dfrac{v^{q1}_2 + v^{q2}_2 + v^{q3}_2 + 3 v^{q1}_2v^{q2}_2v^{q3}_2}
{1+2\left( v^{q1}_2 v^{q2}_2 + v^{q1}_2 v^{q3}_2 + v^{q2}_2 v^{q3}_2\right)} \right]_{P_T/3}
\end{align}
respectively. For hadrons with similiar quark content all coefficients $v^q_2$ are equal and by further neglecting quadratic and cubic terms, one obtains the much celebrated constituent quark number scaling \cite{Fries:2003kq}
\begin{align}
\left[v_2\right]^{\mbox{hadron}}_{P_T}  \approx n_q \left[v_2\right]^{\mbox{quark}}_{P_T/n_q}
\end{align}
Beside all the simplifications, this scaling law is confirmed with great success by the experimental data. A common way to show the good agreement is to scale the data of hadrons by $1/n_q$ and plot it as $P_T/n_q$ vs. $v_2/n_q$. According to the scaling law, all curves for mesons and baryons will lie on one universal curve which would be the quark elliptic flow.

\subsubsection{Hexadecupole flow scaling}
\label{sec:cqns_v4}
With the above simplifications, one can also derive a scaling law for the hadron $v_4$. Again neglecting higher powers and assuming equal $v_2$ and $v_4$ for all quarks, one finds
\begin{align}
 \left[v_4\right]^{\mbox{meson}}_{P_T}  \approx & v^{q1}_4 +v^{q2}_4 + v^{q1}_2 v^{q2}_2\nonumber\\
=& \left[2 v_4 + v^2_2\right]^{\mbox{quark}}_{P_T/2}
\end{align}
and
\begin{align}
  \left[v_4\right]^{\mbox{baryon}}_{P_T} \approx& v^{q1}_4 + v^{q2}_4 + v^{q3}_4 + \left(v^{q1}_2 v^{q2}_2 + v^{q1}_2 v^{q3}_2 + v^{q2}_2 v^{q3}_2\right)\nonumber\\
=& \left[3 v_4 + 3 v^2_2\right]^{\mbox{quark}}_{P_T/3}
\end{align}

Combining the both scaling laws, ratio between $v_2$ and $v_4$ can be expected to be approximately
\begin{align}
 \left[\dfrac{v_4}{v_2^2}\right]^{\mbox{meson}}_{P_T} 
= \dfrac{1}{4} + \dfrac{1}{2} \left[\dfrac{v_4}{v_2^2}\right]^{\mbox{quark}}_{P_T/2} 
\end{align}
and
\begin{align}
  \left[\dfrac{v_4}{v_2^2}\right]^{\mbox{baryon}}_{P_T} 
= \dfrac{1}{3} + \dfrac{1}{3} \left[\dfrac{v_4}{v_2^2}\right]^{\mbox{quark}}_{P_T/3} 
\end{align}

\subsection{The breaking of the CQNS}
\label{sec:cqns_breaking}
Although the CQNS is experimentally well observed, the scaling is broken by the masses of the hadrons and the simplified equations from the upper section ca not account for that. Therefore, I want to emphasize that the scaling laws will only serve as a rough guide and all calculations will be done with the full spatial correlations of the quark density, if not specified otherwise. 

\begin{figure}
 \centering
 \includegraphics{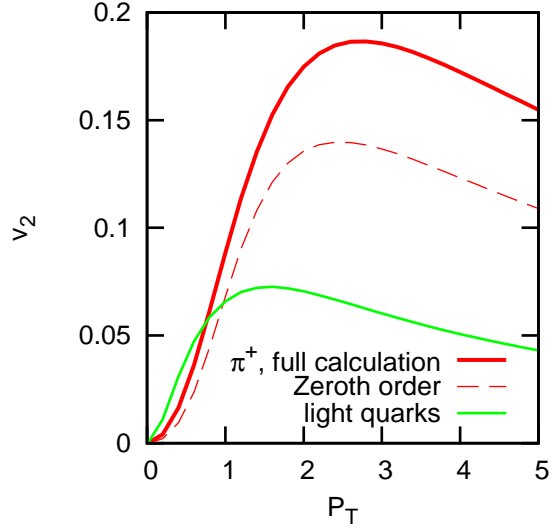}
 \caption[Study of the breaking of constituent quark number scaling]{Study of the CQNS: The elliptic flow of pions from the full calculation compared to zeroth order (CQNS) in the expansion of eq.~\ref{eqn:cqns_full_expansion} together with the light quark flow.}
 \label{fig:cqns_test}
\end{figure}

Let us shortly discuss the scale breaking terms:
Taking the quark-antiquark density from eq.~\eqref{eq:meson_quark_product} the meson elliptic flow can be calculated via
\begin{align}
\left[v_2\right]^{\mbox{meson}}_{P_T} \approx \dfrac{\int_0^1 \rho' \dn \rho' \int_0^{2\pi} \dn \varphi\, f^2(\varphi) I_0^2\left[i(\varphi,\rho,P_T/2)\right] K_1\left[k^{\mbox{meson}}(\varphi,\rho,1/2)\right] V_2 \cos(2\beta)}
{\int_0^1 \rho' \dn \rho' \int_0^{2\pi} \dn \varphi\, f^2(\varphi) I_0^2\left[i(\varphi,\rho,P_T/2)\right] K_1\left[k^{\mbox{meson}}(\varphi,\rho,1/2)\right] V_0}.
\end{align}
$V_0$ and $V_2$ are the contributing terms from the infinite sum. Because of the orthogonality of the cosine, all terms not containing the integral weightings (i.e. $\cos(0\phi)$ for the denominator and $\cos(2\phi)$ for the numerator) have vanished after the $\phi$-integration. With the trigonometric product
\begin{align}
 \cos (nx) \cos (mx) = \dfrac{1}{2}\left(\cos (n-m)x + \cos(n+m)x\right)
\end{align}
an infinite sum of coefficient products $v^w_n v^w_m$ that fulfill $n+m=d$ or $\vert n-m\vert=d$, contributes to
\begin{align}
V_d=& \left(\sum_{n+m=d}^\infty v^w_n v^w_m + \sum_{\vert n-m=0 \vert}^\infty v^w_n v^w_m\right) 2 \int \dn \phi \cos^2(d\phi).
\label{eqn:cqns_full_expansion}
\end{align}
This leads to
\begin{align}
V_0=& 4 \pi \left(2 \left(v^w_0\right)^2 + \sum_{n}^\infty \left(v^w_n\right)^2\right)
\nonumber\\
=&2 \pi \left(1+2\sum_{n}^\infty \left(v^w_n\right)^2\right)
\end{align}
and
\begin{align}
V_2=& 2 \pi \left(4 v^w_0 v^w_2 + v^w_1 v^w_1 + 2\sum_{n}^\infty v^w_n v^w_{n+2}\right)
\end{align}
In fig.~\ref{fig:cqns_test} you see the comparison of the full calculation to zeroth order. In zeroth order, the terms contain only the $v^w_2$ contributions:
\begin{align}
V_2 = 2\pi\left(2v^w_2\right)
\qquad V_0=2\pi\left(1 +2v^w_2 v^w_2 \right).
\end{align}
This would correspond to the CQNS. Obviously, the simple approximation of zeroth order is not sufficient when calculating with the full momentum space correlations. Therefore, the higher terms can not be discarded and a simple, analytical form for quark number scaling does not exist in this framework.

Additionally to the contribution from the higher order flow components, the bessel functions give further deviations from the scaling law: In the high $P_T$ region the additional powers of $I_0$ give an enhancement for the elliptic flow, so it will be higher than $n_q v^q_2(P_T/n_q)$. The higher $n_q$ the higher the enhancement will be. On the other hand, in the low $P_T$ region where $I_0$ is approximately one, the scaling will be broken by $K_1$. The argument $k^{\mbox{hadron}}(\varphi,\rho,1/n_q)$ is the sum over the transverse masses of the quarks at $P_T/n_q$. So for $P_T \rightarrow 0$ the transverse mass is equal to the quark mass and the argument is proportional to the sum of the quark masses. Since $K_1$ is a monotonic decreasing function, hadrons with heavy quarks have a lower elliptic flow at low $P_T$ as compared to hadrons with light quarks. Similiarly, baryons will have a lower elliptic flow than mesons when they have quarks with comparable masses.

This explicit mass scaling is experimentally observed and can not be explained within the simplifications for an analytical CQNS.

\chapter{Results and predictions from recombination}
\label{cha:results}
In this chapter I will present my results from recombination with the discussed quark density and freeze-out hypersurface. The chapter is organized as follows:
\begin{itemize}
 \item Before I will present the results, I would like to summarize the parameters in the model and how they are determined.
\item Then I will show some effects of the parameters on the observables.
\item The third section will contain results of the transverse momentum spectra.
\item And in the last section, I will discuss the flow coefficients $v_2$ and $v_4$. There, I study the ratio of these coefficients to extract reasonable parameter values and then discuss the $p_T$, centrality and $\sqrt{s}$ dependence.
\end{itemize}

\section{Overview of the parameters}
\label{sec:parameters}
\subsection{Recombination parameters}
\subsubsection{Quark masses}
The bare quark masses (current quark masses) for the light quarks are much smaller then the hadrons they constitute, with $m_u=1.5-3.3\MeV$, $m_d=3.5-6.0\MeV$ and $m_s=70-130\MeV$. So while the mass of $u$ and $d$ make up only about 0.15\% of the proton mass, the main contribution comes from the gluons and the virtuell quark-antiquark pairs. To account for the dynamically generated mass, the quarks are assigned so called constituent masses.

Since in the recombination approach there are no dynamical elements like gluons or pair production, it would be consistent to assume that the recombining quarks are already surrounded by this virtuell cloud. Therefore, one has to use constituent quark masses.

These constituent masses are chosen to fit the hadron masses calculated from mass formulas within the quark model. The best fit is achieved when using different masses for mesons and baryons since they have different nuclear size. I take the values from Gasiorowicz \cite{Gasiorowicz:1981jz} as
\begin{align}
m_u=&m_d=310 \MeV, & m_s &= 483 \MeV &&\mbox{for mesons,}\\
m_u=&m_d=363 \MeV , & m_s &= 538 \MeV &&\mbox{for baryons.}
\end{align}

Mesonic systems consisting of heavy quarks like charm or bottom can be described non-relativisticly due to the large mass of the quarks. That is why I take the constituent masses as the half of the meson masses. That means
\begin{align}
m_c=&m_{J/\psi}/2 = 1.548\GeV\\
m_b=&m_{\Upsilon}/2 = 4.730\GeV
\end{align}

\subsubsection{High-\texorpdfstring{$p_T$}{p\_T} damping}
To account for the fact that the elliptic flow has some maximum and is then slowly droping at higher $P_T$, I introduced a phenomenological damping factor in section~\ref{sec:high-pt_flow}. It will be chosen to fit the high-$p_T$ elliptic flow data.

Since a non-zero freeze-out eccentricity will give an enhancement to the high-$p_T$ elliptic flow, the value for $p_0$ will be coupled to $e_f$.
\subsection{Freeze-out hypersurface parameters}
\subsubsection{Transverse freeze-out area}
The transverse freeze-out area $A_T=\pi \rho_0^2$ effects the total multiplicity. I fix $\rho_0=11 \fm$ to fit the experimental data on the invariant yield of different hadrons in central collisions (see Fig.~\ref{fig:yields}). As already noted, the multiplicity in peripheral collisions is assumed to scale with the initial transverse area which depends on the impact parameter. 

\subsubsection{Impactparameter}
The impact parameter $b$ is not a real free parameter, but there are some uncertainties in relating it to the centrality which is the experimentalists impact measure. To do it without additional calculations from the Glauber, I take the functional form \cite{Broniowski:2001ei}
\begin{align}
 c(N) = \dfrac{\pi b^2(N)}{\sigma_{\mbox{inel}}} \quad\mbox{with }\, \sigma_{\mbox{inel}} = 7.05 \mbox{b}=705 \fm^2
\end{align}
where $c(N)$ is the centrality of events with a multiplicity higher than $N$. Despite its simplicity, the results are almost equal to the Glauber model for binary collisions. Only for ultra peripheral collision ($b>14 \fm$) there are small deviations. 

\subsubsection{Freeze-out eccentricity}
While the initial eccentricity is fixed by the impact parameter, the eccentricity at freeze-out, which originates from the initial one, depends on the expansion dynamics. Due to the induced momentum asymmetry, the eccentricity of the fireball will become smaller as it expands. But there is no need to expect it to be zero at freeze-out.

The numerical eccentricity $e_f$ models the shape of the transverse freeze-out area. The consequences from a non-zero $e_f$ are discussed in section~\ref{sec:geometrical_contrib} and the functional form of the dependence on $e_i$ can be found in section~\ref{sec:geometrical_contrib_strength}.

\subsubsection{Time dependent hypersurface}
By assuming an elliptic freeze-out in the previous section, the transverse part of the hypersurface will be treated quite general. And the assumed boost invariance of the longitudinal expansion in section~\ref{sec:freeze-out} is generally accepted and well established.

An open question concerns the time dependence of the hypersurface. As already said in section~\ref{sec:integration_measure} I let $\tau$ be a function of the radial coordinate $\rho$. The functional form of the dependence and its implications are discussed in section~\ref{sec:tau_rho_correlation}.

Together with $\rho_0$, the mean freeze-out time
\begin{align}
 \tau_0 = \dfrac{\int \tau(\rho)\rho\, \dn \rho}{\int \rho \, \dn\rho}.
\end{align}
 will enter only as an overall normalization factor. Therefore, I take it to be constant with $\tau_0 = 5 \fm$ \cite{Fries:2003kq}.

\subsection{Blast wave parameters}
The parameters for the blast wave model are the temperature $T$, the baryo-chemical potential $\mu_B$, the transverse flow rapidity $\eta_T^0$ and its radial profile parameter $d_\rho$. The first two depend on the phase boundary and will be determined within the MIT bag model, the rapidity has to be extract from fits to experimental data and the profile will be adjusted to a reasonable value. An additional strange fugacity $\gamma_s=\gamma_{\bar s}=0.8$ is introduced to fit the invariant yields (sec.~\ref{sec:yields}) and ratios (sec~\ref{sec:hadron_ratios}) of the strange hadrons.

\begin{figure}[!h]
 \centering
 \subfigure[\label{fig:mu_T_vs_E-fit}]{\includegraphics[scale=1]{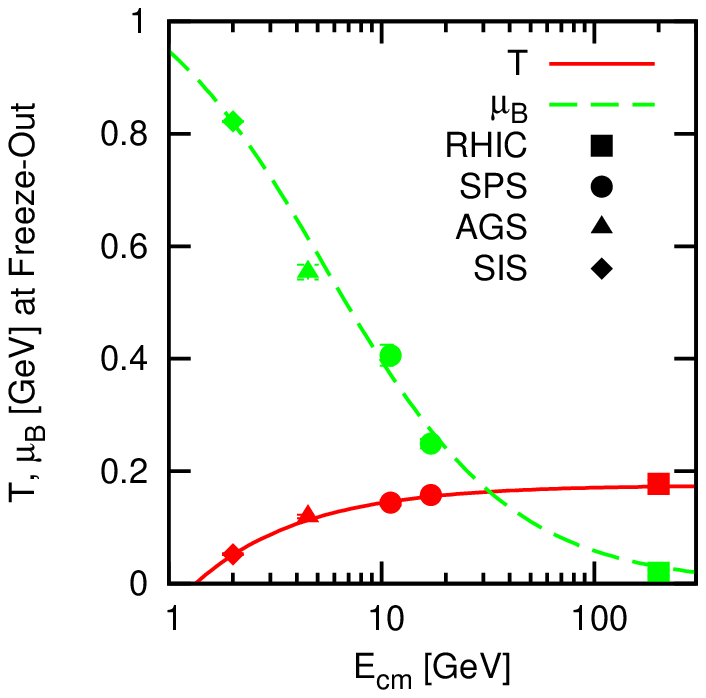}}
 \subfigure[\label{fig:mu_T_vs_E-plane}]{\includegraphics[scale=1]{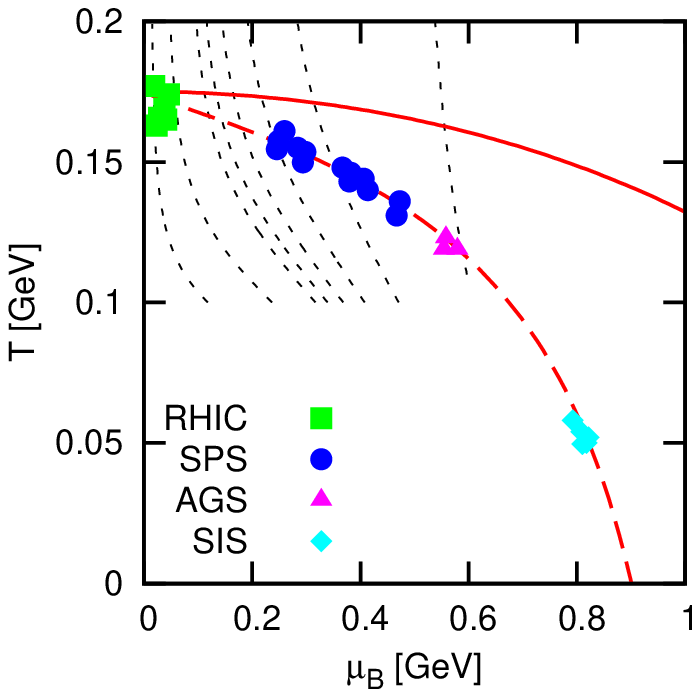} }
 \caption[Temperature $T$ and baryo-chemical potential $\mu_B$]{Temperature $T$ and baryo-chemical potential $\mu_B$\newline 
a) $T$ and $\mu_B$ at freeze-out as a function of $\sqrt{s}$ for different colliders. The lines are fits from eqs.~\eqref{eq:T_vs_E} and~\eqref{eq:mu_vs_E}.\newline
b) Comparison of the phase boundary (solid) from the MIT bag model and the freeze-out curve (dashed) from a fit to different colliders. The dotted lines are the paths along an isentropic expansion.}
 \label{fig:mu_T_vs_E}
\end{figure}
\subsubsection{Phase boundary}
The thermodynamic quantities $T$ and $\mu_B$ at the phase boundary are calculated via the MIT bag model from sec.~\ref{sec:hic-mit_bag}.
To choose reasonable values for the different colliders, one needs the dependence on the center-of-mass (cm) energy $\sqrt{s}$ for either $T$ or $\mu_B$.

One can extract the values for these quantities within the thermodynamical model by fits to experimental data. From the values in \cite{Cleymans:2006zz} I choose the ansatz
\begin{align}
T(\sqrt{s})=&T_C \left(1-(\sqrt{s}/b)^a\right)\label{eq:T_vs_E}\\
\mu_B(\sqrt{s})=&\dfrac{c}{1+(\sqrt{s}/d)}\label{eq:mu_vs_E}
\end{align}
with $a=-0.85$, $b=1.33 \GeV$, $c=1.12 \GeV$ and $d=5.48 \GeV$.
The fit is shown in fig.~\ref{fig:mu_T_vs_E}.

But since these are the values at freeze-out, one has to link the baryo-chemical potential from there to the phase boundary. By applying an isentropic expansion within a simple hadrongas model, I establish a connection of the conditions from the phase boundary (pb) to the freeze-out (fo).

The isentropic path of the system in the $T-\mu_B$-plane is shown in fig.~\ref{fig:mu_T_vs_E-plane}. It shows that the baryo-chemical potentials can be simply related by
\begin{align}
\mu_B^{\mbox{pb}} = 0.938 \cdot \mu_B^{\mbox{fo}}
\end{align}
Together with eq.~\eqref{eq:mu_vs_E} and eq.~\eqref{eqn:mit_bag_model} I can express $T$ and $\mu_B$ at the phase boundary as a function of $\sqrt{s}$.

\subsubsection{Transverse expansion}
\begin{figure}
\centering
 \includegraphics{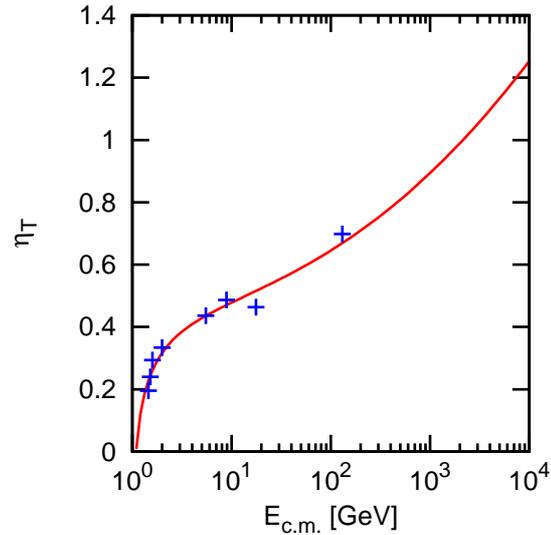}
 \caption[Parameterisation of the transverse flow rapidity]{Parameterisation of the transverse flow rapidity $\eta_T$ (full line) as function of $\sqrt{s}$. The data at kinetic freeze-out (crosses) are taken from \cite{Xu:2001zj}.}
\label{fig:eta_T_vs_E}
\end{figure}

The mean transverse flow rapidity $\eta_T^0$ will be fitted to flow rapidities extracted from the experimental data \cite{Xu:2001zj}
at kinetic freeze-out. The data showed in \cite{Xu:2001zj} is for mean transverse flow velocity $\beta_T$, but I use the rapidity $\eta_T=\tanh \beta_T$ to avoid that the velocity becomes greater than $c$ in the fit.

To fit these values I choose 
\begin{align}
 \eta_T^{\rm Freeze}(\sqrt{s}) =& a + b x + c x^2 + d \ln(x)\\
\mbox{with }x=&ln\left(\sqrt{s}\right) \nonumber
\end{align}
with the constants $a = 0.418, \; b = -0.064,\; c=0.012,\; d=0.170$.
As these values are extracted at freeze-out, I scale the obtained transverse rapidities by a constant 
factor $k=0.85$ to obtain the transverse flow at the hadronization surface.
Using these parameters, the value for $v_T=0.54$ at $\sqrt{s}=200\GeV$ (RHIC) agrees with the value from \cite{Fries:2003kq}. 

For $\sqrt s =5.5 \TeV$ (LHC) I obtain a transverse flow velocity of $v_T=0.75$ also in line with previous 
estimates \cite{Fries:2003fr}. Fig. (\ref{fig:eta_T_vs_E}) depicts the fit (line) and the available 
data on $\eta_T$ (crosses).

\clearpage
\section{Influence of the parameters and parameterisations}
\subsection{Delta-shaped wavefunctions}
\begin{figure}[hb]
 \centering
\includegraphics[scale=0.8]{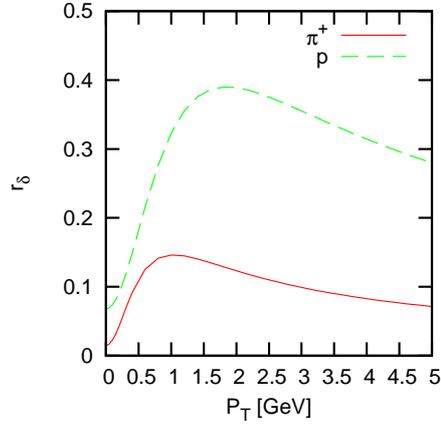}
\caption[Relative deviations $r_\delta$ of delta-shaped wavefunctions]{Relative deviations $r_\delta$ of delta-shaped wavefunctions from realistic light cone wavefunctions in eq.~\ref{eqn:realistic_wavefunctions}.}
\label{fig:delta_wavfunction}
\end{figure}

The ansatz in eq.~\eqref{eqn:realistic_wavefunctions} for the local light cone wavefunction with the product of the momentum fractions $x$ is motivated by the asymptotic form of the pertubative pion distribution amplitude \cite{Fries:2003kq}. This function is very broad and so the quark momenta are strongly smeared around $P/n_q$.

The other extreme case is the delta-function approximation of eq.~\eqref{eqn:delta-shaped_wavefunction} which is already used in the derivation of the CQNS, where the hadron consists of $n_q$ quarks with all having the same momentum $P/n_q$. The influence of this strong simplification compared to the realistic light cone wavefunctions is shown in Fig.~\ref{fig:delta_wavfunction}, where I have given the relative deviations $r_\delta=(dN_\delta-dN)/dN$ for pions and protons.

\subsection{Effects of the blast wave parameters}
\subsubsection{Baryo-chemical potential}
The influence of the baryo-chemical potential $\mu_B$ on the results is quite neglectable. First of all, it only enters as a factor for each quark in the invariant yield. Second, it drops out of the flow coefficients completely at least for boltzmann statistics. Finally the value of $\mu_B$ at RHIC and higher energies is much smaller then the temperature (Fig.~\ref{fig:mu_T_vs_E-fit}). Therefore the fugacity $\gamma=\exp{\mu_B/T}\approx 1$ gives only small corrections.

\subsubsection{Temperature}
The temperature $T$ sets the slope of the thermal quark spectrum and also enters the invariant yields. So the temperature is an important parameter, but it varies mainly at low c.m. energies. For $\sqrt{s}\gtrsim10\GeV$ it quickly approaches the critical temperature $T_C=175\MeV$ and for RHIC energies and beyond it is essentially constant (Fig.~\ref{fig:mu_T_vs_E-fit}). So for the energy range, where recombination can be considered to play an important role, the energy dependence of temperature is not significant.

\subsubsection{Transverse flow rapidity}
\label{sec:beta_T_dependence}
The main parameter with the biggest impact on the results is the transverse flow rapidity $\eta_T=\atanh(\beta_T)$, since it is expected to show strong variations as the fit from fig.~\ref{fig:eta_T_vs_E} suggests. Although the functional form is very uncertain, the general trend of a rising rapidity can be expected. The exact shape of the dependence is therefore not that important, as long as it fits the value at LHC which is generally expected to be $v_T=0.7-0.8 \mbox{c}$.

The influence of $\eta_T$ can be studied by looking at the slope of the $p_T$ spectra. The bessel functions (eq.~\eqref{eqn:phi_integrated_yield_quark}) can be expanded in exponentials $\exp{-m_T/T_{\mbox{eff}}}$ with an effective temperature that sets the slope. The inverse temperature can then be found by
\begin{align}
 \dfrac{\dn}{\dn m_T} \ln\left[\dfrac{\dn N}{p_T \dn p_T}\right] = -\dfrac{1}{T_{\mbox{eff}}}
\end{align}

For central collisions, a fixed radial coordinate $\rho$ and $\partial_\rho \tau = 0$, this expression can be calculated analytically \cite{nucl-th/9307020}:
\begin{align}
 -\dfrac{1}{T} &=\dfrac{\dn}{\dn m_T} \ln\left[K_1\left(\dfrac{m_T \cosh\eta_T}{T}\right) I_0\left(\dfrac{p_T \sinh\eta_T}{T}\right) m_T\right] \nonumber\\
 &= \dfrac{1}{m_T}+\dfrac{I_1}{I_0}\dfrac{m_T \sinh\eta_T}{p_T T}-\dfrac{K_0}{K_1}\dfrac{ \cosh\eta_T}{T}\nonumber\\
&\overset{m_T\rightarrow\infty}{=} \dfrac{\cosh\eta_T-\sinh\eta_T}{T} = -\dfrac{1}{T}\sqrt{\dfrac{1-\beta_T}{1+\beta_T}}
\end{align}
So at high $m_T$ (or $p_T$) the effective temperature is blue shifted as
\begin{align}
 T_{\mbox{eff}}= T \sqrt{\dfrac{1+\beta_T}{1-\beta_T}} > T
\end{align}
so the slope at high transverse momenta is less steep for a non-zero transverse expansion velocity $\beta_T$.

\begin{figure}
 \centering
 \includegraphics{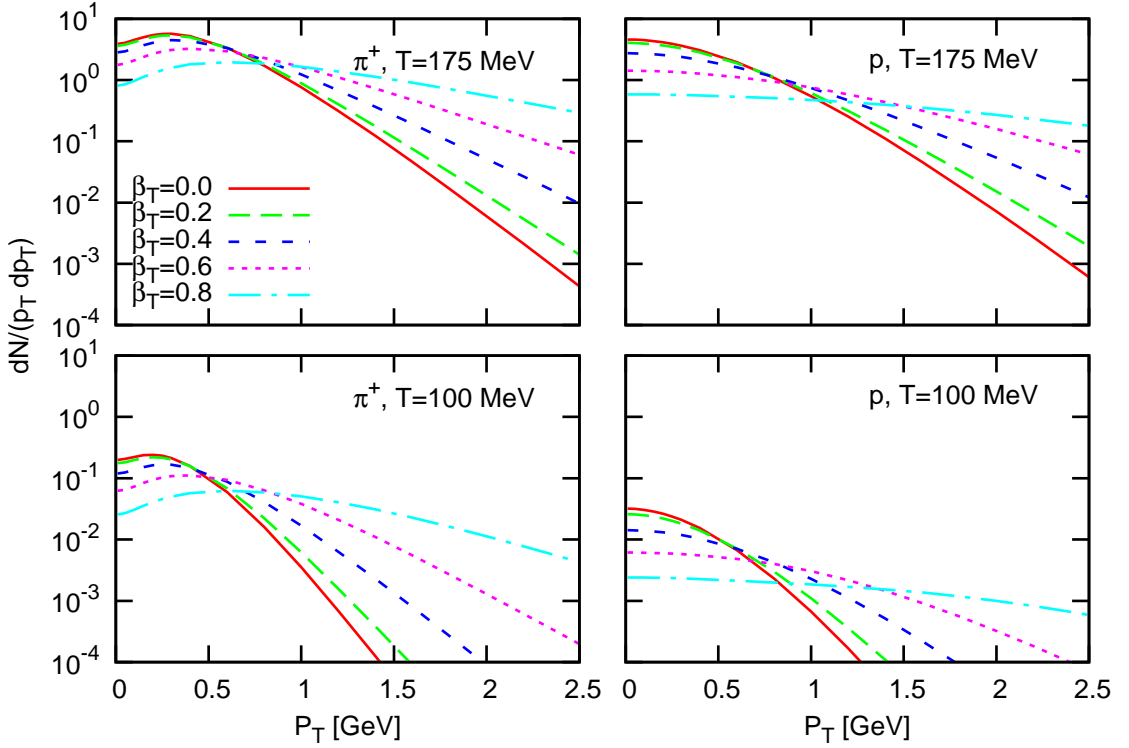}
 \caption{The invariant yield of pions and protons for different transverse flow velocities $\beta_T$.}
 \label{fig:yield_beta_dependence}
\end{figure}
\begin{figure}
 \centering
 \includegraphics{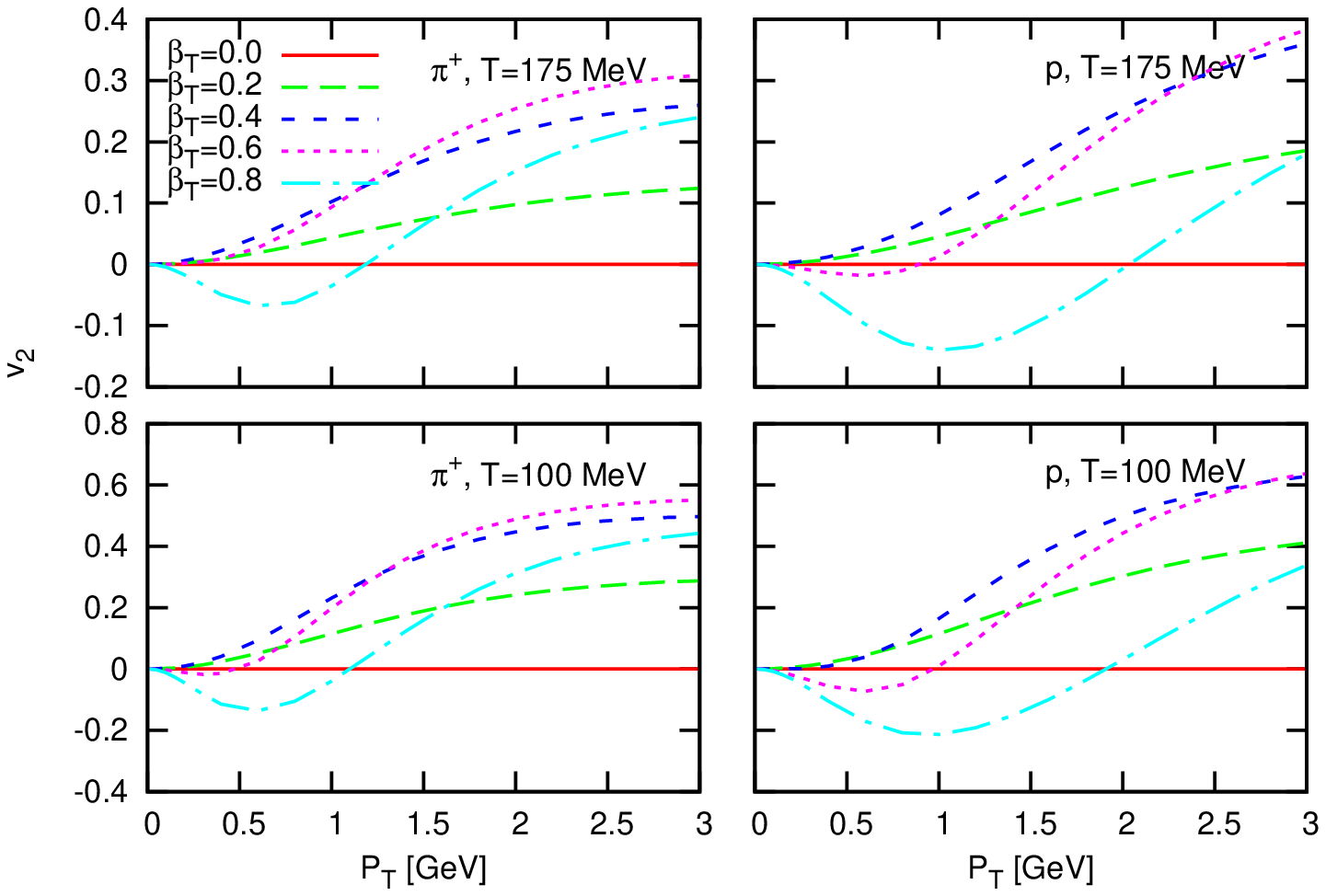}
 \caption{The elliptic flow of pions and protons for different transverse flow velocities $\beta_T$.}
 \label{fig:v2_beta_dependence}
\end{figure}

At low $m_T$ the case is less simple. Therefore, I show the invariant yield for pions and protons in Fig.~\ref{fig:yield_beta_dependence} for different values of $\beta_T$ and also for two different temperatures. The same comparison for the elliptic flow is depicted in Fig.~\ref{fig:v2_beta_dependence}

\subsubsection{Radial profile of the transverse rapidity}
\label{sec:d_rho_dep}
The second parameter in the transverse rapidity $\eta_T$, beside the mean rapidity $\eta_T^0$, is $d_\rho$ which models the radial growth in eq.~\eqref{eqn:transverse_rapidity} as $\left(\rho/\rho_0\right)^{d_\rho}$. In the derivation of the CQNS, I already used the appproximation of a radially constant transverse rapidity i.e. $d_\rho=0$. This is a very unrealistic assumption since there is no pressure gradient in the center and so there should be no expansion. Thus, one expects $\eta_T$ to be a smoothly rising function of the radial coordinate with a dependence somewhere between square-root and quadratic. Therefore, reasonable values for $d_\rho$ should be between 0.5 and 2.

\begin{figure}
 \centering
 \includegraphics{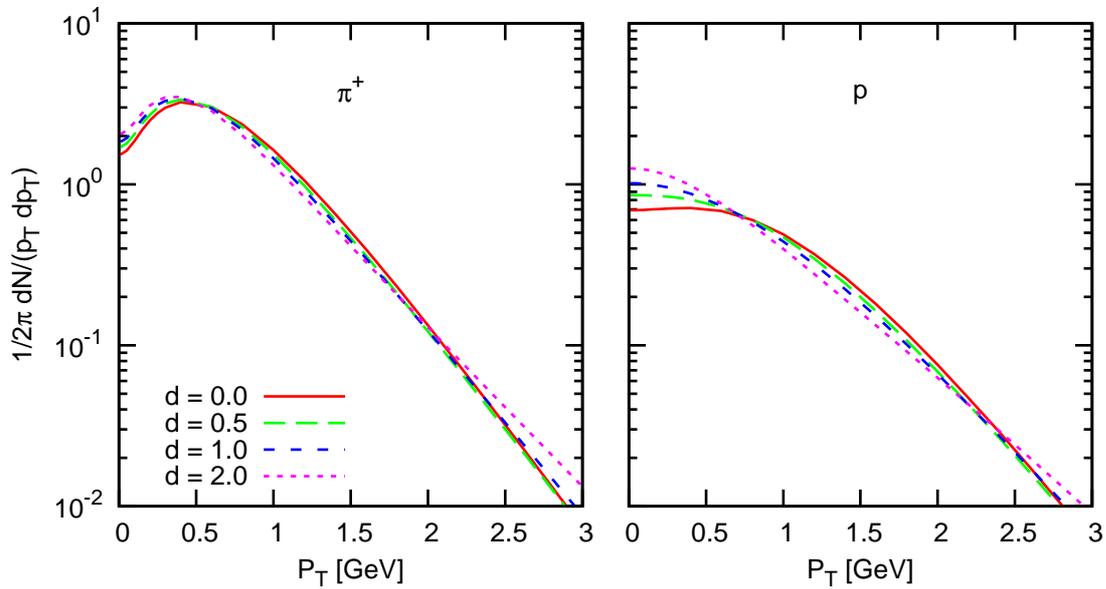}
 \caption{The invariant yield of pions and protons for different radial profiles of the transverse rapidity.}
 \label{fig:yield_d-rho_dependence}
\end{figure}

\begin{figure}
 \centering
 \includegraphics{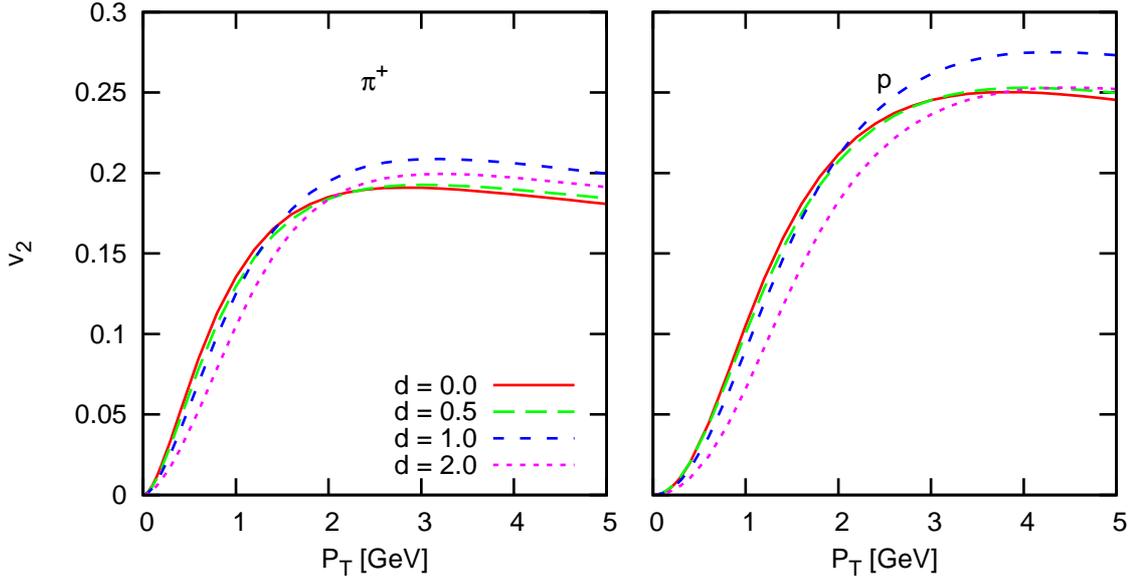}
 \caption{The elliptic flow of pions and protons for different radial profiles of the transverse rapidity.}
 \label{fig:v2_d-rho_dependence}
\end{figure}

The yield and elliptic flow of pions and protons are compared for different values of $d_\rho$ in Figs.~\ref{fig:yield_d-rho_dependence} and \ref{fig:v2_d-rho_dependence}. For the pion yield the modifications are neglectable, but at low $p_T$ the proton yield shows a strong enhancement for higher values of $d_\rho$. On the other hand, the elliptic flow decreases with increasing $d_\rho$ for both particles at low to mid $p_T$.

The real dependence can be quite non-trivial and can only be studied in dynamical calculations. To have a possibilty to determine a reasonable value, I will compare the results for the yield to experimental data (section~\ref{sec:yields}). The elliptic flow in fact is also very sensitive to the choice of $d_\rho$, but it also depends very much on the freeze-out parameters as can be seen in the next sections. Therefore, a fit to $v_2$ data would be too involved and uncertain. That is why I will use $d_\rho=1$, since it is the best fit for the yield and also a reasonable compromise between square-root and quadratic dependence.


\subsection{Separating flow and non-flow effects}
The information about the $\phi$-dependence of the invariant yield is stored in the fourier coefficients $v_n$. There are two different mechanisms that generate the azimuthal asymmetric particle yields. So before going into a detailed analysis of the results from $v_2$ and $v_4$, I will analyse the respective strength of both contributions.
The fourier coefficients $v_n$ can be separated in two different contributions:
\begin{itemize}
 \item The flow contribution comes from the asymmetry in the expansion velocity which is parameterized in eq.~\eqref{eqn:asymmetry_expansion}. The asymmetry depends on the initial numerical eccentricity $e_i$. This is the contribution which results in the CQNS and therefore the more prominent one.
\item The geometrical contribution comes from the asymmetry of the transverse freeze-out area which depends on the numerical freeze-out eccentricity $e_f$. Because of its spatial, geometrical origin it is nearly independent of the particle species.
\end{itemize}
Both effects develop from the the same origin, namely the initial eccentricity in non-central collisions. While the strength of the flow part is fixed by the impact parameter, the geometrical contributions depend on the functional dependence $e_f=e_f(e_i)$.

Hence, I will first discuss the geometrical and then the flow contributions and after that study the freeze-out eccentricity.

\subsubsection{Geometrical contributions}
\label{sec:geometrical_contrib}
To focus on the geometrical contributions, I turn off the flow contributions by setting $e_i=0$. Without an initial spatial anisotropy, there is no expansion velocity asymmetry. This normally also means that the freeze-out eccentricity is zero, so I will set it manually to a finite value.

\begin{figure}
 \centering
 \includegraphics[scale=0.8]{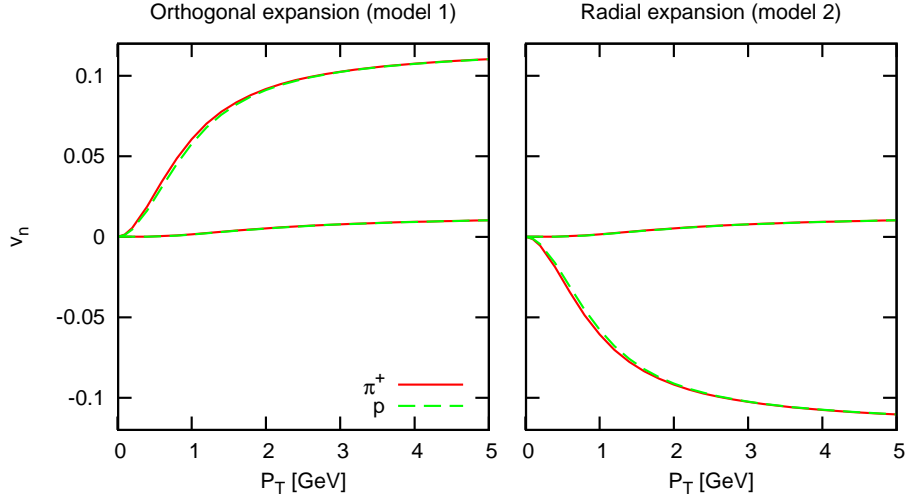}
 \caption[Purely geometrical contributions to the elliptic flow from an elliptical freeze-out area]{Purely geometrical contributions to the elliptic flow of pions (red curve) and protons (green curve) from an elliptical freeze-out area with $e_f = 0.62$.\newline
Left: The system expands orthogonal to the surface, which yields a positive $v_2$ (upper curve) and a positive $v_4$ (lower curve).\newline
Right: The system expands radial. The $v_2$ (lower curve) is negative, but with same amplitude, $v_4$ (upper curve) is equal to the other case.
}
 \label{fig:flow_components_spatial_contribution}
\end{figure}

As already explained in section~\ref{sec:surface_flow} for an elliptic freeze-out, the radial direction is not equal to the one perpendicular to the surface. Thus, we have to investigated the contributions for both models since non should be excluded a priori. Fig.~\ref{fig:flow_components_spatial_contribution} shows $v_2$ and $v_4$ contributions for pions and protons in both scenarios. Due to the geometrical nature, the contributions are nearly equal for pions and protons, so I will not show any calculations for other particle species.

What can be seen directly is that the $v_4$ is equal for both cases, while the $v_2$ switches sign when going from perpendicular (model 1) to radial expansion (model 2). This is the consquence of the transformation behavior of the integration measure derived in section~\ref{sec:surface_flow}. As stated therein, the flow of model 1 is equal to the flow from model 2 with the ellipse rotated by $90^\circ$. Reversing this rotation is just switching the x- and y-axis. For $v_2$, which compares the flow of x- to y-direction, this is equal to changing the sign, but $v_4$, which compares x- and y-direction to the diagonal direction, is not affected by this rotation.

To distinguish between the models, first one has to predict the orientation of the freeze-out ellipse. Then, by comparing the geometrical together with the flow contributions to experimental data, the correct model can be identified. On the other hand, this means that without the knowledge of the ellipse' orientation, there is no way to exclude one of the models.

In section~\ref{sec:geometrical_contrib_strength} I will present experimental evidence for the direction of the elongation.

\subsubsection{Flow contributions}
\label{sec:flow_contrib}
\begin{figure}
 \centering
\includegraphics{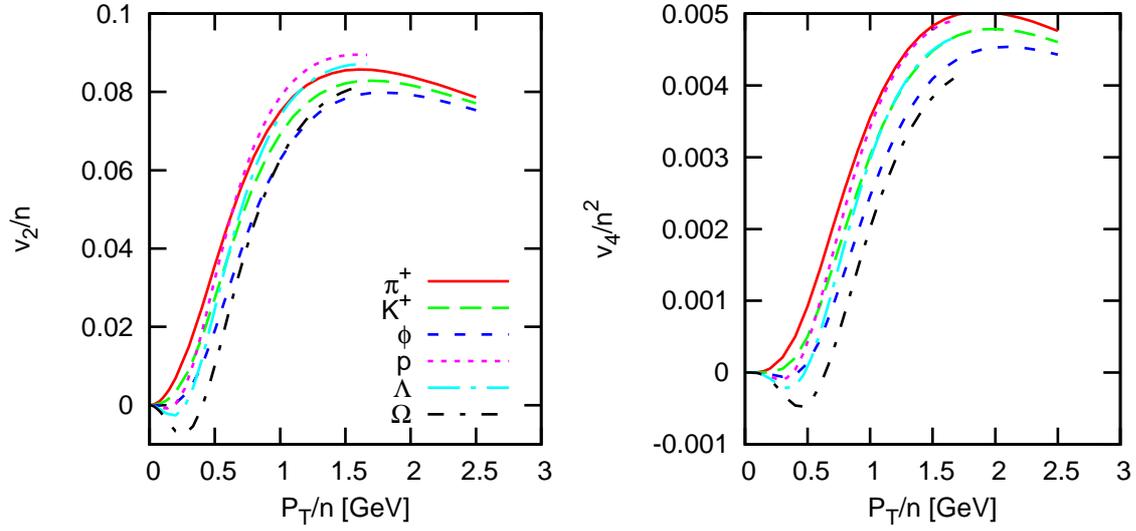}
\caption[Scaled elliptic and hexadecupole flow for different hadrons]{Scaled flow coefficients of different hadrons for a circular freeze-out.\newline
Left: Scaled elliptic flow. Right: Scaled hexadecupole flow.}
\label{fig:flow_contrib_vn}
\end{figure}

The flow contributions are studied by setting $e_f = 0$. These contributions are sensitive to the quark content and are the source of the CQNS discussed in sec.~\ref{sec:cqns_v2}. To simplify the comparison to the geometrical contributions, I also set $\partial_\rho \tau = 0$. The modifcations arising from $\tau=\tau(\rho)$ will be discussed later.
Fig.~\ref{fig:flow_contrib_vn} shows the scaled elliptic and hexadecupole flow for different hadrons.

The general trend of the CQNS is clearly visible which is one of the major successes of recombination. Despite the simplifications made in section~\ref{sec:cqns_v2}, the CQNS is at the heart of the formalism and therefore persists even when calculating with the full space-momentum correlations. Although the very simple scaling law is nice, it should not be taken to far, because the scaling is explicitly broken as discussed in section~\ref{sec:cqns_breaking}. Most obvious is the additional mass scaling where heavier particles have a smaller and even negative flow coefficient.

\begin{figure}
 \centering
\includegraphics{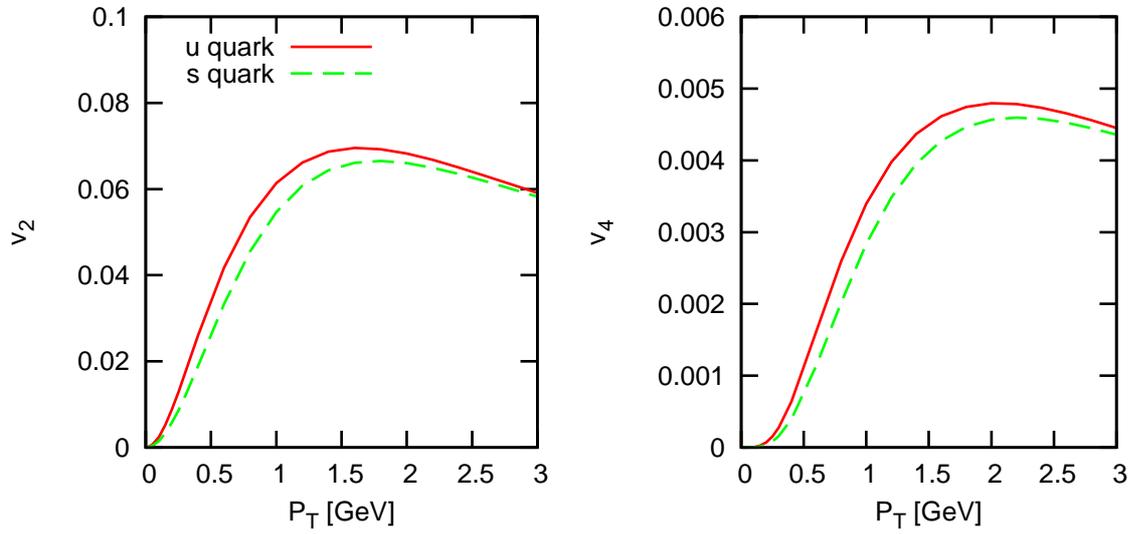}
\caption[Elliptic and hexadecupole flow for light quarks]{Flow coefficients of light quarks for a circular freeze-out. \newline
Left: Elliptic flow. Right: Hexadecupole flow.}
\label{fig:flow_contrib_quarks}
\end{figure}

The mass scaling between mesons and baryons comes from the sum of the constituent masses as explained in section~\ref{sec:cqns_breaking}. And the mass scaling among different mesons/baryons comes from the quark flow (Fig.~\ref{fig:flow_contrib_quarks}) where only the different masses lead to a different elliptic flow. This is also true for $v_4$. To have an approximate comparison to the scaled elliptic flow, Fig.~\ref{fig:flow_contrib_vn} shows a scaled hexadecupole flow with $v_4/n^2$ vs. $P_T/n$. The $1/n^2$ is inspired by an observed constant ratio $v_4/(v_2)^2$ (for details see sec.~\ref{sec:flow_ratios}).

\subsubsection{Relative strength of both contributions}
\label{sec:geometrical_contrib_strength}
While the flow contributions are fixed by the impact parameter, the geometrical contributions depend on the orientation of the freeze-out ellipse and its eccentricity. To relate the initial and the freeze-out eccentricity, I will use a simple linear ansatz
\begin{align}
e_f = c_e \cdot e_i
\label{eqn:relation_eccent_initial_final}
\end{align}
with the parameter $c_e < 1$.

To obtain a first approximation, let us compare the results from the previous chapter with a circular freeze-out ($c_e=0$) to actual data (see Fig.~\ref{fig:diff_v2_cent_circular} on page \pageref{fig:diff_v2_cent_circular}). As you can see the elliptic flow is generally underpredicted at low $p_T$. Since $p_0$ only models the high-$p_T$ damping, it has no effect to change it. The only possibility to raise the pion $v_2$ high enough would be to lower the quark mass to a few $\MeV$, which not only is very questionable, but also fails to reproduce the baryons.

Hence, to match the data, one seems to need some additional positive contributions from the geometry. This is achieved by setting $c_e \approx 0.7$. That means the freeze-out eccentricity should be about 70\% of the initial eccentricity. This also increases the high $p_T$ region which is compensated by decreasing $p_0$ to $0.65$.

This is a first hint that $e_f$ is not necessarily zero and needs to be incorporated. This is supported by STAR data \cite{nucl-ex/0312009}, where the initial and final eccentricity was calculated via the Glauber model and HBT interferometry. Their results indicate that the elliptic flow cannot quench its initial asymmetry completely.

As discussed in sec.~\ref{sec:asymmetry_parameters}, the numerical eccentricity $e$ is compared to the glauber eccentricity $\eps$ via
\begin{align}
 \eps =& \dfrac{1}{2}\dfrac{e^2}{2-e^2}\quad \mbox{or respectively}\\
e =& \sqrt{\dfrac{4\eps}{2\eps+1}}
\end{align}

\begin{figure}
 \centering
 \includegraphics{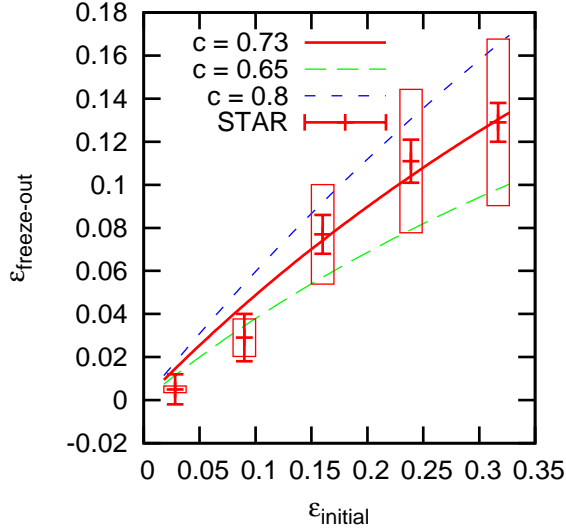}
 \caption[Comparison of the freeze-out eccentricity]{Comparison of the freeze-out eccentricity from eq.~\eqref{eqn:fit_eccent_initial_final} for different $c_e$ to STAR calculations \cite{nucl-ex/0312009}, where the freeze-out eccentricity has been obtained with azimuthally-sensitive HBT and the initial eccentricity was calculated within a Glauber model.
Uncertainties on the precise nature of space-momentum correlations lead to 30\% systematic errors on $\eps_f$.
}
 \label{fig:fit_eccent_initial_final}
\end{figure}

Together with the linear scaling of $e_f = c_e e_i$ this relates the initial and the freeze-out eccentricity as 
\begin{align}
 \eps_f =& \dfrac{1}{2}\dfrac{c^2 e_i^2}{2-c^2 e_i^2} =\dfrac{1}{2}\dfrac{c^2\dfrac{4\eps_i}{2\eps_i+1}}{2-c^2\dfrac{4\eps_i}{2\eps_i+1}}\nonumber\\
=& \dfrac{c^2\eps_i}{2\eps_i+1-2c^2\eps_i}.
\label{eqn:fit_eccent_initial_final}
\end{align}

Fig.~\ref{fig:fit_eccent_initial_final} shows the results of eq.~\eqref{eqn:fit_eccent_initial_final} to the STAR calculations, which gives $c_e=0.731 \pm 0.015 $ for the best fit case. This good agreement supports the simple linear ansatz in eq.~\eqref{eqn:relation_eccent_initial_final} and confirms the first estimate of $c_e \approx 0.7$. But the large systematic errors do not allow a firm conclusion, so $\eps\approx 0.65-0.8$ is consistent with errors.

This data also clearly indicates a positive $\eps_f$ which means that the orientation has not changed compared to the initial ellipse and it is still elongated out-of-plane. From what I already said in sec.~\ref{sec:geometrical_contrib}, I conclude that the transverse expansion of the system is perpendicular to the surface and drop the surface flow model no. 2 in favor of model 1 (sec.~\ref{sec:surface_flow}).

\subsection{Correlation between \texorpdfstring{$\tau$}{tau} and \texorpdfstring{$\rho$}{rho}}
\label{sec:tau_rho_correlation}
So far I only studied the elliptic flow with a constant freeze-out time. To be a bit more general I assume a linear dependence on the radial coordinate as
\begin{align}
 \tau = \lambda \rho + \tau' \quad\mbox{and therefore}\quad \partial_\rho \tau = \lambda.
\end{align}
For a given mean freeze-out time $\tau_0$ it follows
\begin{align}
 \tau' = \tau_0 -\dfrac{2}{3}\rho_0 \lambda
\end{align}
With $\lambda=0$ a constant freeze-out time $ \tau=\tau_0$ is recovered. This is a quite simple ansatz, but it suffices to study a general dependence.

\begin{figure}[p]
 \centering
 \includegraphics{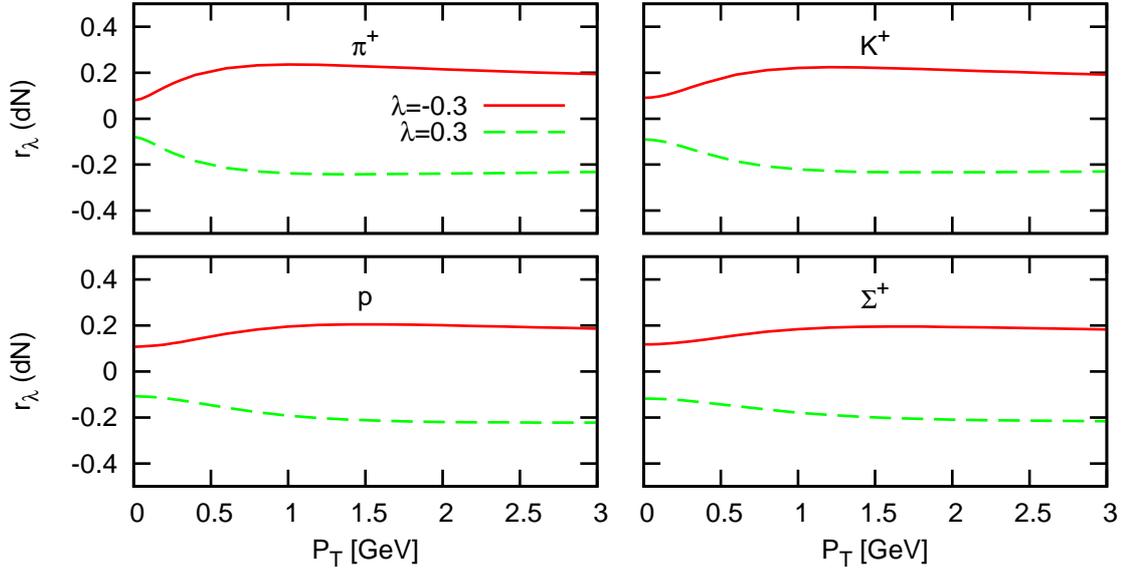}
 \caption[The relative deviations $r_\lambda (dN)$ of the invariant yields with $\tau-\rho$ correlations]{The relative deviations $r_\lambda (dN)$ of the invariant yields for $\lambda=\pm 0.3$ to $\lambda = 0$ in central collisions ($b=0\fm$).}
 \label{fig:tau_rho_yield}
\end{figure}

The dependence of the yield on $\lambda$ is shown in Fig.~\ref{fig:tau_rho_yield} for different hadrons. The relative deviations $r_\lambda (dN)=\left(dN_{\lambda} -dN_{\lambda = 0}\right) / dN_{\lambda=0}$ is similiar for mesons and baryons and approximately independent of $p_T$ with about -0.2 for $\lambda=0.3$ and 0.2 for $\lambda=-0.3$.

\begin{figure}[p]
 \centering
 \includegraphics{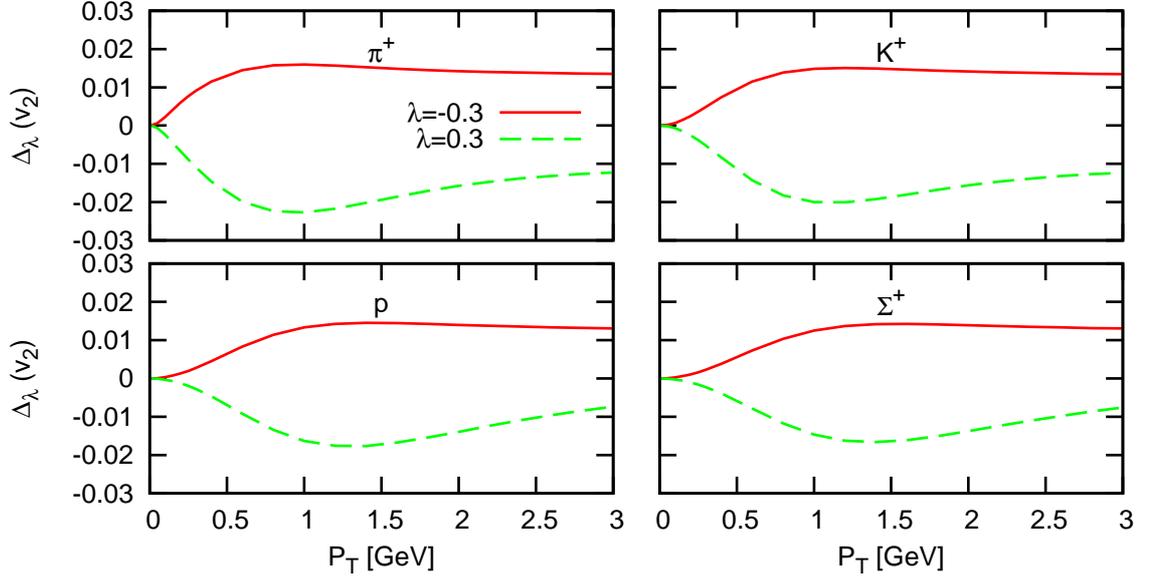}
 \caption[The deviations $\Delta_\lambda (v_2)$ of the elliptic flow with $\tau-\rho$ correlations]{The deviations $\Delta_\lambda (v_2)$ of the elliptic flow for $\lambda=\pm 0.3$ to $\lambda = 0$ in mid-central collisions ($b=8\fm$) for an elliptic freeze-out ($c_e=0.7$).}
 \label{fig:tau_rho_vn}
\end{figure}

The deviations $\Delta_\lambda (v_2)=v_2^{(\lambda)} - v_2^{(\lambda=0)}$ of the elliptic flow for different $\lambda$ are shown in Fig.~\ref{fig:tau_rho_vn} for an elliptic freeze-out with $c_e=0.7$. For a circular freeze-out the deviations are neglectable. So non-zero $\lambda$ has the largest effects on $v_2$ at low $p_T$. This behaviour is similar for the hexadecupole flow.

\clearpage
\section{Transverse momentum spectra}
In this section I show hadron production for Au+Au collisions at $\sqrt{s}=200 \GeV$ from recombination. For the LHC predictions of Pb+Pb collisions at $\sqrt{s}=5.5 \TeV$ I will use the same nuclear size $\rho_0$, since the radius only enters an overall factor and there is no data yet available to compare to.
\subsection{Yields}
\label{sec:yields}

\begin{figure}
 \centering
\includegraphics[width=\textwidth]{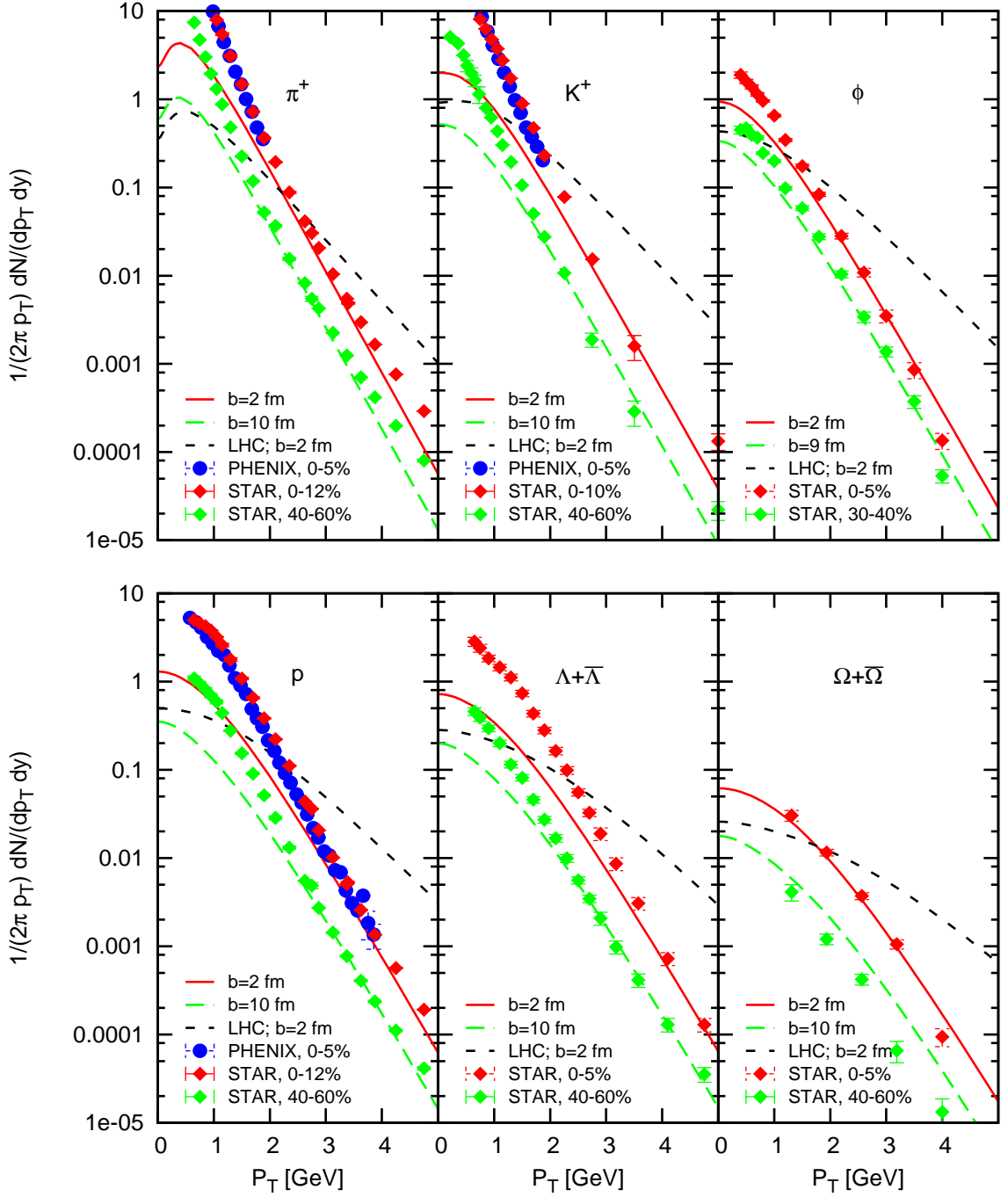}
\caption[Transverse momentum spectra of hadrons for Au+Au collisions at $\sqrt{s}=200\GeV$]{Transverse momentum spectra of hadrons for Au+Au collisions at $\sqrt{s}=200\GeV$ compared to pion, kaon and proton data \cite{nucl-ex/0209027} from PHENIX (blue points), pion and proton data \cite{nucl-ex/0606003}, kaon data \cite{Ming:2008zz}, $\phi$ data \cite{nucl-ex/0703033}, $\lambda$ and $\omega$ data \cite{nucl-ex/0606014} from STAR (red and green points).}
\label{fig:yields}
\end{figure}

Fig.~\ref{fig:yields} shows the transverse momentum spectra of light and strange hadrons compared to data from PHENIX \cite{nucl-ex/0209027} and STAR \cite{nucl-ex/0606003,Ming:2008zz,nucl-ex/0703033,nucl-ex/0606014} for different centralities. In the mid $p_T$ range, the results agree with the data. But below $2\GeV$ and especally below $1\GeV$, the data is significantly underpredicted for hadrons with light quark content. This is due to the energy violation in the recombination formalism and also the contributions from resonance decay are not incorporated. For hadrons with only strange quark content, namely $\phi$ and $\Omega$, the predictions are in good agreement even down to very low $p_T$, since for the massive particles the energy violation is not that strong and the contributions from resonance decay are minor. The black lines correspond to LHC predictions for an impact parameter of $b=2\fm$.

\begin{figure}
 \centering
\includegraphics{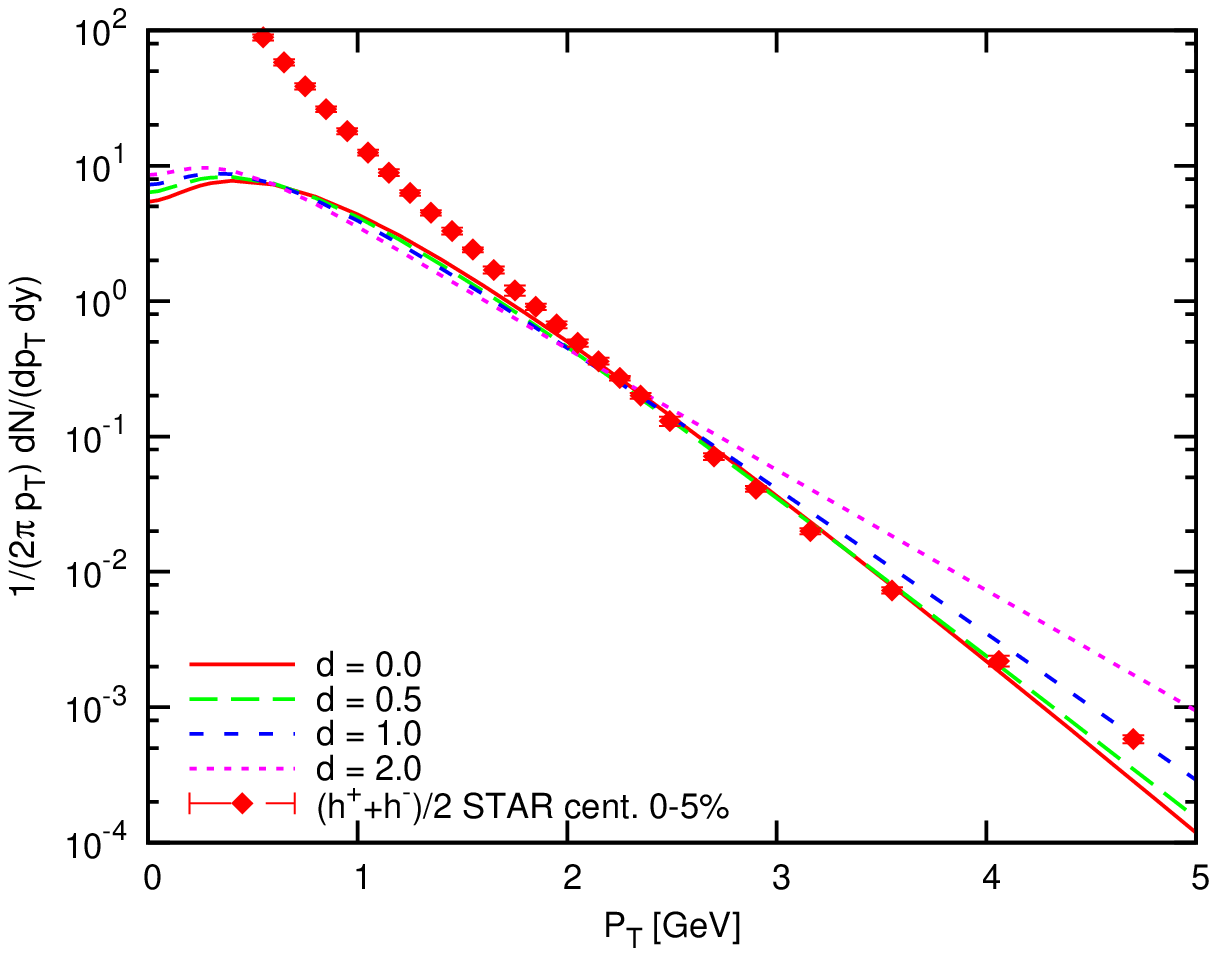}
\caption[Charged hadron yield for different radial profiles of the transverse rapidity]{Comparison of the charged hadron yield for different radial profiles of the transverse expansion rapidity ($d_\rho$) to data from STAR \cite{nucl-ex/0305015}.}
\label{fig:d_rho_yield_compare}
\end{figure}

Fig.~\ref{fig:d_rho_yield_compare} shows the transverse momentum spectrum of charged hadrons compared to data from STAR \cite{nucl-ex/0305015} for different values of $d_\rho$, the parameter for the radial profile of the transverse rapidity. While the results are equal at low $p_T$, they differ for $p_T > 3 \GeV$ and the best fit is achieved for about $d_\rho=0.5-1$. But the large deviations from the data at low $p_T$ do not allow for a firm conclusion on that parameter. Another analysis is given in sec.~\ref{sec:flow_ratios-d_rho}, where I study the dependence of the flow ratios on this parameter. For now I will use $d_\rho=1$.
The other parameters are $c_e=0.73$ and $\lambda=-0.3$ as discussed in the previous sections. The negative $\lambda$ gives a constant factor of about 1.2 while the freeze-out eccentricity does not affect the transverse momentum spectra, since it is chosen to preserve the total transverse freeze-out area.

\subsection{Mean transverse momentum}
\begin{figure}
 \centering
 \includegraphics{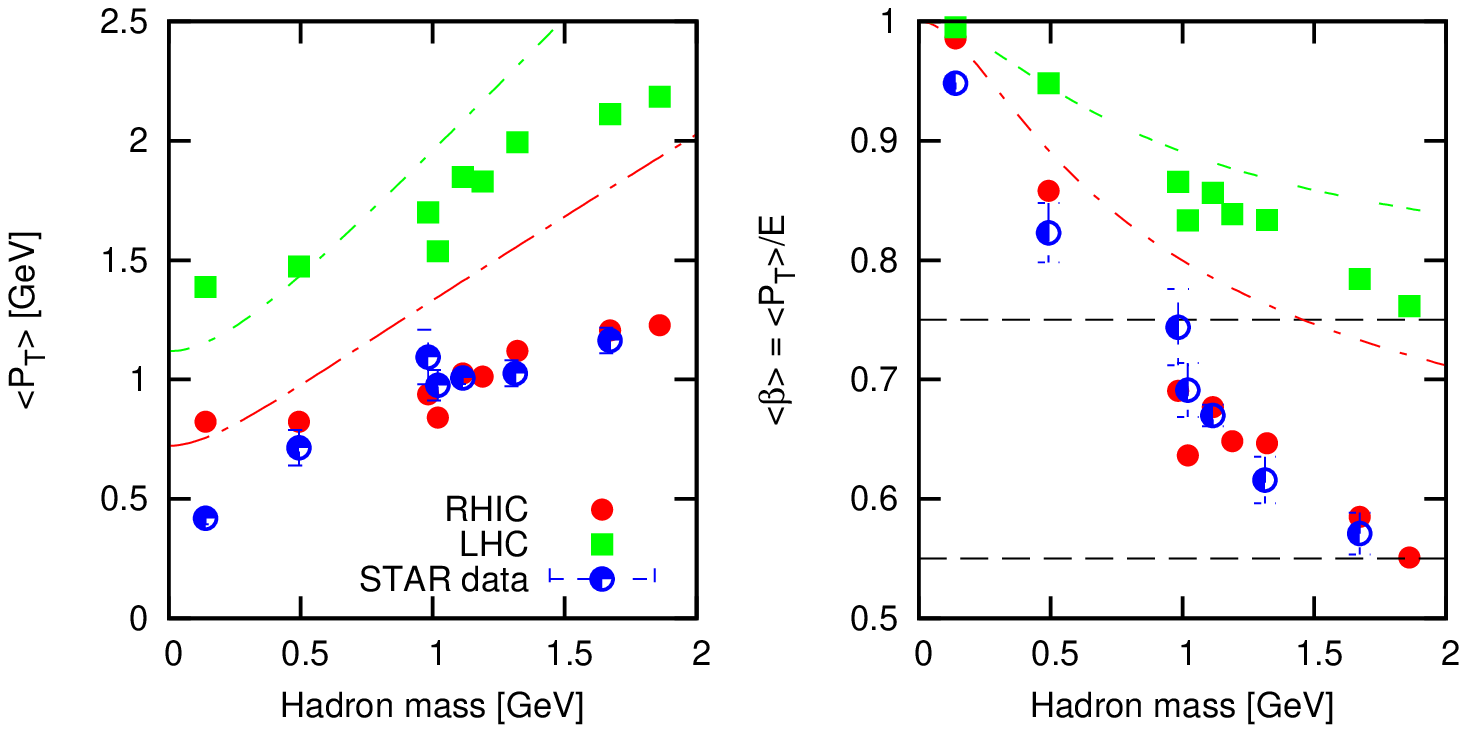}
 \caption[Mean transverse momentum and velocity of hadrons as a function of the hadron mass]{Mean transverse momentum (Left) and velocity (Right) of hadrons $\left\langle \beta \right\rangle = \left\langle P_T\right\rangle / E$ as a function of the hadron mass at RHIC (red) and LHC (green). Data points (blue) are from STAR \cite{0809.4737} and the black lines are the theoretical expectation within the equipartition theorem.}
 \label{fig:mean_pt_hadronmass}
\end{figure}
Fig.~\ref{fig:mean_pt_hadronmass} shows the mean transverse momentum $\langle p_T \rangle$ as a function of the hadron mass in the left panel. The results for RHIC (red points) are compared to data from STAR \cite{0809.4737} (blue points). At low hadron mass, namely for pions and kaons, the results for the mean $p_T$ overestimate the data, since the low $p_T$ yield is underestimated as shown in the previous paragraph. For protons and heavier hadrons, the deviations for the yield are smaller and therefore the predicitions of the mean $p_T$ are in much better agreement.

Additionally, in the right panel I show the mean transverse velocity $\left.\langle v_T \rangle\right\vert_{y=0} =\dfrac{\langle p_T \rangle}{E} =\dfrac{\langle p_T \rangle}{\langle m_T \rangle} $ at midrapidity calculated from the mean $p_T$. As one can see, this quantity softens the deviations for the light mesons. For masses of about $2 \GeV$ the velocity approaches the value of the mean transverse expansion $\beta_T=0.55$ (black lines) at RHIC. Similiar the predictions for the LHC (green points) approach $\beta_T=0.75$.

The figure also depicts calculations from the equipartition theorem (red/green lines), which states
\begin{align}
 \left\langle p_m \dfrac{\partial H}{\partial p_n}\right\rangle = \delta_{mn} k_B T
\end{align}
with the temperature $T$ and the Boltzmann constant $k_B$ (which is equal to 1 in natural units). Using the energy $H=E=p_\mu u^\mu$ from eq.~\eqref{eqn:local_frame_energy} in the local rest frame, this leads to
\begin{align}
T=\left\langle p_z \dfrac{\partial H}{\partial p_z}\right\rangle 
&= p_z^2 \dfrac{\cosh(\eta_T)}{E}\\
&\nonumber\\
2T= p_T^2 \dfrac{\cosh(\eta_T)}{E} - p_T \cos(\varphi-\Phi) \sinh(\eta_T)=&\left\langle p_T \dfrac{\partial H}{\partial p_T}\right\rangle \nonumber\\
 &= \left\langle p_x \dfrac{\partial H}{p_x}\right\rangle + \left\langle p_y \dfrac{\partial H}{\partial p_y}\right\rangle
\end{align}
These equations are coupled via $E=\sqrt{p_T^2+p_z^2+m^2}$ so they need to be solved numerically.

Interestingly, the values for hadron masses above $1\GeV$ are larger then the results from recombination, if one uses the same temperature $T=0.175 \GeV$ and transverse velocity $\beta_T=0.55$ or respectively $\beta_T=0.75$. Therefore, the mean velocity $\langle \beta_T \rangle$ reaches these ``limiting'' values much later. When using $T$ and $\beta_T$ as fit parameters, the equipartition theorem reproduces the recombination results for $T=0.324 \GeV$, $\beta_T=0.0$ for RHIC and $\beta_T=0.45$ for LHC.

\subsection{Hadron ratios}
\label{sec:hadron_ratios}
\begin{figure}
 \centering
 \includegraphics{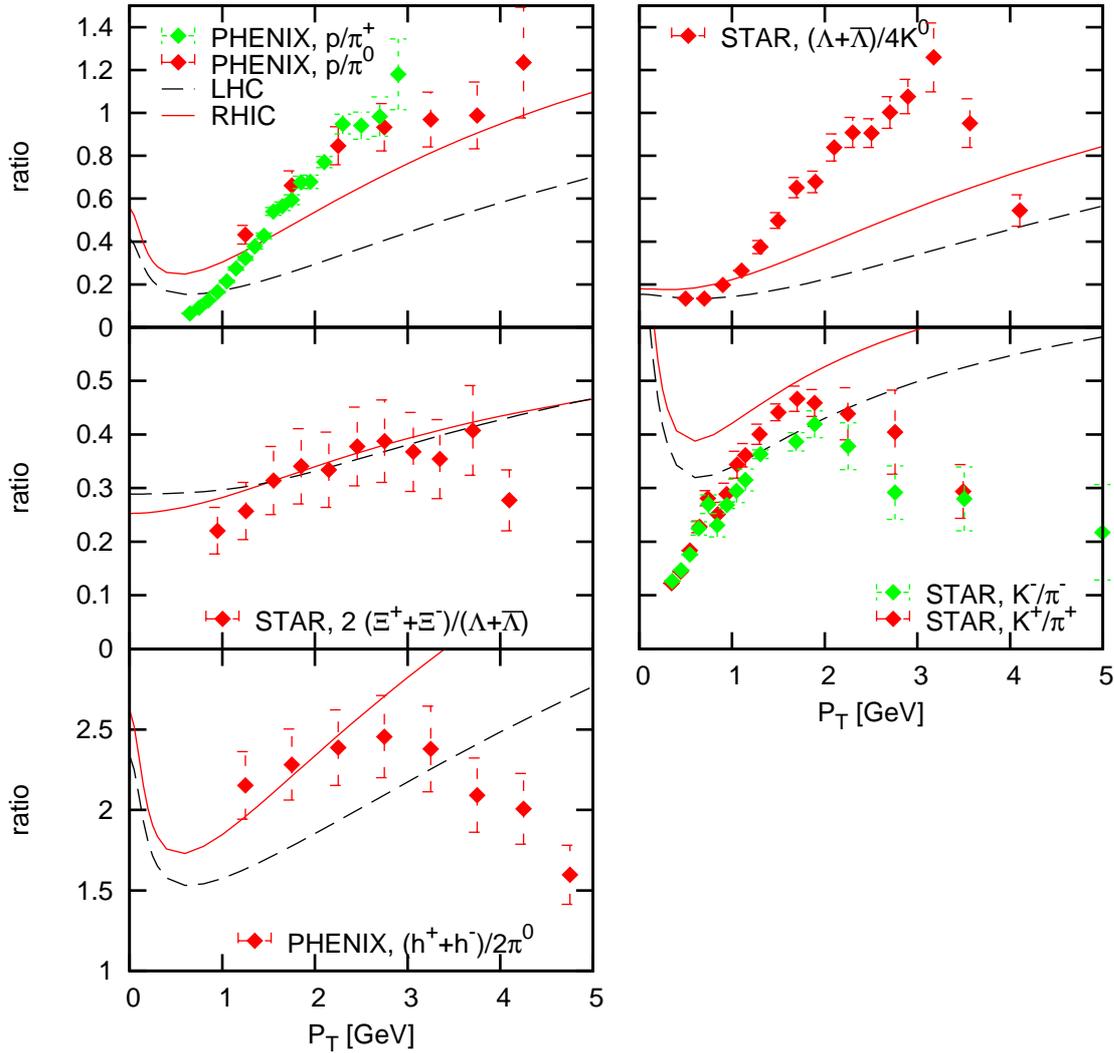}
 \caption[Invariant yield ratios for different hadrons as a function of $P_T$]{Invariant yield ratios for different hadrons as a function of $P_T$ compared to data from PHENIX \cite{nucl-ex/0305036} and STAR \cite{Long:2004cf,Ming:2008zz}.}
 \label{fig:hadron_ratios}
\end{figure}

The large proton/pion ratio at mid $p_T$ was one of the puzzles that fragmentation failed to describe and thus, it is one of the motivations for studying recombination. So, Fig.~\ref{fig:hadron_ratios} shows the ratios of the invariant yields from different hadrons compared to data from PHENIX \cite{nucl-ex/0305036} and STAR \cite{Long:2004cf,Ming:2008zz}.
The low $p_T$ region is mainly overestimated which can be attributed to the failed description of the $p_T$ spectra at low $p_T$, due to the energy violation. The mid $p_T$ results are between a mediocre and good agreement, apart from the $\lambda$ to $K^0$ ratio which is about a factor of 2 smaller. The main point is the prediction of the large proton to pion ratio of about 1. The high $p_T$ region is badly predicted and would require the combined treatment of recombination and fragmentation \cite{Fries:2003kq}.

The black lines correspond to LHC predictions, but with the same strange fugacity $\gamma_s=0.8$ which would have to be extracted from fits and can be expected to be larger.

\clearpage
\section{Flow coefficients \texorpdfstring{$v_2$}{v\_2} and \texorpdfstring{$v_4$}{v\_4}}
In this section, I study the elliptic and hexadecupole flow at RHIC and LHC. I will start with the eccentricity dependence of the mean flow coefficients. The findings will motivate the investigation of the flow ratio $v_4/(v_2)^2$ which will support the findings in sec.~\ref{sec:geometrical_contrib_strength} about the freeze-out eccentricity.

This ratio also gives the opportunity to study a $\tau-\rho$ correlation and the radial profile of the transverse expansion rapidity. The obtained values for the corresponding parameters in the sections~\ref{sec:flow_ratios-lambda} and \ref{sec:flow_ratios-d_rho} will be used in the comparison to experimental data of the differential flow coefficients.

\subsection{Mean flow coefficients}
The mean elliptic or hexadecupole flow $\left\langle v_n \right\rangle$ at midrapidity is the $p_T$ integrated average folded by the transverse momentum distribution:
\begin{align}
 \left\langle v_n \right\rangle = \dfrac{\int \dn p_T\, v_n(p_T) \dfrac{\dn N}{\dn p_T}}{\int \dn p_T\, \dfrac{\dn N}{\dn p_T}}
\end{align}

\subsection{Eccentricity dependence}
\label{sec:eccent_scaling}
The strength of the flow coefficients $v_n$ depends on the impact parameter or respectively the eccentricity. To study the dependence, I will compare the mean, $p_T$-integrated flow $\left\langle  v_n \right\rangle$ to the eccentricity $\eps$. Looking at equation eq.~\eqref{eqn:asymmetry_expansion}, one expects the mean $v_2$ to scale mainly linear with $\eps$ and $v_4$ to have a larger quadratic contribution.

\begin{figure}
 \centering
 \includegraphics{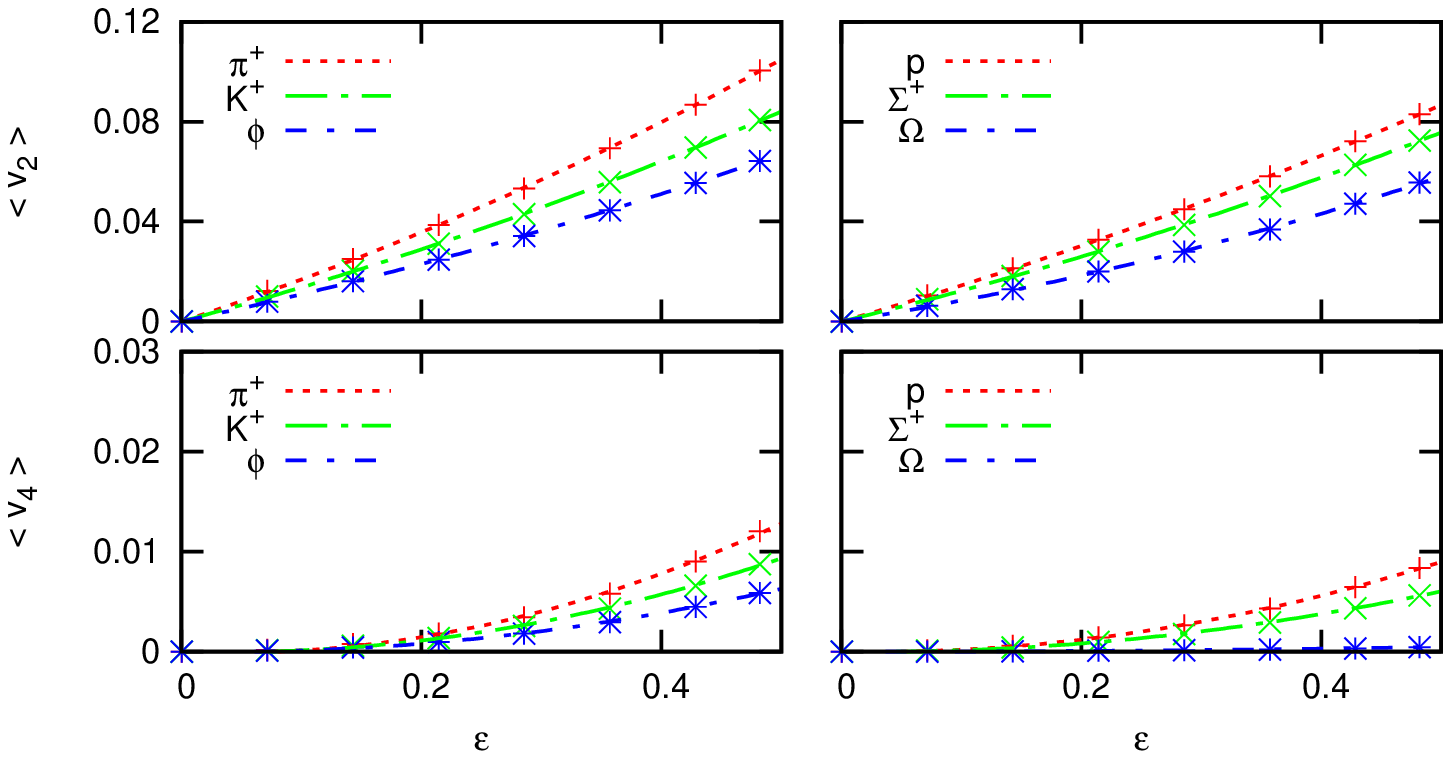}
 \caption[Mean flow $\left\langle v_n\right\rangle $ as a function of the eccentricity $\eps$]{Mean flow $\left\langle v_n\right\rangle $ as a function of the eccentricity $\eps$ for mesons (left) and baryons (right) from a circular freeze-out ($c_e=0$). The lines are fitted by $\left\langle v_n\right\rangle (\eps)=a\eps+b\eps^2$ with the values from Tab.~\ref{tab:flow_eccent_fits}.\newline
Upper: Mean elliptic flow. Lower: Mean hexadecupole flow.}
 \label{fig:vn_eccent_0.0}
\end{figure}

\begin{figure}
 \centering
 \includegraphics{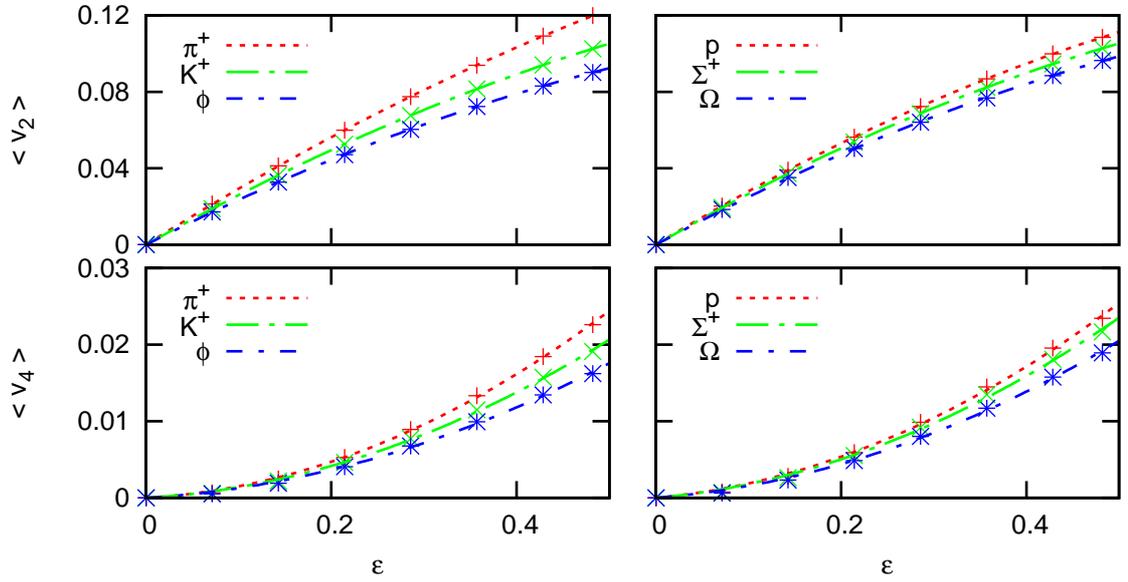}
 \caption[Mean flow $\left\langle v_n\right\rangle $ as a function of the eccentricity from an elliptic freeze-out]{Mean flow $\left\langle v_n\right\rangle $ from an elliptic freeze-out with $c_e=0.73$. Tab.~\ref{tab:flow_eccent_fits} shows the fit values.}
 \label{fig:vn_eccent_0.7}
\end{figure}

Therefore, the calculations are fitted by $\left\langle v_n\right\rangle (\eps)=a\eps+b\eps^2$ and shown in Figs.~\ref{fig:vn_eccent_0.0} and \ref{fig:vn_eccent_0.7}. The fit values support the expected behaviour and can be found in Tab.~\ref{tab:flow_eccent_fits}.

\begin{table}[hb]
\centering
\caption[Fit values for the function $\left\langle v_n\right\rangle (\eps)=a\eps+b\eps^2$]{Fit values for the function $\left\langle v_n\right\rangle (\eps)=a\eps+b\eps^2$ as shown in Figs.~\ref{fig:vn_eccent_0.0} and~\ref{fig:vn_eccent_0.7}.}
\label{tab:flow_eccent_fits}
\begin{scriptsize}

\begin{tabular}{lllllll}
\toprule
& $\pi^+$ & $K^+$ & $\phi$ & $p$ & $\Sigma^+$ & $\Omega$ \\
\midrule
\multirow{2}{*}{$v_2 \mbox{ mit } c_e=0.0$}
& $a=$1.58e-01 & 1.28e-01 & 1.02e-01 & 1.37e-01 & 1.15e-01 & 7.57e-02\\
& $b=$1.04e-01 & 7.95e-02 & 6.39e-02 & 7.26e-02 & 7.20e-02 & 8.06e-02\\
\cmidrule{1-7}
\multirow{2}{*}{$v_2 \mbox{ mit } c_e=0.73$}
& 3.05e-01 & 2.72e-01 & 2.47e-01 & 2.94e-01 & 2.80e-01 & 2.63e-01\\
& -1.17e-01 & -1.23e-01 & -1.25e-01 & -1.43e-01 & -1.40e-01 & -1.31e-01\\
\cmidrule{1-7}
\multirow{2}{*}{$v_4 \mbox{ mit } c_e=0.0$}
& -4.92e-03 & -3.13e-03 & -1.75e-03 & -2.00e-03 & -1.21e-03 & -7.62e-05\\
& 6.11e-02 & 4.36e-02 & 2.86e-02 & 4.00e-02 & 2.65e-02 & 2.06e-03\\
\cmidrule{1-7}
\multirow{2}{*}{$v_4 \mbox{ mit } c_e=0.73$}
& 6.67e-03 & 7.05e-03 & 7.08e-03 & 1.08e-02 & 1.05e-02 & 9.36e-03\\
& 8.43e-02 & 6.85e-02 & 5.61e-02 & 8.02e-02 & 7.31e-02 & 6.33e-02\\
\cmidrule{1-7}
\end{tabular}
\end{scriptsize}
\end{table}

\subsubsection{Fluctuations}
For a fixed impact parameter, the eccentricity is expected to fluctuate \cite{0708.0800,0706.4266}. Since the flow components scales with powers of $\eps$, the fluctuations in the $p_T$-integrated $\left\langle v_n\right\rangle$ can be related to the eccentricity fluctuations. To study these fluctuations, I take the fits from table~\ref{tab:flow_eccent_fits}, where I used
\begin{align}
\left\langle v_n\right\rangle(\eps) = a \eps +b \eps^2.
\end{align}
I assume a gaussian distribution of the eccentricity around the mean expected eccentricity $\overline{\eps}$ as
\begin{align}
 p(\eps) = N \exp{-\dfrac{1}{2} \dfrac{\left(\eps-\overline{\eps}\right)^2}{\sigma^2(\eps)}}
\end{align}
with a width $\sigma$ that generally can depend on the mean eccentricity. The mean of the elliptic flow distribution can then be calculated as
\begin{align}
\overline{\langle v_n \rangle} = b\left(\sigma^2(\eps)+\bar{\eps}^2\right)+a\bar{\eps}
\end{align}
and the fluctuations are 
\begin{align}
\sigma\left(\left\langle v_n\right\rangle\right) = \sigma(\eps) \sqrt{2b^2(2\overline{\langle v_n \rangle} - b \sigma^2(\eps))+a^2}
\end{align}
where $\left\langle x\right\rangle$ is the $p_T$-integrated mean and $\overline{x}$ denotes the fluctuation mean.

The relative fluctuations for a purely linear dependence ($b=0$) simplify to
\begin{align}
 \dfrac{\sigma(\left\langle v_n\right\rangle)}{\overline{\langle v_n \rangle}}= \dfrac{\sigma(\eps)}{\overline{\eps}}.
\label{eqn:fluct_b0}
\end{align}
For a purely quadratic dependence ($a=0$) the relative fluctuations scale with $\sqrt{b}$ as
\begin{align}
 \dfrac{\sigma(\left\langle v_n\right\rangle)}{\overline{\langle v_n \rangle}} \approx 2 \sqrt{b} \left(\dfrac{\sigma(\eps)}{\left\langle \eps\right\rangle}-\dfrac{3}{4}\left(\dfrac{\sigma(\eps)}{\left\langle \eps\right\rangle}\right)^3\right)
\label{eqn:fluct_a0}
\end{align}

Looking at the fit values in table~\ref{tab:flow_eccent_fits}, the quadratic contributions for the elliptic flow are small compared to the linear ones ($a > b$) and vice versa for the hexadecupole flow ($b > a$).

\begin{figure}
 \centering
\includegraphics{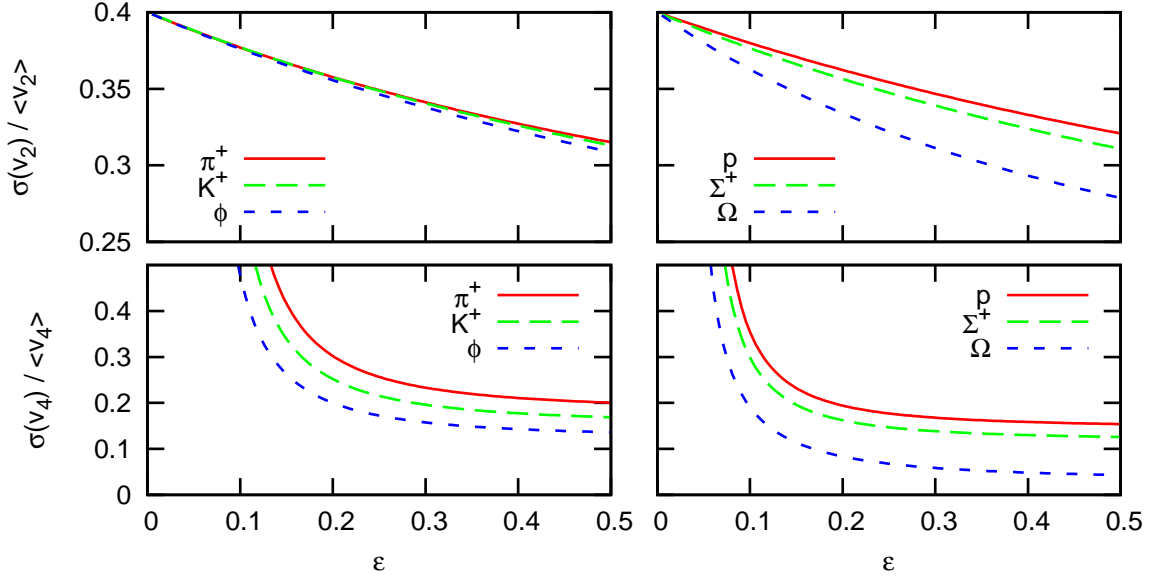}
\caption[The relative flow fluctuations for a circular freeze-out]{The relative flow fluctuations for a circular freeze-out ($c_e=0$) and $\sigma(\eps) / \langle \eps\rangle \approx 0.4$.}
\label{fig:vn_fluct_0.0}
\end{figure}

\begin{figure}
 \centering
\includegraphics{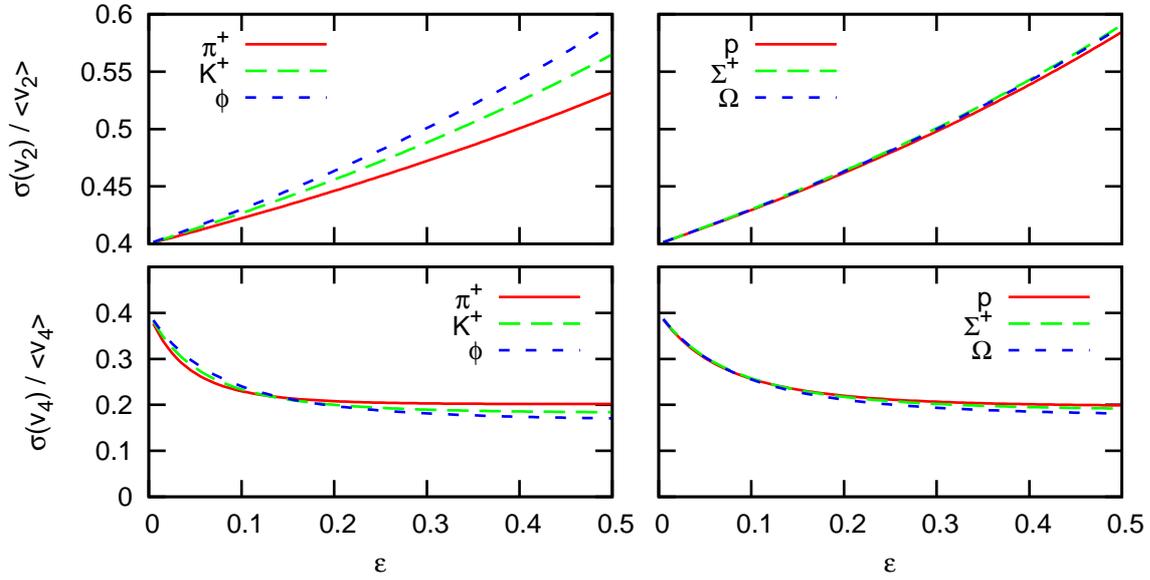}
\caption[The relative flow fluctuations for an elliptic freeze-out]{The relative flow fluctuations for an elliptic freeze-out ($c_e=0.73$) and $\sigma(\eps) /\langle \eps\rangle \approx 0.4$.}
\label{fig:vn_fluct_0.7}
\end{figure}

The relative fluctuations of the eccentricity are predicted to be approximately constant with $\dfrac{\sigma(\eps)}{\overline{\eps}} \approx 0.4$ \cite{0707.0249, nucl-ex/0702036, nucl-ex/0612021}. As can be seen in Figs.~\ref{fig:vn_fluct_0.0} and \ref{fig:vn_fluct_0.7}, the relative fluctuations of the elliptic flow is almost independent of the particle type and also comparable to the fluctuations of the eccentricity as eq.~\ref{eqn:fluct_b0} suggests. On the other hand the relative hexadecupole fluctuations are much smaller and approch the value from eq.~\ref{eqn:fluct_a0} for large $\eps$. The divergence of $\sigma(v_4) / \overline{\left\langle v_4\right\rangle}$ at low $p_T$ is due to the negative fit value of $a$, so that the denominator becomes zero at a finite $\eps$.

\subsection{Flow ratio}
\label{sec:flow_ratios}
As was shown in the last section, for a circular freeze-out the different flow coefficients $v_n$ will scale with the eccentricity mainly as $\eps^{n/2}$. Therefore ratios of these coefficients are an interesting probe to study, since the flow contributions will be insensitive to the initial geometry. The geometrical contributions from an elliptic freeze-out on the other hand can have great influence on these ratios. So they offer the possibility to verify the statements about the freeze-out eccentricity. One can also take a closer look on the influence of a $\tau-\rho$ correlation ($\lambda \not=0$) and the radial profile of transverse expansion ($d_\rho$).

\begin{figure}
 \centering
\includegraphics{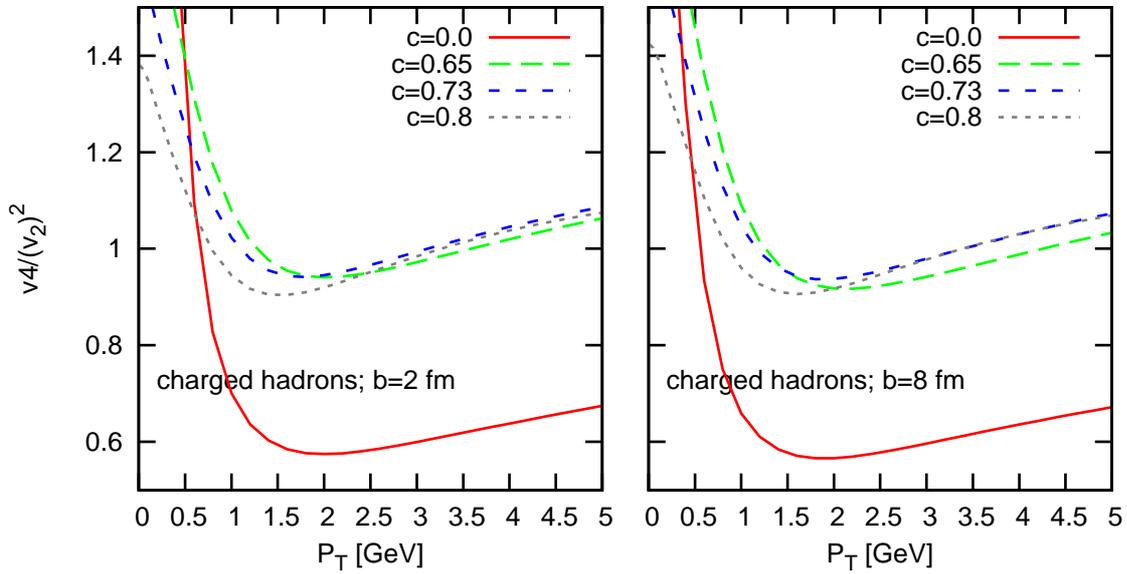}
\caption[Comparison of the ratio $v_4/(v_2)^2$ from an ellipsoidal freeze-out for different strength]{The ratio $v_4/(v_2)^2$ for charged hadrons from an ellipsoidal freeze-out with different $c_e$ and two impact parameters.}
\label{fig:v42_ratio_study}
\end{figure}
I will focus on the ratio $\dfrac{v_4}{v_2^2}$, that I already discussed shortly in framework of the CQNS (sec.~\ref{sec:cqns_v4}). For a quick approximation of the value for the quarks I compare the prefactors from the transverse rapidity (eq.~\eqref{eqn:asymmetry_expansion}): $v_2 \sim \eps$ and $v_4\sim\dfrac{1}{4}\eps^2$, which yields $\left[\dfrac{v_4}{v_2^2}\right]^{\mbox{quark}} \approx\dfrac{1}{4}$. This will turn out to be too small. Inserting a more realistic value of 1 in the scaling law from section~\ref{sec:cqns_v4} gives constant values of
\begin{align}
\left[\dfrac{v_4}{v_2^2}\right]^{\mbox{meson}}
\approx \dfrac{1}{4} + \dfrac{1}{2} = \dfrac{3}{4}
\end{align}
and
\begin{align}
  \left[\dfrac{v_4}{v_2^2}\right]^{\mbox{baryon}}
\approx \dfrac{1}{3} + \dfrac{1}{3} = \dfrac{2}{3}
\end{align}
So the simple cqns formulas suggest a ratio around 0.7. But these values are much smaller than the experimental values from STAR \cite{nucl-ex/0701044} (with $\sim 1.2$) and PHENIX \cite{0804.4864} (with $\sim 0.9$ ). Also the full calculations give a similiar value of about 0.7 from mid to high $p_T$. That indicates that we have to consider contributions from geometrical effects.

Fig.~\ref{fig:v42_ratio_study} shows the flow ratio for charged hadrons. The lines correspond to different percentages $c_e$ of the freeze-out eccentricity to the initial one. The left figure corresponds to an impact parameter of $b=2\fm$, the right one to $b=8\fm$. The good agreement between both supports the predicted independence from the eccentricity.

At low $p_T$ the $v_4/(v_2)^2$ ratio is much greater than $1$ for all percentages of the freeze-out eccentricity. At mid $p_T$ the effect of an elliptic freeze-out is most clearly visible. While for $c_e=0$ the ratio is about 0.6, it rises to about $1$ for $c_e=0.73$. The three curves with $c_e=0.65,0.73$ and $0.8$, which are the estimates from sec.~\ref{sec:geometrical_contrib_strength} within errors, only differ at low $p_T$. For these three values the ratio is in the range of the experimental data. So to compare the results to experimental data, I will take the best fit case of $c_e=0.73$ and in the next two paragraphs I look at the dependence on two other parameters, namely $\lambda$ and $d_\rho$.

\subsubsection{Influence of a \texorpdfstring{$\tau-\rho$}{tau-rho} correlation}
\label{sec:flow_ratios-lambda}
\begin{figure}
 \centering
\includegraphics{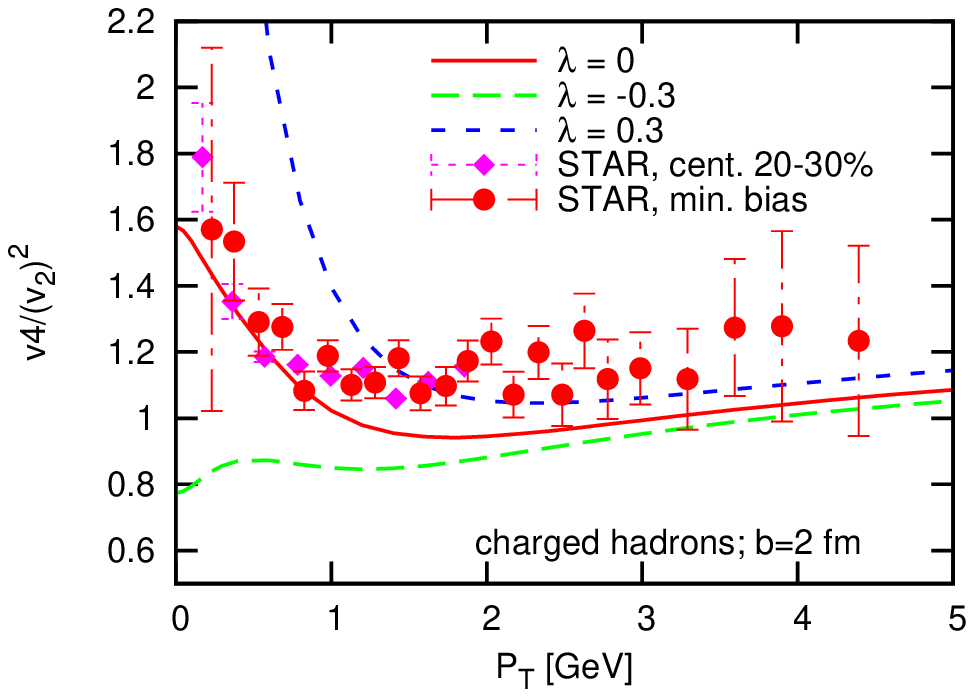}\\
\includegraphics{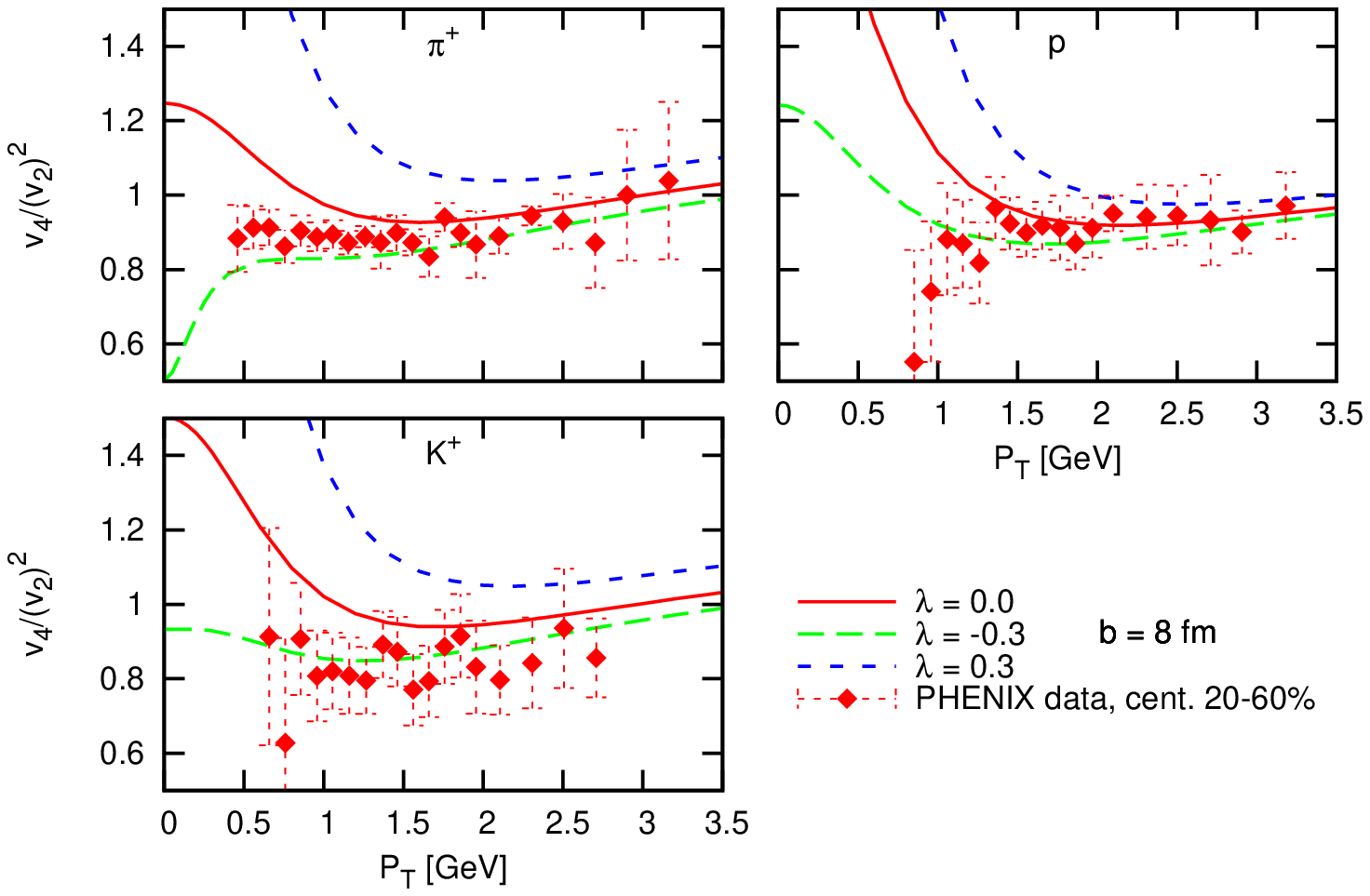}
\caption[The flow ratio $v_4/(v_2)^2$ for an ellipsoidal freeze-out compared to STAR and PHENIX data with different $\tau-\rho$ correlations]{The flow ratio $v_4/(v_2)^2$ for an ellipsoidal freeze-out with $c_e=0.73$. The influence of a $\tau-\rho$ correlation is depicted with $\lambda=0,\pm0.2$.
Upper: Ratio at $b=2\fm$ for charged hadrons compared to data from STAR \cite{nucl-ex/0701044,nucl-ex/0403019}.
Lower: Ratio at $b=8\fm$ for pions, kaons and protons compared to data from PHENIX \cite{0804.4864}.}
\label{fig:v42_ratio}
\end{figure}

This is dependence is studied by varying the parameter $\lambda$ (see section~\ref{sec:tau_rho_correlation}). Therefore I fix $d_\rho=1$, which means a linear growth if the transverse rapidity. The upper panel in Fig.~\ref{fig:v42_ratio} shows the ratio for charged hadrons at $b=2\fm$ for $\lambda=0,\pm0.3$ compared to data from STAR \cite{nucl-ex/0701044, nucl-ex/0403019}. A constant freeze-out time, i.e $\lambda=0$, seems to give the best agreement. This different to the lower panel with $v_4/(v_2)^2$ at $b=8\fm$ for pions, protons and kaons, where at high $p_T$ the value is independent of $\lambda$, but the predictions for $\lambda=0$ at low $p_T$ overestimate the data. A very good agreement at least for the mesons can be found with $\lambda=-0.3$. A (negatively) larger $\lambda$ for baryons could increase the agreement at low $p_T$, but the very sharp drop of the proton ratio can not be accounted for. And it also would enhance the discrepancy to the charged hadron data from STAR.

A negative $\lambda$ describes a fireball, where the outer particles freeze-out earlier ($\tau_o$ than the inner ones ($\tau_i$). With $\lambda=-0.3$ and a mean freeze-out time $\tau_0=5\fm$, this corresponds to $\tau_o=4.1\fm$ and $\tau_i=6.8\fm$. The data does not allow to draw a firm conclusion about the value of $\lambda$, since it also could (or even should) depend on the particle species or on their in-medium cross sections respectively. But the case of a large positive $\lambda$, where the $\tau_o> \tau_i$, can be discarded.

\subsubsection{Influence of the radial rapidity profile}
\label{sec:flow_ratios-d_rho}
The radial profile of the transverse rapidity is set by the parameter $d_\rho$ (see section~\ref{sec:d_rho_dep}). So this time I use a constant $\lambda=0$.

\begin{figure}
 \centering
\includegraphics{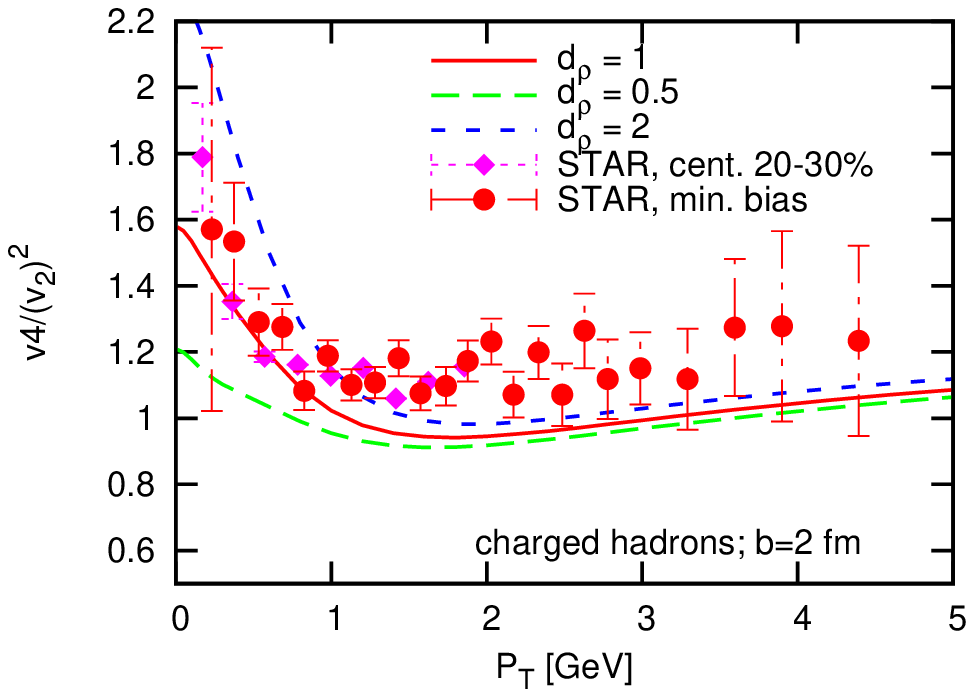}\\
\includegraphics{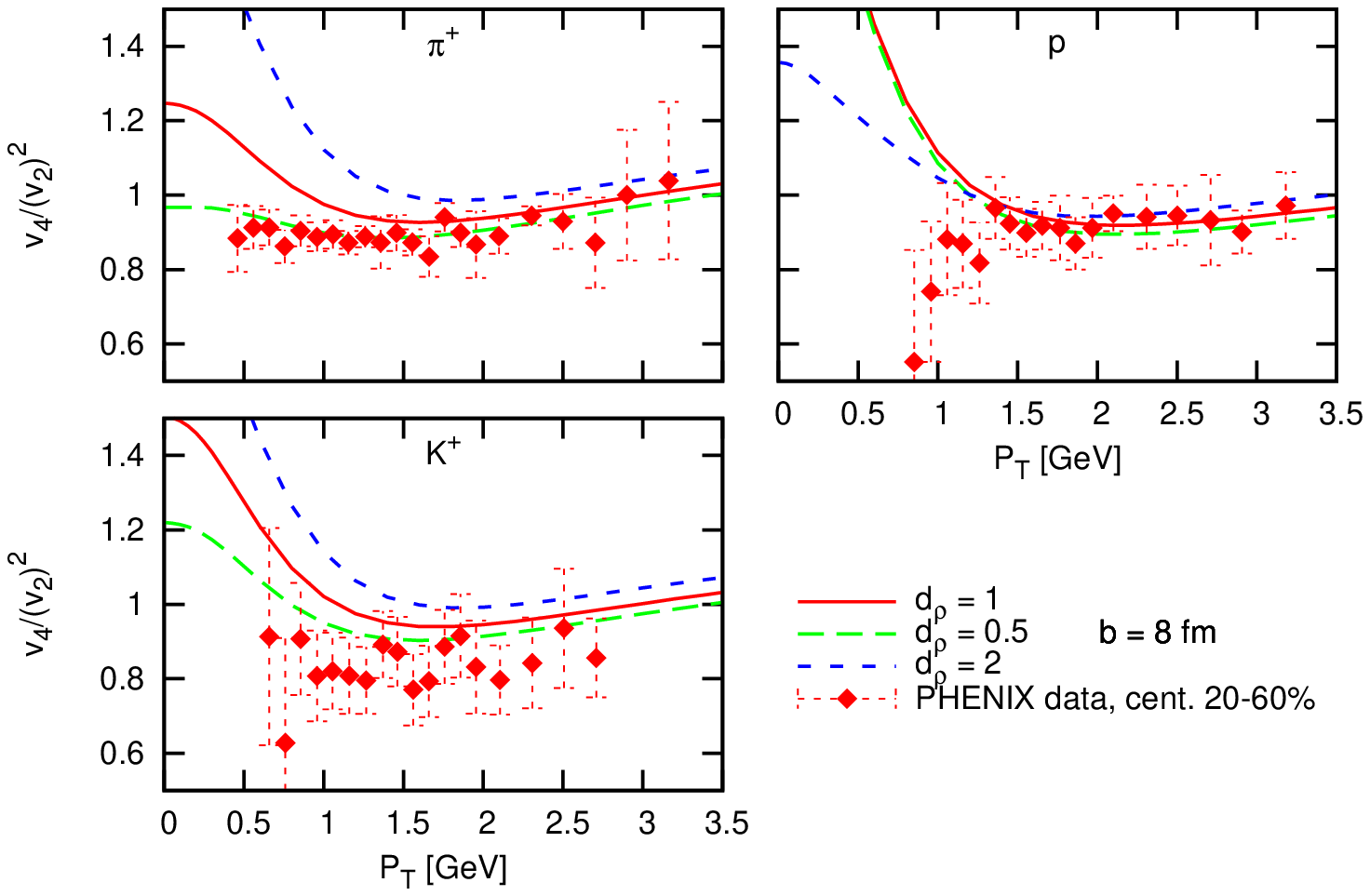}
\caption[The flow ratio $v_4/(v_2)^2$ for an ellipsoidal freeze-out compared to STAR and PHENIX data with different radial profiles of the transverse rapidity]{The flow ratio $v_4/(v_2)^2$ for an ellipsoidal freeze-out with $c_e=0.73$. The behaviour for different radial profiles of the transverse rapidity is depicted with $d_\rho=0.5,1$ and $2$.
Upper: Ratio at $b=2\fm$ for charged hadrons compared to data from STAR \cite{nucl-ex/0701044,nucl-ex/0403019}.
Lower: Ratio at $b=8\fm$ for pions, kaons and protons compared to data from PHENIX \cite{0804.4864}.}
\label{fig:v42_ratio-d_rho}
\end{figure}

The behaviour for different $d_\rho$ is similiar to the previous paragraph. At high $p_T$ all three curves give the same result and only at low $p_T$ the values are modified. The data on charged hadrons (Fig.~\ref{fig:v42_ratio-d_rho}, upper panel) is best reproduced with $d_\rho=1$ (and $\lambda=0$) as in the previous paragraph. The data from PHENIX on pions and kaons seems to favour a smaller $d_\rho$ of about 0.5, while proton data at low $p_T$ points to $d_\rho > 2$, where $d_\rho=2$ still gives a large overestimation.

So generally both parameters can be used to fine tune the $v_4/(v_2)^2$ ratio and this would preclude a determination of these parameters from this observable. But there are two reason to fix $d_\rho$ at a value of $1$ and use $\lambda$ as a free parameter:
\begin{itemize}
 \item The low $p_T$ pion data could be explained by a lower value of $\lambda$ for baryons. A larger value of $d_\rho$ on the other hand is not only unreasonable, but also largely fails to predict the meson data. And since $d_\rho$ is a blast-wave parameter is would be inconsistent to let it depend on the particle species.
\item The outer particles in the shell of the fireball suffer less collisions, since they can expand into the vacuum, and therefore one expects them to freeze-out earlier than the particles in the inner fireball. From that, one would expect a negative $ \lambda < 0$, which is supported by hydrodynamical calculations \cite{nucl-th/0602039, nucl-th/0603065}.
\end{itemize}

\subsection{Differential flow}
\label{sec:elliptic_flow_results}
\begin{figure}
\centering
\includegraphics{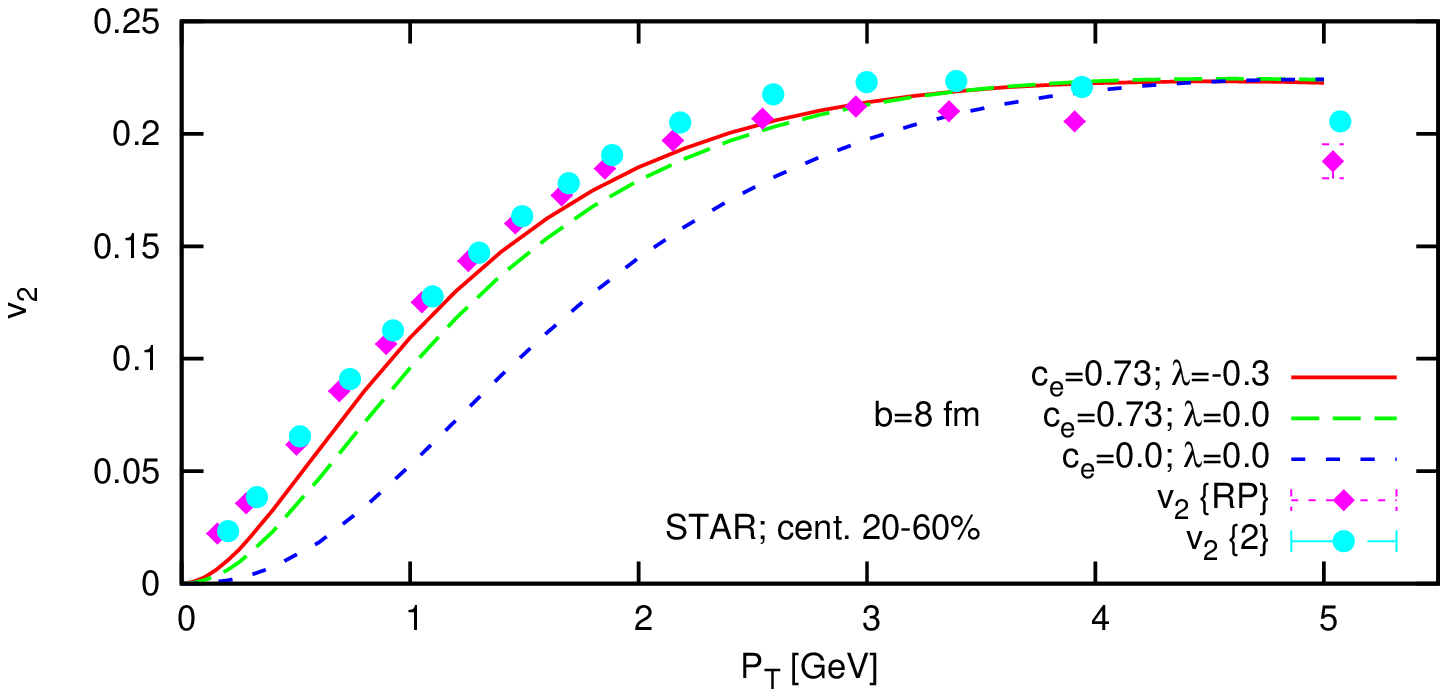}
 \caption[Elliptic flow of charged hadrons as a function of $p_T$ for $c_e=0.0$ and $c_e=0.73$]{Elliptic flow of charged hadrons as a function of $p_T$ for $c_e=0$ (blue line) and $c_e=0.73$ (red line ($\lambda=-0.3$), green line ($\lambda=0$)) for an impact parameter $b=8\fm$ compared to data from STAR, cent. 20-60\% \cite{Tang:2003kz} (points).}
\label{fig:diff_v2_ch}
\end{figure}

In this section I will show differential elliptic and hexadecupole flow $v_n (p_T)$ compared to data from RHIC. I will use the parameters as extracted from the previous sections as $c_e=0.73$, $d_\rho = 1$ and $\lambda=-0.3$ if not specified otherwise. To show the relevance of these parameters, Fig.~\ref{fig:diff_v2_ch} compares the elliptic flow of charged hadrons with an impact parameter of $b=8\fm$ for three different freeze-out scenarios: circular freeze-out with $c_e=0, \lambda=0$ (blue line), elliptic freeze-out with $c_e=0.73,\lambda=0$ (green line) and elliptic freeze-out with $c_e=0.73$, $\lambda=-0.3$ (red line). Again, a circular freeze-out can be safely discarded. An elliptic freeze-out with a radial decrease of the freeze-out time ($\lambda < 0$) is the best description of the STAR data \cite{Tang:2003kz} and only gives a slight underestimation at low $p_T$. The data points correspond to the event plane (squares) and two-particle cumulant method (circles) for the determination of the elliptic flow.

\begin{figure}
\centering
\includegraphics{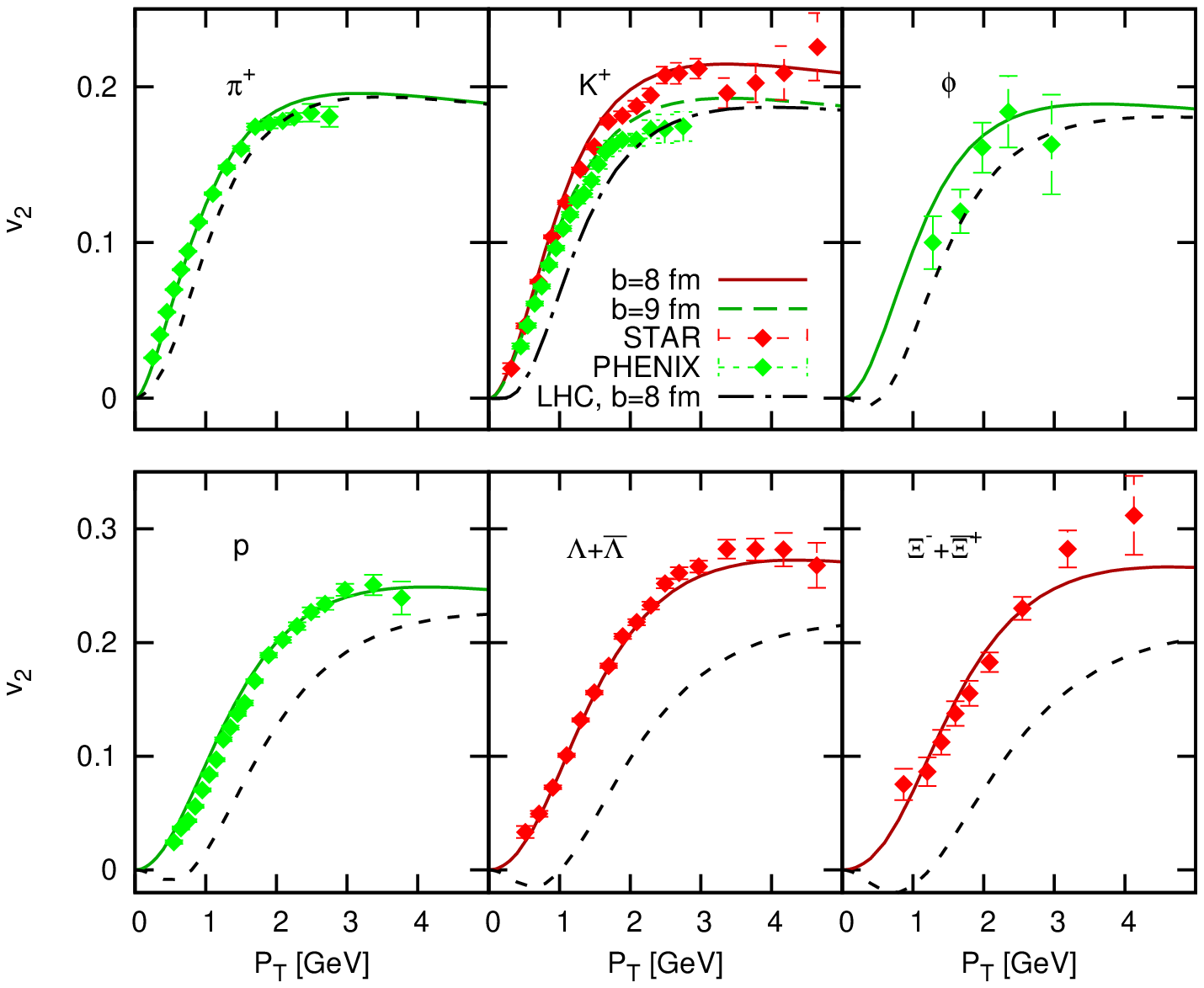}
 \caption[Elliptic flow of identified hadrons as a function of $p_T$ with $c_e=0.73$ and $\lambda=-0.3$]{Elliptic flow of identified hadrons as a function of $p_T$ with $c_e=0.73$ and $\lambda=-0.3$ for impact parameters of $9\fm$ (red lines) and $8\fm$ (green lines) compared to data from STAR, cent. 40-80\% \cite{0801.3466} (red points) and PHENIX, cent. 20-60\% \cite{nucl-ex/0703024} (green points). The black lines correspond to the predictions for LHC ($\sqrt{s}=5.5\TeV$) with $b=8\fm$.}
\label{fig:diff_v2_cent}
\end{figure}

Fig.~\ref{fig:diff_v2_cent} compares the differential elliptic flow $v_2 (p_T)$ of identified hadrons to data from RHIC from different centralities. The red data points are from STAR with a centrality of 40-80\% and the green ones from PHENIX with 20-60\% centrality. They are compared to an impact parameter of $9\fm$ (red lines) and $8\fm$ (green lines) respectively.

\begin{figure}
\centering
\includegraphics{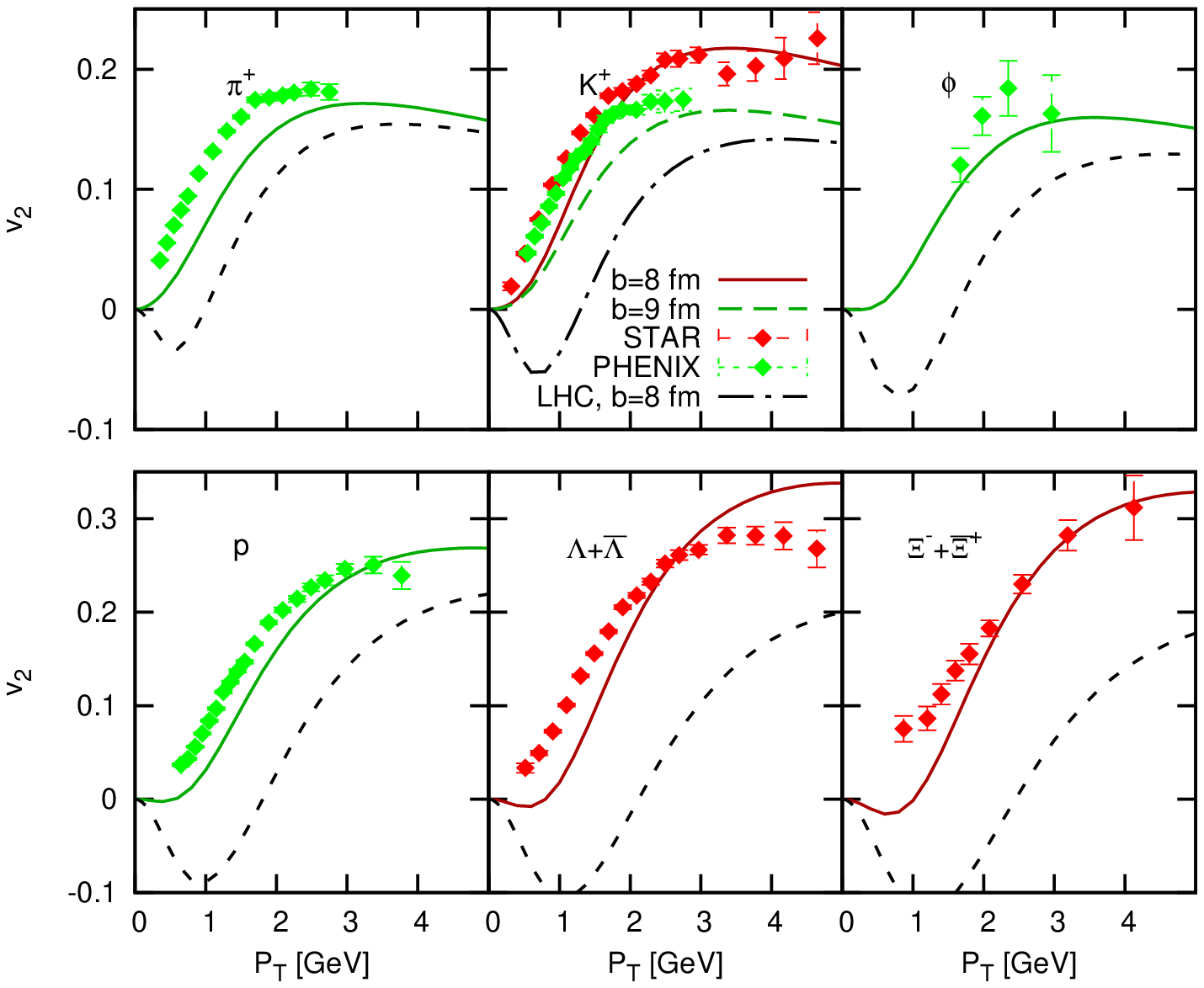}
 \caption[Elliptic flow as a function of $p_T$ with $c_e=0.0$ and $\lambda=0$]{Elliptic flow as a function of $p_T$ \emph{for a circular freeze-out} with $c_e=0.0$ and $\lambda=0$ for impact parameters of $9\fm$ (red lines) and $8\fm$ (green lines) compared to data from STAR, cent. 40-80\% \cite{0801.3466} (red points) and PHENIX, cent. 20-60\% \cite{nucl-ex/0703024} (green points). The black lines correspond to the predictions for LHC ($\sqrt{s}=5.5\TeV$) with $b=8\fm$.}
\label{fig:diff_v2_cent_circular}
\end{figure}

The agreement is quite good. While the behaviour at high $p_T$ changes with the phenomenological parameter $p_0$, the precise predictions at mid and low $p_T$ confirm the relevance of the parameter $c_e$. To show that a large freeze-out eccentricity of about 70\% of the initial one is not only need for the flow ratio from the previous section, but also for the differential elliptic flow, Fig.~\ref{fig:diff_v2_cent_circular} gives the same comparison for $c_e=0$ and $\lambda=0$.

So for a circular freeze-out the low and mid $p_T$ range is generally underpredicted for all hadrons and the high $p_T$ region only agrees, because of a larger $p_0$. As can be seen, a remaining freeze-out eccentricity can not be neglected.

The predictions for LHC (black lines) are also shown. A first striking observation is that the values are generally smaller. A similar pattern was also observed within a parton transport approach, if the viscosity was set to the ADS/CFT limit \cite{Molnar:2007an}. %NEW: Also cite{Chaudhuri:2008je}
 It seems as the elliptic flow reaches a maximum somewhere between these two center of mass energies which is due to the increased transverse expansion velocity from $0.55$ (RHIC) to $\beta_T=0.75$ (LHC). I have already shown this strong dependence on $\beta_T$ in section~\ref{sec:beta_T_dependence}.
At low $p_T$ the elliptic flow becomes even negative for (multi-)strange particles which will be exploited in the following sections~\ref{sec:charm_flow} and \ref{sec:sqrt-s_v2}. A discussion about the negative flow then follows in section~\ref{sec:negative_flow}.

\begin{figure}
\centering
\includegraphics{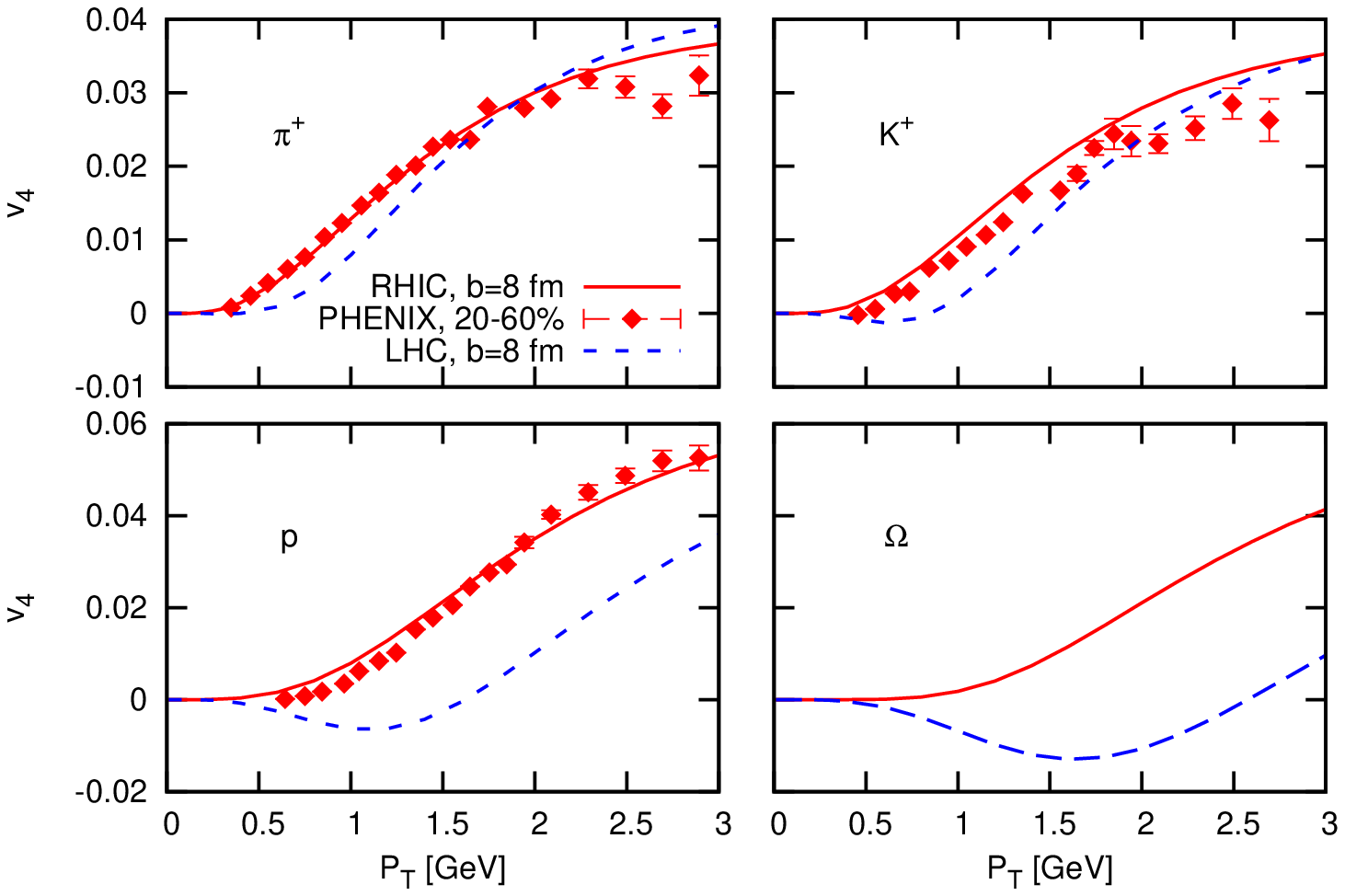}
 \caption[Hexadecupole flow as a function of $p_T$ with $c_e=0.73$ and $\lambda=-0.3$ for different hadrons]{Hexadecupole flow as a function of $p_T$ with $c_e=0.73$ and $\lambda=-0.3$ for an impact parameter of $8\fm$ (lines) compared to data from PHENIX, cent. 20-60\% \cite{0804.4864} (points). The blue lines correspond to the predictions for LHC ($\sqrt{s}=5.5\TeV$) with $b=8\fm$. }
\label{fig:diff_v4}
\end{figure}

For the fourth fourier coefficient $v_4$, experimental data is rare up to now. Fig.~\ref{fig:diff_v4} shows the hexadecupole flow of different hadrons for an elliptic freeze-out compared to the available data from PHENIX \cite{0804.4864}. The pion results agree very well with the data while the kaon and proton data is slightly overestimated at low $p_T$. Also the predictions for the LHC (blue lines) are shown.

\subsection{Deuteron flow}

Due to their small binding energy, deuterons are expected not to survive the early phase, but to be produced by proton-neutron coalescence at freeze-out \cite{nucl-th/9701053, nucl-ex/0409006}. Thus, the invariant yield can be written as
\begin{align}
 E_d \dfrac{\dn^3 N}{\dn^3 p_d} = B_2 \left(E_p \dfrac{\dn^3 N}{\dn^3 p_p}\right)
\end{align}
with $p_d=2p_p$ and a coalescence parameter $B_2$ which will depend on the freeze-out volume. Hence, the elliptic flow is related by
\begin{align}
 v_2^{(d)}(P_T) = 2 v_2^{(p)}\left(\dfrac{P_T}{2}\right).
\end{align}
But by inserting the CQNS for the proton elliptic flow, the expected scaling of the deuteron flow with the quark flow is obtained:
\begin{align}
 v_2^{(d)}(P_T) = 6 v_2^{(q)}\left(\dfrac{P_T}{6}\right).
\end{align}
Thus, the general behaviour of the deuteron elliptic flow seems to be independent of the production process: $uud+udd\rightarrow p+n \rightarrow d$ (after the freeze-out of $p$ and $n$) or $uuuddd \rightarrow d$ (at hadronization).

\begin{figure}
\centering
\includegraphics{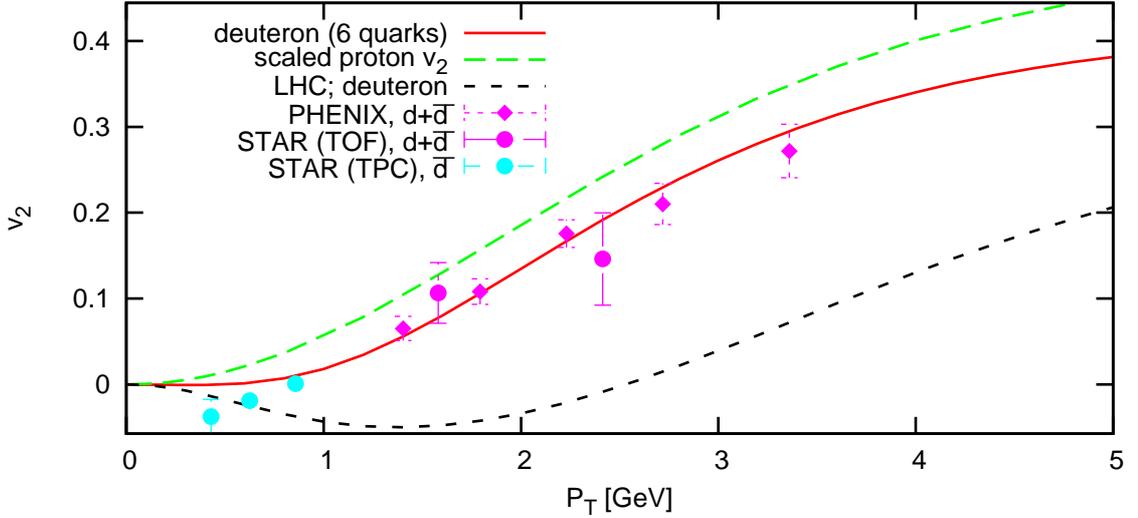}
 \caption[Elliptic flow of deuterons from 6 recombining quarks or scaled proton $v_2$]{Elliptic flow of deuterons from 6 recombining quarks (red line) or scaled proton $v_2$ (green line) compared to PHENIX \cite{nucl-ex/0703024} and STAR \cite{nucl-ex/0701057} data. The black lines correspond to the LHC prediction.}
\label{fig:deuteron_v2}
\end{figure}

Fig.~\ref{fig:deuteron_v2} compares both cases with realistic calculations at $b=8\fm$ to data from PHENIX \cite{nucl-ex/0703024} (cent. 20-60\%) and STAR \cite{nucl-ex/0701057} (minimum bias). The red line corresponds to the scenario of 6 recombining quarks at hadronization which agrees very well with the data. Only at very low $p_T$, it does not follow the STAR data of a slightly negative $v_2$. The green line depicts the scaled proton elliptic flow $2 v_2^{(p)}\left(P_T/2\right)$ which does not give the same results but follows the same trend as expected. It overestimates the data which may be due to the ``delta-shaped wavefunction'' approximation $p_p=p_n=p_d/2$. So, while the study of the transverse momentum spectra requires a dynamical treatment with hadronic degrees of freedom, the elliptic flow can be described by quark recombination and thus follows the CQNS. Additionally, the black line shows the prediction for the LHC for the quark recombination scenario.

\subsection{Heavy quark flow}
\label{sec:charm_flow}
A large elliptic flow implies, from a hydrodynamical point of view, a rapid thermalization of the fireball and a strong collective flow created at the QGP stage. This assumption is well established and the large flow of light quarks is reflected in the good agreement with the $v_2$ data. Heavy quarks on the other side have a much larger mass then the light quarks and therefore it is an open question, if charm or even bottom quarks participate in the collective expansion.

The recently published data on $J/\psi$ from PHENIX offer the possibility to investigate whether also the charm quark does locally equilibrate and therefore follows the flow of the light quarks. The most prominent feature of this preliminary data, despite its large errors, is the negative $v_2$ value at $p_T=1.5 \GeV$. Within the CQNS this implies also a negative elliptic flow for the charm quark. Since the multistrange $\Omega$ also shows a slightly negative $v_2$ for a mean transverse expansion velocity of $\beta=0.55c$, the much heavier charm quark can be expected to show an even more pronounced behaviour, when assuming a similar transverse expansion.

\begin{figure}
\centering
\includegraphics{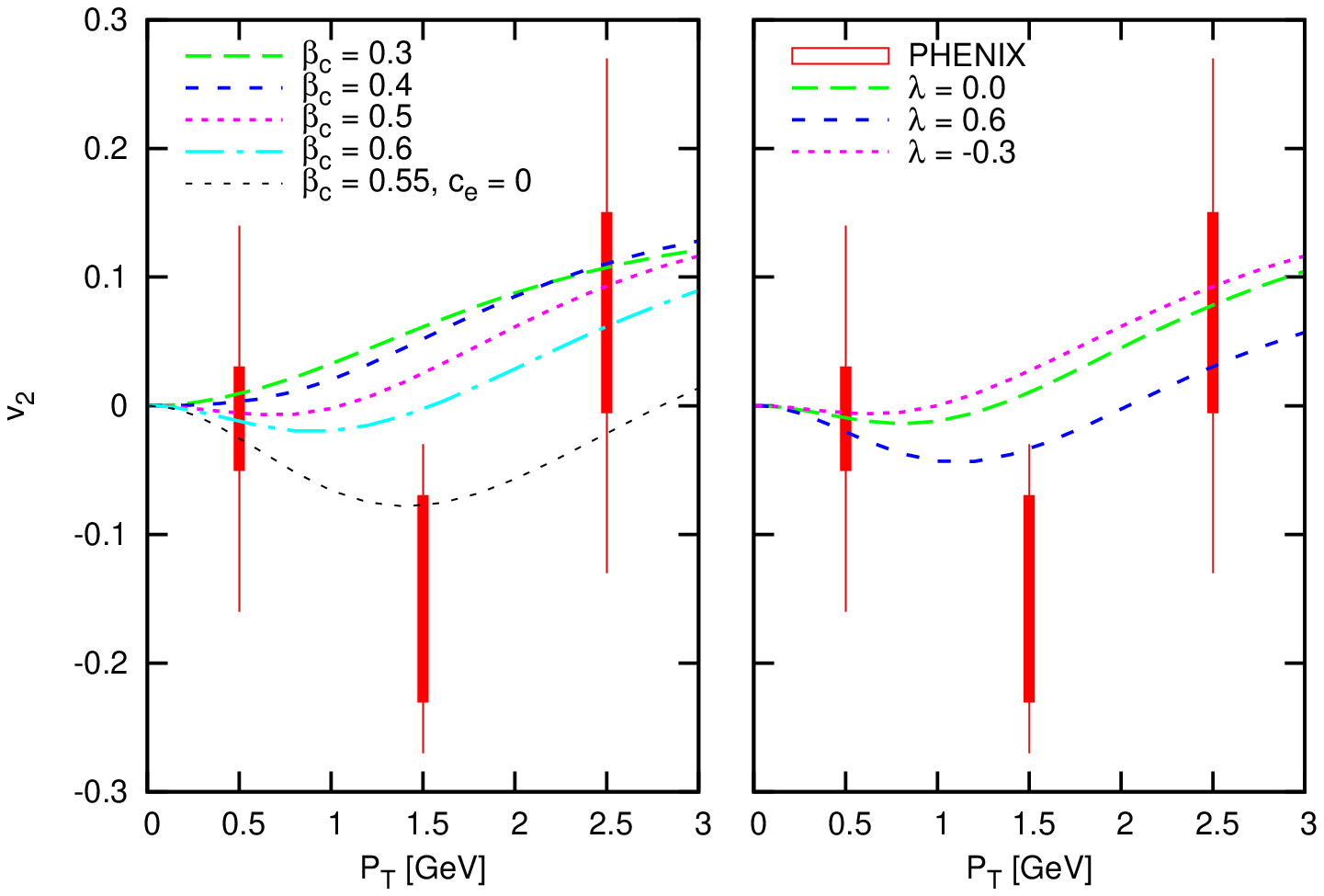}
 \caption[Estimating the charm flow from the elliptic flow of $J/\psi$]{Estimating the charm flow from the elliptic flow of $J/\psi$ compared to data from PHENIX \cite{Silvestre:2008tw}.\newline
Left: Comparison for increasing mean expansion velocities $\beta$ from top to bottom with $c_e=0.73$ and $\lambda=0$. The lowest line (black) corresponds to $\beta=0.55c$ and $c_e=0$.\newline
Right: Comparison for different $\lambda$ with $\beta=0.55c$ and $c_e=0.73$.}
\label{fig:jpsi_v2}
\end{figure}

The elliptic flow for $J/\psi$ is depicted in Fig.~\ref{fig:jpsi_v2} together with the data from PHENIX \cite{Silvestre:2008tw}. The left panel is calculated with the parameter $\lambda=0$ and shows different expansion velocities and the right panel compares $\beta=0.55c$ for different values of $\lambda$.

Looking at the left panel, the lines from top to bottom correspond to $\beta=0.3,\,0.4,\,0.5$ and $0.6c$ with $c_e=0.73$, while the last line (black) is for $\beta=0.55$ from a circular freeze-out ($c_e=0$). So a large charm flow that is equal to the light quark flow seems to produce large out-of plane elliptic flow ($v_2 < 0$, black line), but this is compensated by the positive geometrical contributions from the elliptic freeze-out. Nevertheless, small transverse expansion velocities seem to be discarded due to the large positive $v_2$ at $p_T=1.5\GeV$, while a thermalized charm with $\beta\approx 0.5-0.6c$ is at least consistent with zero flow at $p_T=1.5\GeV$. I would like to mention that the discussion from \cite{Krieg:2007bc} is for a circular freeze-out only, since we had employed only the simpified equations from \cite{Fries:2003kq}.

From the flow ratio section I predicted a negative $\lambda$. This delayed freeze-out of the particles in the inner fireball can be explained by in-medium rescattering. But the $J/\psi$ is expected to have a low cross-section, i.e. it can escape the system quite undisturbed. So the $J/\psi$ will freeze-out early, while the system is still expanding. Therefore one could assume a zero or even positive $\lambda$. This is depicted in the right panel of Fig.~\ref{fig:jpsi_v2} for a mean expansion velocity $\beta=0.55c$, where the data clearly seems to favor a positive $\lambda$ (blue line).

The large errorbars do not allow to draw any firm conclusions, but I want to state that the data is consistent within errors with a charm quark flow of $\beta=0.55c$ equal to the light quarks and a positive $\lambda\approx0.6$ ($\tau_i<\tau_o$). At least one can extract a \emph{lower} bound on the expansion velocity of about $0.4c$. So if more precise data will still support the negative $v_2$, I conclude from this observation that charm quarks reach a substantial amount of local kinetic equilibration.

Another possibility is to study $D$-meson elliptic flow. There, the positive light quark $v_2$ will compete with the negative one from the charm quark. So far, there is no data on $D$-meson elliptic flow available, but in the near future the Heavy Flavor Tracker (HFT) from STAR will close this gap. Preliminary, one could compare to $v_2$ data of electrons from heavy flavor decay. This is no direct probe of the charm flow, since there will be contributions from $B$-mesons and the decay kinematics might smear out the charm flow signal. However, calculations within transport theory \cite{Greco:2003vf} predict the $D$-meson elliptic flow to be similiar to the heavy-electron $v_2$.

\begin{figure}
\centering
\includegraphics{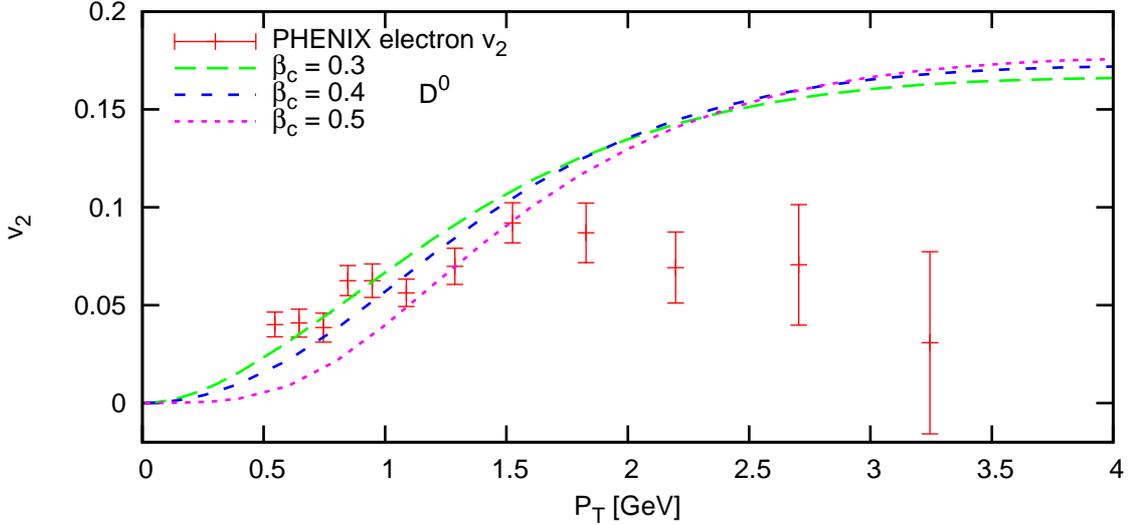}
 \caption[Electron elliptic flow data from heavy flavor decay comapared to $D_0$ $v_2$]{Electron elliptic flow data from heavy flavor decay from PHENIX \cite{nucl-ex/0611018} compared to $D_0$ $v_2$ for different charm transverse velocities $\beta_c = 0.3c,0.4c,0.5c$ (lines, top to bottom) with $c_e=0.73$ and $\lambda=0$.}
\label{fig:heavy-e_v2}
\end{figure}

Fig.~\ref{fig:heavy-e_v2} compares the electron elliptic flow data from PHENIX \cite{nucl-ex/0611018} to $D_0$ for different charm transverse velocities $\beta_c = 0.3c,0.4c,0.5c$ (lines, top to bottom) with $c_e=0.73$ and $\lambda=0$.
The $D_0$ elliptic flow agrees with the electron $v_2$ data at low $p_T$, but does not allow for a distinction between the different transverse velocities. While the data follows the curve for a $\beta_T\approx 0.3$ below $p_T=1 \GeV$, above $1\GeV$ it is better predicted by a large expansion velocity of $\beta_T \approx 0.5$. But at high $p_T$, the $v_2$ of the electrons is generally about a factor of 2 smaller then that of the $D_0$. That might be due to an early onset of the fragmentation regime for the charm. Again, there is no clear conclusion, but the data does not seem to contradict a large charm transverse velocity, which is comparable to the light quark velocity of $\beta_T = 0.55c$, as far as the heavy electron $v_2$ can be compared to the $D_0$ elliptic flow.

\begin{figure}
\centering
\includegraphics{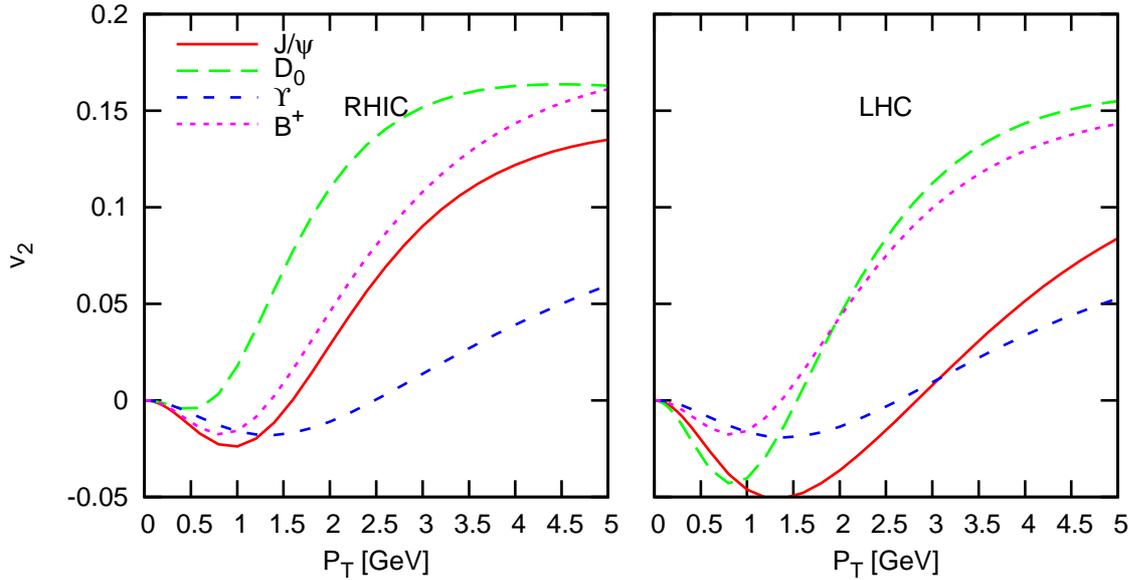}
 \caption[Elliptic flow of heavy mesons with charm and bottom quark content at RHIC and LHC]{Elliptic flow of heavy mesons with charm and bottom quark content at RHIC (left) and LHC (right) for $b=8\fm$, $c_e=0.73$ and $\lambda=0.3$.}
\label{fig:heavy_hadrons_v2}
\end{figure}

As a finally comparison, Fig.~\ref{fig:heavy_hadrons_v2} shows the elliptic flow for charm and bottom mesons at RHIC (left) and LHC (right) for $c_e=0.73$ and $\lambda=0.3$.

\subsection{Centrality dependence}
\label{sec:cent_dep}
\begin{figure}
\centering
\includegraphics{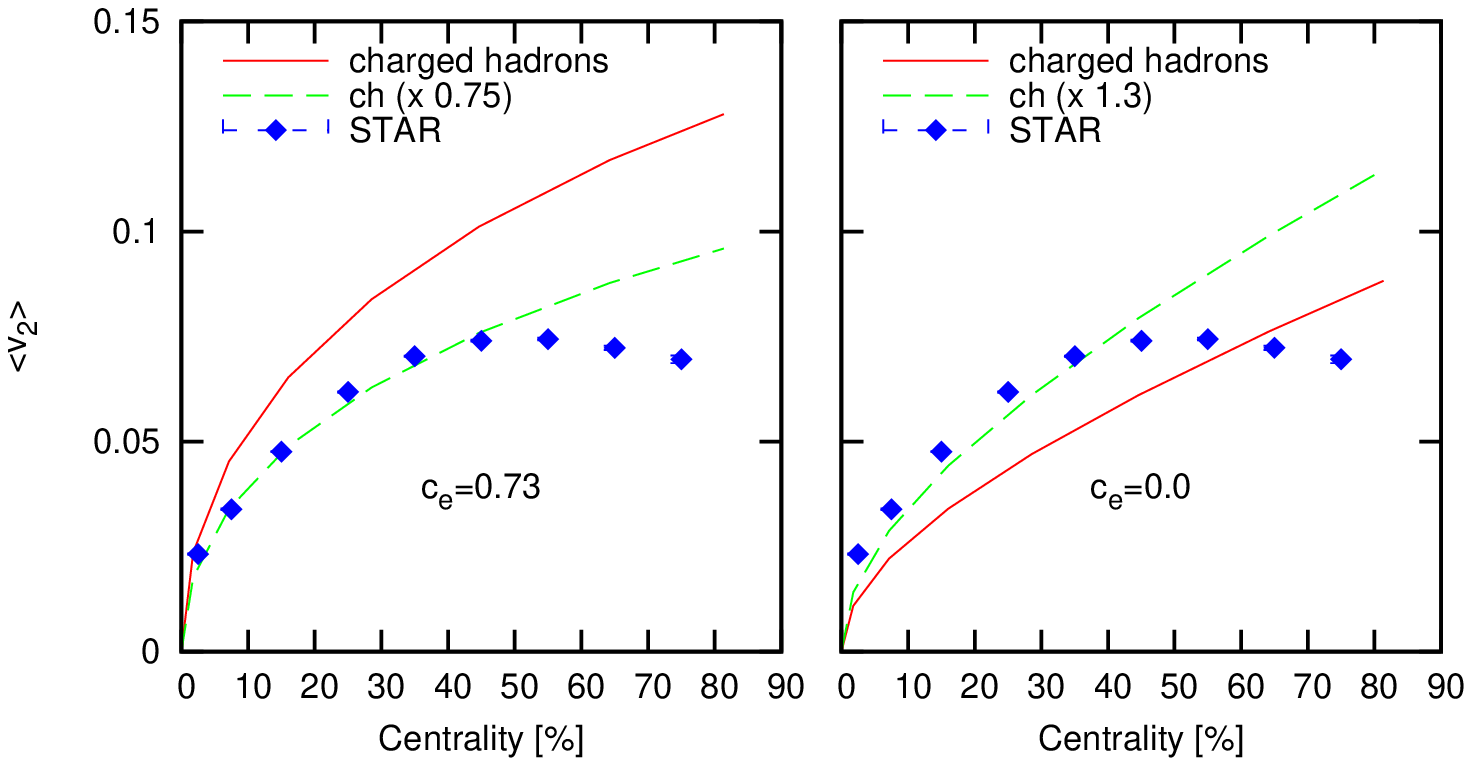}
 \caption[Mean elliptic flow for charged hadrons as a function of the centrality]{Mean elliptic flow for charged hadrons as a function of the centrality compared to data from STAR \cite{0801.3466}.\newline
Left: $\left\langle v_2 \right\rangle$ from an elliptic freeze-out. Right: $\left\langle v_2 \right\rangle$ from a circular freeze-out.}
\label{fig:mean_v2_cent}
\end{figure}

Fig.~\ref{fig:mean_v2_cent} shows the mean elliptic flow of charged hadrons as a function of the centrality. The left panel is for an elliptic freeze-out and the right for a circular one. Although the differential $v_2$ shows very good agreement with the data as shown above, the results for $\langle v_2\rangle$ (left panel, red line) generally overestimate the data from STAR \cite{0801.3466}. This is due to the yield at low $p_T$, which is to small because of energy violation and the absence of resonance feed-down as discussed in section~\ref{sec:yields}. Therefore the mean $v_2$ weighted with the yield favors larger $v_2$ values. I account for that by scaling the results with a constant factor of $0.75$ (green line), so that they are in agreement with the data for the central and mid-central collisions, but for peripheral collisions the data is still overestimated. This is due to the used eccentricity $b/(4R_A)$ (eq.~\eqref{eqn:eccent_definition}), which follows the glauber eccentricity for central and mid-central collisions, but gives wrong results for peripheral collisions (Fig.~\ref{fig:eccent_compare}).

The mean $v_2$ from a circular freeze-out fails to describe the data and underestimates it as expected. Only to have a comparison between both cases, I will scale the mean elliptic flow value for $c_e=0$ by a constant factor of 1.3 (right panel, green line) so that it fits the data for central to mid-central collisions. In the next section I will use both factors implicitly.

\begin{figure}
\centering
\includegraphics{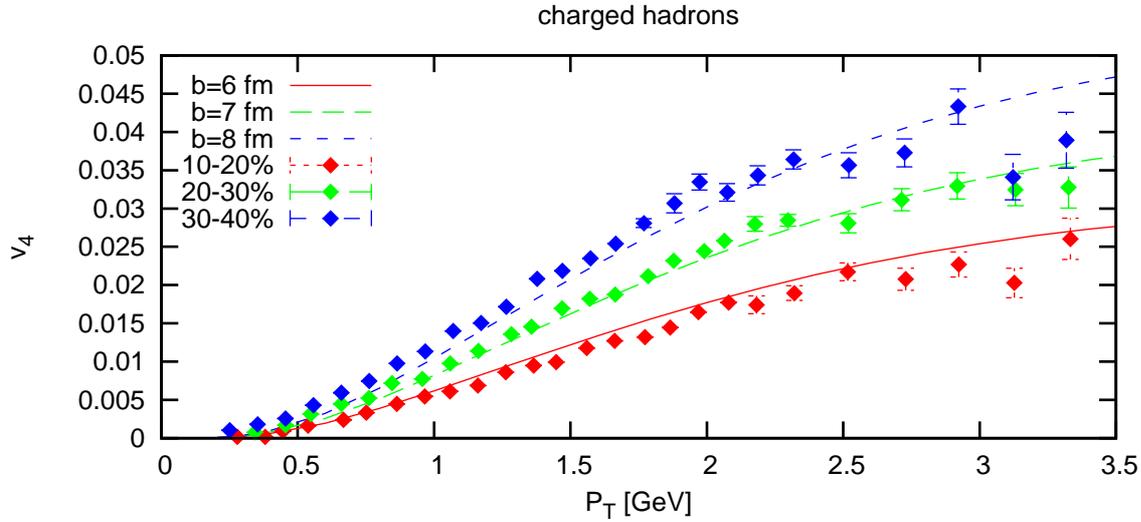}
 \caption[Hexadecupole flow as a function of $p_T$ with $c_e=0.73$ and $\lambda=-0.3$ for different centralities]{Hexadecupole flow as a function of $p_T$ with $c_e=0.73$ and $\lambda=-0.3$ for impact parameter of $6,7$ and $8\fm$ (lines) compared to data from PHENIX (points) of different centrality classes \cite{0805.4039}.}
\label{fig:cent_v4}
\end{figure}

Due to the lack of similiar data for the hexadecupole flow, I compare the results to $v_4$ of charged hadrons for different centrality classes from PHENIX \cite{0805.4039} in Fig.~\ref{fig:cent_v4}. The agreement with the data is quite good within the uncertainty of relation between the impact parameter and the centrality. But an underestimation of the mean $\langle v_4 \rangle$ for peripheral collisions can be expect similiar to the mean elliptic flow.

\subsection{\texorpdfstring{$\sqrt{s}$}{sqrt(s)} dependence}
\label{sec:sqrt-s_v2}
\begin{figure}
 \centering
 \includegraphics{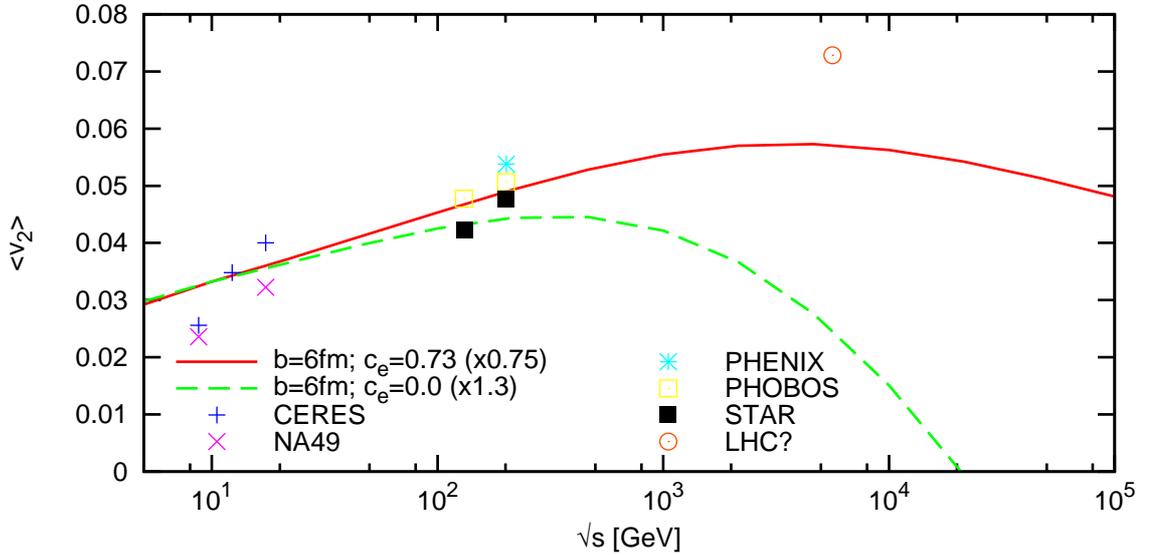}
 \caption[Comparison of $\left\langle v_2 \right\rangle$ for charged hadrons as a function of center of mass energy]{Comparison of $\left\langle v_2 \right\rangle$ for charged hadrons as a function of center of mass energy for Au+Au/Pb+Pb reactions at an impact parameter of $b=6\fm$. The calculations are scaled with the constant factors from Fig.~\ref{fig:mean_v2_cent} as discussed in sec.~\ref{sec:cent_dep}. Data points and the extrapolation to LHC are taken from \cite{Borghini:2007ub}.}
 \label{fig:sqrt_s-v2}
\end{figure}

The effect of a negative elliptic flow at low $p_T$ at LHC will be visible even for hadrons with light quark content, like protons as shown in Fig.~\ref{fig:diff_v2_cent}. Therefore, it is interesting to look at the mean elliptic flow as a function of the center of mass energy $\sqrt{s}$. To compare the results from an elliptic and circular freeze-out, Fig.~\ref{fig:sqrt_s-v2} uses the correction factors from the previous section.

\begin{figure}
 \centering
 \includegraphics{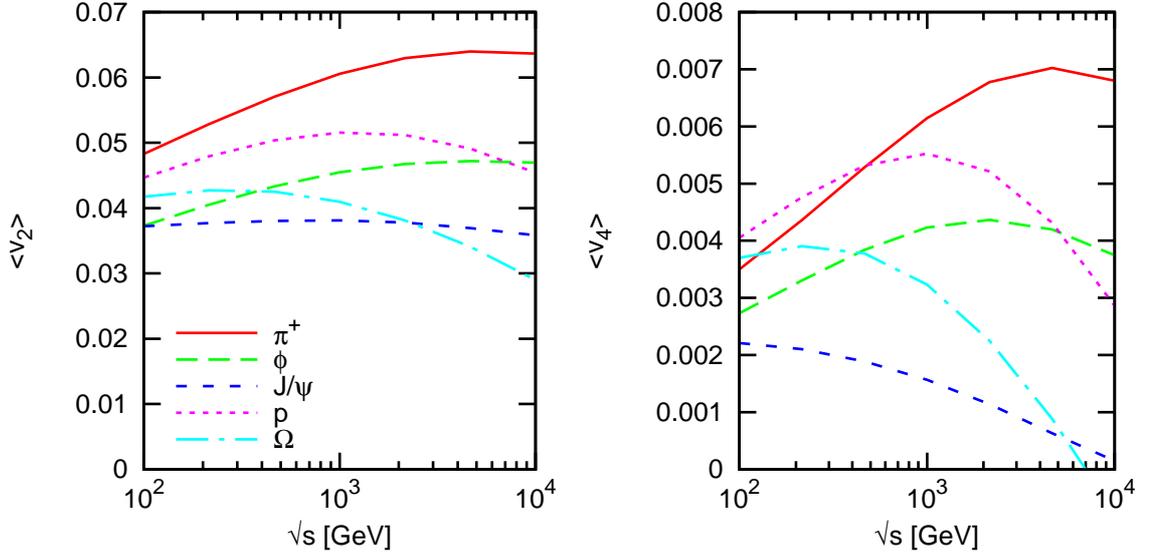}
 \caption[Comparison of $\left\langle v_2 \right\rangle$ and $\left\langle v_4 \right\rangle$ for different hadrons from $\sqrt{s}=100\GeV$ to $10\TeV$]{Comparison of $\left\langle v_2 \right\rangle$ and $\left\langle v_4 \right\rangle$ for different hadrons from $\sqrt{s}=100\GeV$ to $10\TeV$ at an impact parameter of $b=6\fm$ and with $c_e=0.73$.}
 \label{fig:sqrt_s-v2_ih}
\end{figure}

The shown value for LHC in Fig.~\ref{fig:sqrt_s-v2} is just a linear interpolation, but the general expectation in the heavy ion community is an monotonic increase from RHIC energies to LHC and beyond, or at least a saturation at some finite value. In this framework, the value of $\langle v_2\rangle$ depends on the interplay of the strength of the negative $v_2$ and the increasing mean $p_T$, since $\langle v_2\rangle \approx v_2\left(\langle p_T\rangle\right)$. But as can be seen in Fig.~\ref{fig:sqrt_s-v2}, the increase in the mean $p_T$ does not seem to compensate the increasing effect of the negative $v_2$ at low $p_T$. Depending on the used freeze-out eccentricity, the maximum of the mean elliptic flow is reached somewhere between RHIC and LHC energies, while a larger eccentricity (a larger $c_e$) delays the extremal point to higher c.m. energies. For the best fit case of $c_e=0.73$ the value at LHC is maximal and then starts to drop; for a circular freeze-out with $c_e=0$ it would have already reached its maximum at RHIC. This may mean that even if this striking prediction is correct, there will be no clear sign of a decreasing mean elliptic flow for charged hadrons at LHC. But the mean elliptic and hexadecupole flow of identified hadrons can be a much better probe. As can be seen in Fig.~\ref{fig:sqrt_s-v2_ih}, the $v_2$ and $v_4$ of pions, $\phi$'s and protons reach their maximum at LHC energies. And while the $v_2$ of the $J/\psi$ and $\Omega$ only indicate a small decrease, the $v_4$ of these heavy hadrons will show a visible drop when going from $\sqrt{s}=200\GeV$ (RHIC) to $5.5\TeV$ (LHC).

\begin{figure}
 \centering
 \includegraphics{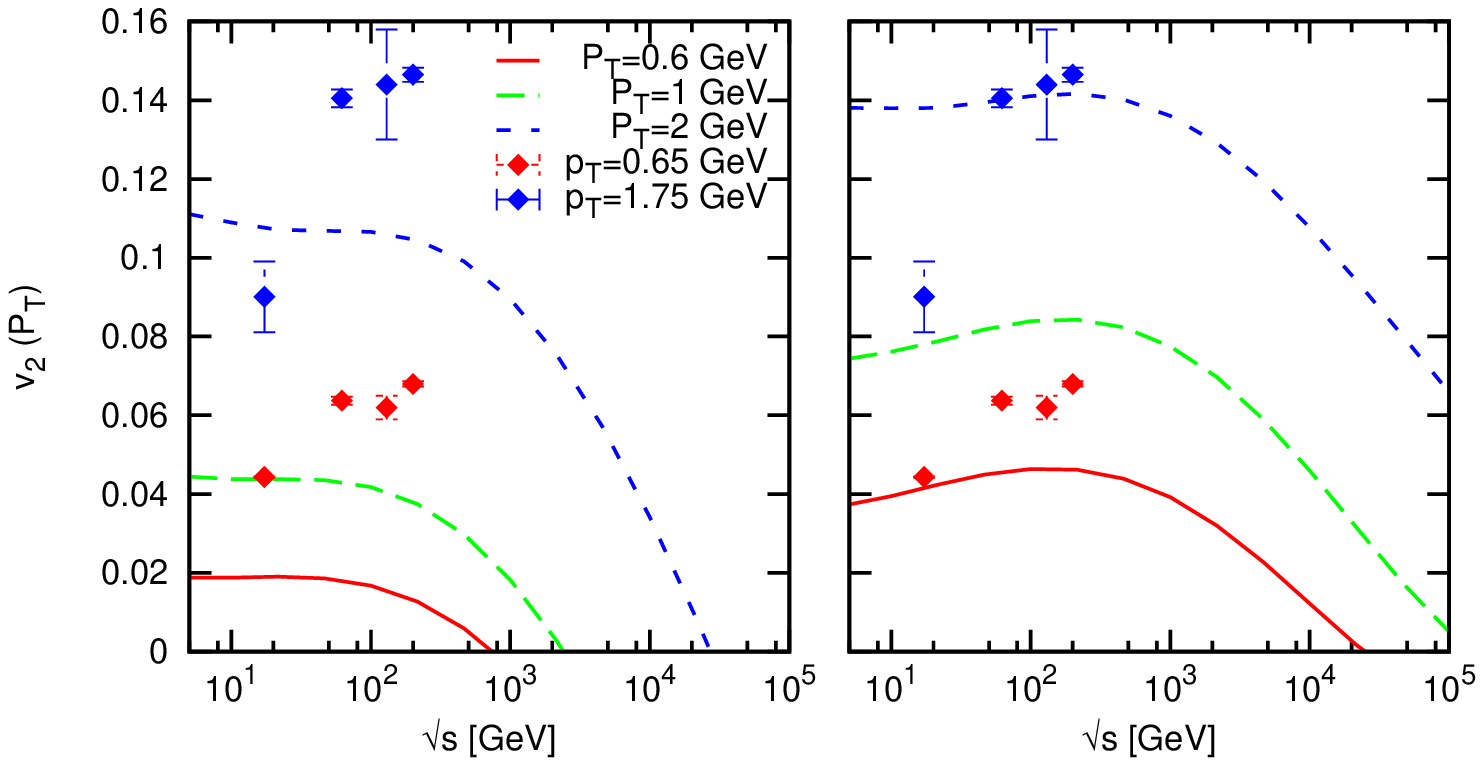}
 \caption[Elliptic flow $v_2$ of charged hadrons for a fixed $p_T$ as a 
function of center of mass energy]{Elliptic flow $v_2$ of charged hadrons at $b=6$ and fixed $p_T$ as a function of center of mass energy for $p_T = 0.6, 1 \mbox{ and } 2 \GeV$ compared to data from PHENIX \cite{nucl-ex/0411040}.}
 \label{fig:sqrt_s-v2_pt}
\end{figure}

A rather critical assumption in this context is the applicability of the recombination approach 
for the elliptic flow for small transverse momenta on the order of $p_T<1 \GeV$. I want to 
emphasise that the result of the decreasing mean 
elliptic flow  $\langle v_2\rangle$ at LHC is not affected by the validity of this assumption, 
because $\langle v_2\rangle \approx  v_2\left(\langle p_T\rangle\right)$ and $\langle p_T\rangle > 1 \GeV$ in LHC regime. 
However, to show the robustness of the prediction Fig.~\ref{fig:sqrt_s-v2_pt} depicts the elliptic flow
at a fixed $p_T$ as a function of $\sqrt{s}$ with an impact parameter $b=6\fm$. With $p_T=0.6, 1 \mbox{ and } 2\GeV$ the $v_2$ exhibits the same drop as the mean elliptic flow from Fig.~\ref{fig:sqrt_s-v2}. The data points are taken from \cite{nucl-ex/0411040} with a centrality of 13-26\%. They confirm the observation of an elliptic flow saturation at $\sqrt{s}=200\GeV$.

\begin{figure}
 \centering
 \includegraphics{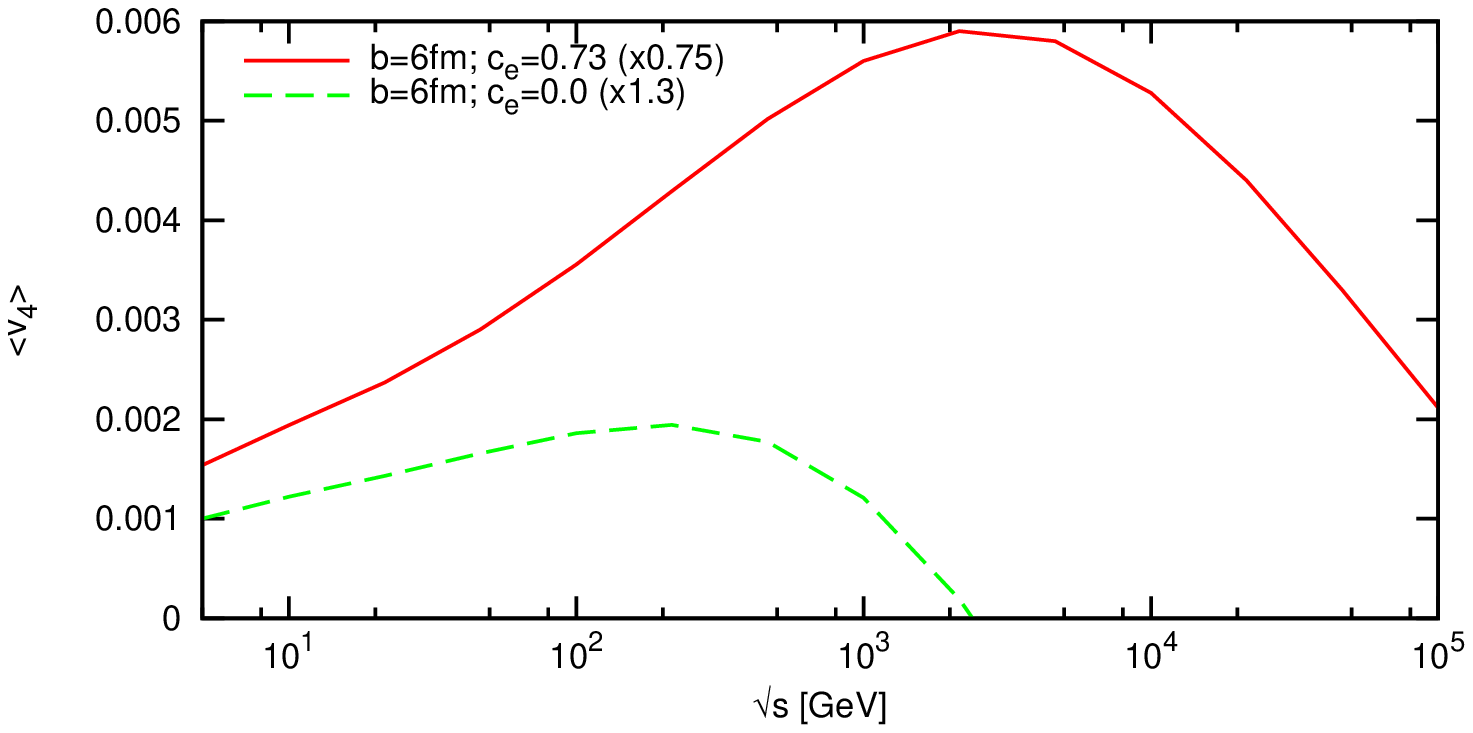}
 \caption[Comparison of $\left\langle v_4 \right\rangle$ for charged hadrons as a function of center of mass energy]{Comparison of $\left\langle v_4 \right\rangle$ for charged hadrons as a function of center of mass energy for Au+Au/Pb+Pb reactions at an impact parameter of $b=6\fm$.}
 \label{fig:sqrt_s-v4}
\end{figure}

As can be seen in Fig.~\ref{fig:sqrt_s-v4}, the mean hexadecupole flow shows a similar dependence on the center of mass energy. In the case of an elliptic freeze-out (red line), the increase from RHIC to LHC is much more pronounced as for the elliptic flow, but also the drop beyond its maximum at LHC is much stronger. In the case of a circular freeze-out (green line), the predicted value is already negative at LHC.

\begin{figure}
 \centering
 \includegraphics{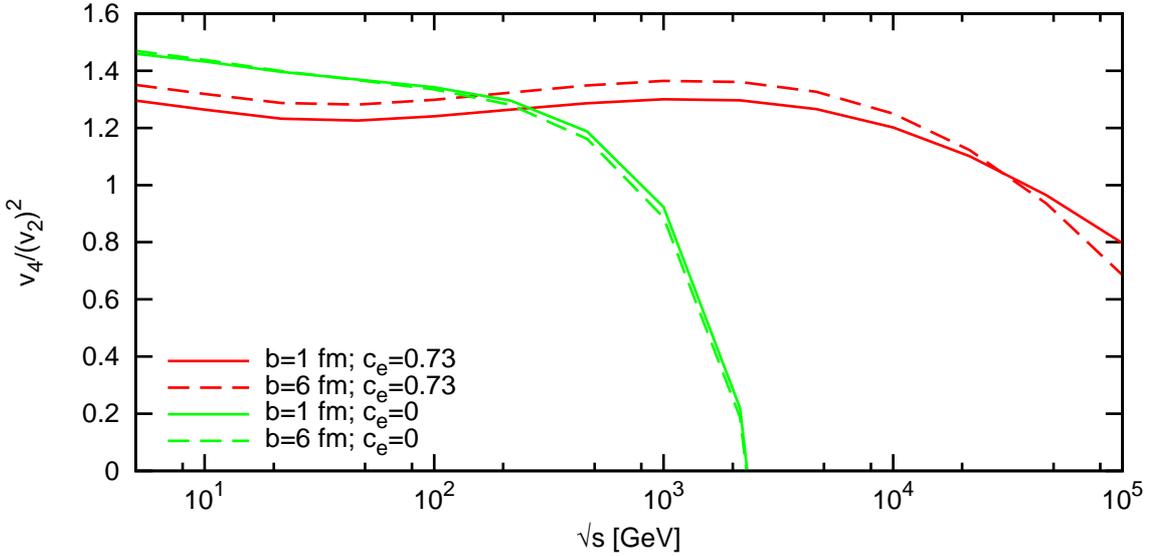}
 \caption[The ratio $\langle v_4 \rangle / \langle v_2 \rangle ^2$ for charged hadrons as a function of center of mass energy]{The ratio $\langle v_4 \rangle / \langle v_2 \rangle ^2$ for charged hadrons as a function of center of mass energy at an impact parameter of $b=1 \fm$ (solid line) and $b=6\fm$ (dashed line). The red lines correspond to an elliptic freeze-out ($c_e=0.73$) and the green lines to a circular ($c_e=0$).}
 \label{fig:sqrt_s-v42}
\end{figure}

Since both flow coefficients show a similiar behaviour as a function of $\sqrt{s}$, I will also take a look at the ratio $\dfrac{\langle v_4 \rangle}{\langle v_2 \rangle ^2}$. Fig.~\ref{fig:sqrt_s-v42} shows this ratio for charged hadrons as a function of the center of mass energy at an impact parameter of $b=1 \fm$ (solid lines) and $b=6\fm$ (dashed lines). The results for both impact parameters are nearly equal, hence this mean ratio is also independent of the centrality, like the $p_T$ dependent ratio in section~\ref{sec:flow_ratios}. For a circular freeze-out (green lines), the ratio is approximately constant between 1.3-1.5 up to RHIC energies and is steeply dropping when going to LHC while for an elliptic freeze-out (red lines), it is constant at about 1.3 up to LHC energies.

\subsection{Analyzing the negative elliptic flow}
\label{sec:negative_flow}
At low center of mass energies, the elliptic flow is generally negative, since the spectators in a non-central collision are blocking the in-plane flow. This behaviour, also known as anti-flow, is not observed at high bombarding energies, since the nuclei are strongly lorentz-contracted. Then the flow asymmetry only depend on the spatial asymmetry as discussed.

The predicted negative elliptic flow at low transverse momenta is different to anti-flow. It has also been found in previous exploratory studies and seem to be a general feature of the blast-wave like flow profile at high transverse velocities \cite{Voloshin:1996nv,Huovinen:2001cy,Voloshin:2002ii,Retiere:2003kf,Pratt:2004zq}. One might argue that this is an artefact of the blast-wave peak and will not survive in more realistic
calculations, however, also
transport simulations indicate slightly negative $v_2$ values for heavy particles at 
very low $p_T$ \cite{Bleicher:2000sx} and the AMPT model predicts a negative $v_2$ at least for heavy charm and bottom quarks \cite{Ko:2007zzc}. Thus, the qualitative behaviour of a negative $v_2$ is a well-known observation. The magnitude, however, of this effect and the particle species affected by it depend on the mass, the amount of transverse flow and the decoupling hyper-surface of this individual particle species. So the surprising observation of a drop in the mean elliptic flow at LHC and an even negative value beyond is not the existence of the elliptic 'anti-flow' but the quantitative 
strength and influence on the light quark sector of this effect at LHC.

To obtain an analytic expression which explains the negative $v_2$ is to consider only the 
in-plane ($\phi=0$) and out-of-plane ($\phi=\pi/2$) directions (similar to the analysis performed 
in \cite{Huovinen:2001cy}). The $\phi$ 
integration breaks down (eq.~\eqref{eqn:elliptic_flow_quarks}) and the elliptic flow is then given by
\begin{align}
v_2^q(p_T) = \frac{\left[I_2 K_1\right]_{\phi=0} - \left[I_2 K_1\right]_{\phi=\pi/2}}
{\left[I_0 K_1\right]_{\phi=0} + I\left[I_0 K_1\right]_{\phi=\pi/2}}
\end{align}
For $p_T \rightarrow 0$ the argument of the Bessel functions $I_n$ (see eq.~\eqref{eqn:quark_bessel_arguments}) goes to zero and $I_n$ becomes 
constant with $I_2 \rightarrow 0$ and $I_0 \rightarrow 1$. Therefore they are independent of the angle. This leads to
\begin{align}
\lim_{p_T \rightarrow 0} v_2^q(p_T) =& \frac{I_2}{I_0} \frac{K_1 \left[m \cosh(\eta_T^0(1+\eps))/T\right] - K_1 \left[m \cosh(\eta_T^0(1-\eps))/T\right]}
{K_1 \left[m \cosh(\eta_T^0(1+\eps))/T\right] + K_1 \left[m \cosh(\eta_T^0(1-\eps))/T\right]}
\end{align}
Since $K_1$ is a monotonically decreasing function, the numerator and thus the elliptic flow is negative (for some small 
transverse momenta). The specific values depend on the mass, the mean flow rapidity $\eta_T^0$ and the temperature. 
For increasing mass (e.g. charmed mesons) or increasing transverse flow rapidity (e.g. LHC and beyond) or decreasing temperature, the elliptic flow will become more negative.

Another way to understand this effect in more detail one could look directly at the thermal 
quark-spectrum with the energy (eq.~\eqref{eqn:local_frame_energy})
\begin{align}
E = m_T \cosh(y-\eta) \cosh \eta_T - p_T \cos(\varphi-\phi) \sinh \eta_T
\end{align}
For simplicity let us look at midrapidity ($y=0$) and for high $\eta$ the spectrum is very low, so I consider the region around $\eta = 0$. For high c.m.-energies, when the source is highly boosted transversally, the particles will mainly be emitted in the direction in which the fireball flies, so one can simplify even more and set $\varphi = \phi$. Because in this case there is no longitudinal momentum I can replace the momentum with $\eta_T^q$, the transverse rapidity of the parton, as
\begin{align}
p_T=&m \sinh \eta_T^q \quad\mbox{and}\\
m_T =& \sqrt{m^2+p_T^2}= m\sqrt{1+\sinh^2 \eta_T^q}\nonumber\\
 =& m \cosh \eta_t^q.
\end{align}
So the energy of the quark in the transverse moving source is given by
\begin{align}
E = m \cosh\left[\eta_T^0(1+ \eps \cos(2\phi))-\eta_T^q \right]
\end{align}
For a fixed quark rapidity $\eta_T^q < \eta_T^0$ (low $p_T$) the energy of the quarks emitted in-plane is higher than the energy of quarks emitted out-of-plane. With the thermal spectrum more energy means less particles, therefore a negative $v_2$. At $\eta_T^q = \eta_T^0$ one would expect the zero-crossing, and above a positive $v_2$.

\chapter{Conclusion}
When studying the elementary building blocks of our universe throughout history, higher and higher energies were needed to resolve the ever smaller structures. Now, physicists are searching for the so-called Quark Gluon Plasma (QGP), a state of matter where the quarks are not confined inside the hadrons but free particles. The transition to this new phase of the strongly interaction matter is probed within Heavy Ion Collisions (HIC) at the Relativistic Heavy Ion Collider (RHIC) and soon with the Large Hadron Collider (LHC).

One major problem in this context is the question if we have created a QGP in a collision, since the hadronization happens on very small timescales, so that only the hadrons from freeze-out can be measured. But also a dynamical description within QCD, of the collision in general and the hadronization process in particular, is not expected to be achieved in the near future. Thus, we depend on phenomenological models.

The description via pQCD with the use of measured, parameterized fragmentation functions is very successful for proton-proton collisions, but in HIC with a dense phase space this approach is only applicable at high $p_T$. With the first results from RHIC which pQCD failed to describe, a phenomenological model named recombination became popular.

One of the most promising observables to study the creation of a QGP is the elliptic flow which is the second fourier coefficient $v_2$ of the invariant yield. It measures the azimuthal asymmetry of the transverse momentum which orginates from the spatial asymmetry in non-central collisions. Since it shows a self-quenching behaviour, the measured elliptic flow of the hadrons is mainly sensitive to the initial stage of the collision. Therefore, it would depend on the partonic elliptic flow from the QGP.

The success of recombination is mainly based on the prediction of a universal scaling law of the elliptic flow which connects the hadron elliptic flow directly with the quark elliptic flow. This predicted constituent quark number scaling (CQNS) is experimentally well observed and the results can describe the data very accurately at low and mid $p_T$. While the qualitative behaviour of the particle species dependence of $v_2$ follows the CQNS law of recombination, the quantitative results depend on the two additional ``ingredients'': the quark density distribution and the hadronization/freeze-out hypersurface.

In this thesis I have shown that these two additionial inputs have strong influence on the flow coefficients. For the quark density distribution I employed the blast-wave model which is inspired by hydrodynamics. It describes a transversally expanding, locally thermalised fireball. A simple geometrical argument to parameterize the pressure gradient suffices to describe not only the elliptic flow $v_2$ but also the next coefficient $v_4$. A result of the blast-wave flow profile is the observed mass scaling of the elliptic flow and especally a negative flow coefficient at low $p_T$ for massive particles and/or high center of mass energies. Such a negative value can also been seen in the preliminary $J/\psi$ elliptic flow data from PHENIX which seems to support the predictions. The most striking prediction of recombination in this framework using the blast-wave model is then a dropping mean $v_2$ at or beyond LHC energies.

The hadronization occurs within a 3-dimensional, time-dependent volume and coincides with the kinematic freeze-out, since there is no hadronic scattering phase. The detailed study of this freeze-out hypersurface would require a dynamical treatment. But to identify a general influence on the spectra, this analytical approach is sufficient, since also in hydrodynamics the process of particle freeze-out is a highly non-trivial problem. As I have shown, a remaining spatial eccentricity at freeze-out and a radial dependence of the freeze-out time have the greatest impact on the flow coefficients, and especally on their ratio $v_4/(v_2)^2$. While a simple circular freeze-out underestimates this flow ratio by a factor of 2, an elliptic freeze-out introduces additional contributions to both flow coefficients and can describe the experimental data very well. The behaviour of the flow ratio at low $p_T$ is then affected by the radial dependence of the freeze-out time.

The presented results in this thesis indicate that recombination is the dominant mechanism for hadron production at mid $p_T$ while the flow coefficients and their constituent quark number scaling are a strong sign for an early partonic stage of the collision. Despite its phenomenological nature, this analytical approach to recombination as the hadronization mechanism in HIC, together with an appropriate quark density and freeze-out hypersurface, describes not only the qualitative constituent quark number scaling of the experimental data, but also the quantitative mass scaling of the elliptic and hexadecupole flow and the large ratio of these coefficients with great detail. Based on this good agreements, the striking predicitions for the LHC should be considered in more detail within a dynamical model.

\clearpage

% Anhang
\bibliographystyle{AlphaDINFirstName}
\bibliography{Diplomarbeit}
\appendix

\chapter{Analytical derivations}
\section{MIT bag model}
\label{app:bag_model}
To derive an analytic expression for the phase boundary within the MIT bag model, I assume an ideal, relativistic gas of quarks and gluons:
\begin{align} 
m_q = m_g = 0 \quad \Rightarrow \quad \vert \mathbf{p} \vert = p = E
\end{align}
Thus, the pressure $P=\dfrac{1}{3}\eps$ is related to the energy density $\eps$, which can be calculated as
\begin{align}
\eps =& g \int_0^\infty E \cdot n(E) \, dE\nonumber\\
 =& \int_0^\infty p \cdot  n(p) \dfrac{d^3p}{(2\pi)^3}\nonumber\\
 =& \dfrac{4\pi}{(2\pi)^3} \int_0^\infty p^3 \dfrac{1}{\exp{\beta(p-B_i \mu_B} \pm 1}
\end{align}
with the degeneracy factor $g$, the inverse temperature $\beta=1/T$, the baryonnumber $B_i$. The $(+)$ sign is for quarks (fermions) and the $(-)$ sign for gluons (bosons). Using the substitution
\begin{align}
x =& \beta(p-\beta B\mu_B) \quad \Rightarrow
p=&\dfrac{x}{\beta}+B \mu_B = T(x+ \beta B\mu_B), \quad \dfrac{dp}{dx} = T
\end{align}
one obtains
\begin{align}
\eps =g\dfrac{T^4}{2\pi^2} \int_{-\beta B\mu_B}^\infty (x+\beta B\mu_B)^3 \dfrac{1}{e^x \pm 1}
\end{align}
For the gluons $g$ with $B_g=0$, it follows
\begin{align}
\varepsilon_g =g_g\dfrac{T^4}{2\pi^2} \int_0^\infty x^3 \dfrac{1}{e^x-1} = g_g \dfrac{\pi^2 T^4}{30}.
\end{align}
For the quarks $q$ und antiquarks $\bar q$ with $B_q=-B_{\bar q}=1/3$ follows
\begin{align}
\eps_q + \eps_{\bar q} =&g_q\dfrac{T^4}{2\pi^2} &&\left[
\int_{-\beta B_q\mu_B}^\infty  \dfrac{(x+\beta B_q\mu_B)^3}{e^x + 1} \dn x
+ \int_{+\beta B_q\mu_B}^\infty  \dfrac{(x-\beta B_q\mu_B)^3}{e^x + 1} \dn x
\right]\nonumber\\
=& g_q\dfrac{T^4}{2\pi^2} &&\left[
\int_0^\infty \dfrac{(x+\beta B_q\mu_B)^3 + (x-\beta B_q\mu_B)^3}{e^x + 1}  \dn x\right.\nonumber\\
&&&\left.+\int_{-\beta B_q\mu_B}^0  \dfrac{(x+\beta B_q\mu_B)^3}{e^x + 1} \dn x
-\int_0^{+\beta B_q\mu_B}  \dfrac{(x-\beta B_q\mu_B)^3}{e^x + 1} \dn x
\right]\nonumber\\
\end{align}
In the first term, the brakets are expanded and in the third term, $e^x$ is pulled out of the denominator and the substitution $x \rightarrow -x$ is applied:
\begin{align}
\eps_q + \eps_{\bar q}=& g_q\dfrac{T^4}{2\pi^2} &&\left[
\int_0^\infty \dfrac{2 x^3+6x(\beta B_q\mu_B)^2}{e^x + 1}  \dn x\right.\nonumber\\
&&&\left.+\int_{-\beta B_q\mu_B}^0  \dfrac{(x+\beta B_q\mu_B)^3}{e^x + 1} \dn x
+\int_{-\beta B_q\mu_B}^0  \dfrac{(x+\beta B_q\mu_B)^3}{e^x + 1} e^x  \dn x
\right]\nonumber\\
=& g_q\dfrac{T^4}{2\pi^2} &&\left[
\int_0^\infty \dfrac{2 x^3+6x(\beta B_q\mu_B)^2}{e^x + 1} +\int_{-\beta B_q\mu_B}^0  (x+\beta B_q\mu_B)^3
 \dn x\right]\nonumber\\
=& g_q &&\left[\dfrac{7\pi^2 T^4}{120}+\dfrac{T^2 (B_q\mu_B)^2}{4}+\dfrac{(B_q\mu_B)^4}{8\pi^2}\right]
\end{align}
The phase transition occurs when the pressure $P$ is equal to the bag pressure $B$:
\begin{align}
B \overset{!}{=}& P = \dfrac{1}{3} \left(\varepsilon_q + \varepsilon_{\bar q} + \varepsilon_g\right)\nonumber\\
\Rightarrow 0 \overset{!}{=}& T^4 \cdot C +T^2 \cdot (2(B_q\mu_B)^2 g_q) + \dfrac{g_q (B_q\mu_B)^4}{\pi^2} - 24B\nonumber\\
\Rightarrow T \overset{!}{=}& \sqrt{\dfrac{1}{C}\left[-(B_q\mu_B)^2g_q+\sqrt{(B_q\mu_B)^4 g_q \left(g_q-\dfrac{C}{\pi^2}\right) + 24 BC}\right]}
\end{align}
with $C=\dfrac{\pi^2}{15}\left(7g_q+4g_g\right)$. The bag pressure $B$ is then fixed by defining the critical temperature $T_C$ at zero chemical potential as
\begin{align}
 T_C := T\Big\vert_{\mu_B=0} = \sqrt[4]{\dfrac{24B}{C}}
\end{align}
which I choose to be $T_C=175\MeV$. Therefore, the critical temperature as a function of the baryo-chemical potential can be written as
\begin{align}
T(\mu_B)= \sqrt{\dfrac{1}{C}\left[-(B_q\mu_B)^2g_q+\sqrt{(B_q\mu_B)^4 g_q \left(g_q-\dfrac{C}{\pi^2}\right) + T_C^4 C^2}\right]}
\end{align}

\section{Integrals of the flow coefficients}
\label{app:integral_flow_coefficients}
Here, I will show how to solve the integrals of the fourier coefficients from eq.~\ref{eqn:def_fourier_coefficients}:
\begin{align}
\tilde{v}_n = & \dfrac{1}{2\pi}\int\dn\phi \cos(n\phi) \dfrac{\dn N}{p_T \dn p_T \dn\phi \dn y}\nonumber\\
=& C \int_0^{\rho_0} \dn \rho \int_0^{2\pi}\dn \varphi \int_{-\infty}^{+\infty}\dn \eta 
\tau(\rho) \rho\nonumber\\
&\times \left[ f^2(\varphi) m_T \cosh(y-\eta) - \dfrac{\partial \tau}{\partial \rho} p_T 
	\left(f \cos(\beta'-\phi)+f' \sin(\beta'-\phi)\right) \right]\nonumber\\
&\times\exp{-\left(m_T \cosh(y-\eta) \cosh \eta_T - p_T \cos(\beta'-\phi) \sinh \eta_T\right)/T}
\end{align}
with $C=\dfrac{g}{(2\pi)^3} \exp{\mu/T}$. Using the following identities for the modified bessel functions
\begin{align}
 I_n (z) =I_{-n}(z) &= \dfrac{1}{2\pi} \int_0^{2\pi} \exp{z \cos \theta} \cos(n\theta) \dn\theta\\
 K_n (z)= K_{-n}(z) &= \dfrac{1}{2} \int_{-\infty}^\infty \exp{-z \cosh t} \cosh(n t) \dn t
\end{align}
and noting that
\begin{align}
 \int_0^{2\pi} \exp{z \cos \theta} \sin(n\theta) \dn\theta \equiv 0
\end{align}
the $\eta$- and $\phi$-integrations can be done analytically:
\begin{align}
 \tilde{v}_n=&2C\rho_0^2\int\dn\rho' \tau(\rho) \rho' \int \dn\varphi \int \dn \phi \nonumber\\
&\times \cos(n\phi) \left[ K_1 f^2(\varphi) m_T - K_0 \dfrac{\partial \tau}{\partial \rho} p_T \left(f \cos(\varphi-\phi)+f' \sin(\varphi-\phi)\right) \right] \nonumber\\
&\times\exp{p_T \cos(\varphi-\phi) \sinh\eta_T}\nonumber\\
=&2C\rho_0^2\int\dn\rho' \tau(\rho) \rho' \int \dn\varphi \int \dn \Phi\, \left(\cos(n\Phi)\cos(n\varphi)+\sin(n\Phi)\sin(n\varphi)\right) \nonumber\\
&\times\left[ K_1 f^2(\varphi) m_T - K_0 \dfrac{\partial \tau}{\partial \rho} p_T \left(f \cos(\Phi)+f' \sin(\Phi)\right) \right] \exp{p_T \cos(\Phi) \sinh\eta_T}\nonumber\\
=& 4\pi C\rho_0^2\int\dn\rho' \tau(\rho) \rho' \int \dn\varphi \nonumber\\
& \Biggl[\cos(n\varphi) K_1 I_n f^2(\varphi) m_T\nonumber\\
 & - K_0 \dfrac{\partial \tau}{\partial \rho} p_T \dfrac{1}{2}\Bigl[ \cos(n\varphi) f(\varphi) \left(I_{n-1}+I_{n+1}\right)+ \sin(n\varphi) f'(\varphi) \left(I_{n-1}-I_{n+1}\right)\Bigr]\Biggr]\nonumber\\
=& 4\pi C\rho_0^2\int\dn\rho' \tau(\rho) \rho' \int \dn\varphi\nonumber\\
& \Biggl[\cos(n\varphi) K_1 I_n f^2(\varphi) m_T\nonumber\\
& - \cos(n\varphi) K_0 \left(I_{n-1}+I_{n+1}\right) \dfrac{\partial \tau}{\partial \rho} p_T \dfrac{1}{2}  f(\varphi) \nonumber\\
& - \Bigl[\cos((n-2)\varphi)-\cos((n+2)\varphi)\Bigr] K_0 \left(I_{n-1}-I_{n+1}\right)\dfrac{\partial \tau}{\partial \rho} p_T \dfrac{\alpha^{-2}-\alpha^2}{8}f^3(\varphi)\Biggr]
\end{align}
In the first step, I used the identity of the $K$ bessel function and substituted $\rho=\rho' \cdot \rho_0$. In the second step, I substituted $\Phi=\varphi-\phi$ and expanded $\cos(n\phi)=\cos(n(\varphi-\Phi))$. In the third step, I applied the trigonometric identities
\begin{align}
 \cos x \cos y = \dfrac{1}{2}\left(\cos(x-y)+\cos(x+y)\right)\\
 \sin x \sin y = \dfrac{1}{2}\left(\cos(x-y)-\cos(x+y)\right) \label{eqn:trig_sin_prod},
\end{align}
so that I could solve the $\Phi$ integral by using the identity for the $I$ bessel function.
In the last step, I used
\begin{align}
 f'(\varphi) = f^3(\varphi) \dfrac{\alpha^{-2}-\alpha^2}{2} \sin(2\varphi).
\end{align}
and again applied eq.~\eqref{eqn:trig_sin_prod} to the sinus terms. Finally, the $\phi$-integrated invariant yield at midrapidity reads
\begin{align}
\dfrac{\dn N}{2\pi P_T \dn P_T \dn y}=&\tilde{v}_0\nonumber\\
=& \dfrac{g}{(2\pi)^3}\exp{\mu/T} 4\pi \rho_0^2 \int_0^{2\pi} \dn\varphi \int_0^1 \dn \rho' \tau(\rho') \rho'\nonumber\\
&\times \Biggl[K_1 I_0 f^2(\varphi) m_T - K_0 I_1 f(\varphi) p_T \partial_\rho \tau \Biggr]
\end{align}
and the numerator of the elliptic flow is calculated as
\begin{align}
 \tilde{v}_2(P_T) =& \dfrac{g}{(2\pi)^3}\exp{\mu/T} 4\pi \rho_0^2 \int_0^{2\pi} \dn\varphi \int_0^1 \dn \rho' \tau(\rho) \rho'\nonumber\\
& \times \Biggl[\cos(2\varphi) K_1 I_2 f^2(\varphi) m_T\nonumber\\
& - \cos(2\varphi) K_0 \left(I_1+I_3\right) \dfrac{\partial \tau}{\partial \rho} p_T \dfrac{1}{2}  f(\varphi) \nonumber\\
& - \left(1-\cos(4\varphi)\right) K_0 \left(I_1-I_3\right)\dfrac{\partial \tau}{\partial \rho} p_T \dfrac{\alpha^{-2}-\alpha^2}{8}f^3(\varphi)\Biggr]
\end{align}
In the case of $\partial_\rho \tau = 0$ (radially constant freeze-out time), this simplifies to
\begin{align}
\dfrac{\dn N}{2\pi P_T \dn P_T \dn y}= m_T \dfrac{g}{(2\pi)^3}\exp{\mu/T} 4\pi \rho_0^2 \tau_0 \int_0^{2\pi} \dn\varphi \int_0^1 \dn \rho' \rho' K_1 I_0 f^2(\varphi)
\end{align}
and
\begin{align}
 v_2(P_T) = \dfrac{\tilde{v}_2}{\tilde{v}_0}=\dfrac{\int\dn\varphi \int \dn \rho' \rho' \cos(2\varphi) K_1 I_2 f^2(\varphi)}{\int\dn\varphi \int \dn \rho' \rho' K_1 I_0 f^2(\varphi)}.
\end{align}

\chapter{Acknowledgements}
I would like to thank everyone who supported me directly or indirectly during my university studies and especally in the creation of this thesis.

So first of all, I want to thank my supervisor Marcus Bleicher, who supported me all throughout my thesis. He encouraged me regarding my public presentations and our joint publications. Without him this thesis would not have been possible. I also thank Horst St\"ocker for the admission in the workgroup and the confidence he has shown me.

\vspace{0.5cm}
Thanks to my colleagues Katharina Schmidt and Jan Michel for the (mostly) fruitful physical discussions and Katharina for the workaday diversion and coffee breaks.

Also a big thank-you to my physics teacher Dirk Rosenauer, who brought me to study physics, even though (or just because) he taught us more about life then about physics.

\vspace{0.5cm}
My deepest gratitudes are expressed at last. My parents, Doris and Gerhard, always believed in me. They supported my curiosity and helped me to achieve my aims. My parents and my sister Tamara taught me the most valueable things in life and gave me self-confidence. Thank you so much.

Finally, I want to put my thankfulness for the most important person into words. My girlfriend Daniela was and is my best friend, strongest supporter, most honest critic, ray of hope and source of happiness ever since. She always built me up with a balance of pressure and motivation and beside all the other things she has done for me, I owe her the successful conclusion of my studies. Thank you, honey.

\chapter{Legal stuff}

\section[Erkl\"arung der Selbstst\"andigkeit (Declaration)]{Erkl\"arung der Selbstst\"andigkeit}
\thispagestyle{empty}
Hiermit versichere ich, die vorliegende Arbeit selbstst\"andig verfasst und keine anderen als die angegebenen Quellen und Hilfsmittel benutzt sowie die Zitate deutlich kenntlich gemacht zu haben.
\vspace{4\baselineskip}\\
Karben, den 17.12.2008 \hfill Daniel Krieg
\vspace{4\baselineskip}\\
\clearpage
\mbox{}\thispagestyle{empty}

\clearpage
\section[Zusammenfassung (german abstract)]{Deutsche Zusammenfassung der Diplomarbeit (German abstract to the thesis)}
Eine der gro\ss en Fragen der Menschheit mit der sich Forscher der verschiedensten Disziplinen besch\"aftigt haben, ist die Frage nach den fundamentalen Bausteinen der Natur bzw. ob es diese \"uberhaupt gibt. Sie reicht zur\"uck bis in die Antike, zu dem Philosophen Empedokles, und den vier griechischen Elementen Erde, Wasser, Luft und Feuer. Im Mittelalter wurden immer mehr der heute bekannten chemischen Elemente klassifiziert und im Periodensystem gruppiert. Anfang des letzten Jahrhunderts stellte man fest, dass das Atom (griech. atomos: unteilbar) eine Substruktur, mit einem Kern aus Proton und Neutronen und einer Schale aus Elektronen, besitzt. Doch auch die Bausteine des Kerns offenbarten in inelastischen Streuexperimenten eine Substruktur. Die F\"ulle der neuen Teilchen genannte Hadronen, die bei diesen Experimenten erzeugt wurden, erforderte auch hier ein Ordnungsschema. Aus dem ``achtfachen Weg'' von Murray Gell-Mann entwickelte sich dann das Quark-Modell, welches die Hadronen in Mesonen, bestehend aus einem Quark und einem Anti-Quark, und Baryonen, bestehend aus 3 Quarks, einteilt.

Seit dem Beginn der Suche nach den elementaren Bausteinen der Natur ben\"otigten die Forscher immer h\"ohere Energien um die immer kleineren L\"angen aufzul\"osen. Nun suchen Physiker nach dem sogenannten Quark-Gluonen-Plasma (QGP), einem Zustand, in dem die Quarks nicht mehr eingesperrt sind in den Hadronen, sondern frei wie in einem Plasma. Mit der Kollision von schweren Ionen am Relativistiv Heavy Ion Collider (RHIC), und bald mit dem Large Hadron Collider (LHC) am CERN, untersucht man den \"Ubergang zu dieser neuen Phase der stark wechselwirkenden Materie.

Ein gro\ss es Problem in diesem Zusammenhang ist die Frage, ob man \"uberhaupt die Phasengrenze \"uberschritten und ein QGP erzeugt hat, denn die Hadronisierung, also der \"Ubergang zur\"uck zur normalen Materie, findet auf extrem kurzen Zeitskalen statt. Daher k\"onnen wir nur die Hadronen vom kinematischen Ausfrieren nachweisen, also die die keine Wechselwirkung mehr erfahren. Aber auch eine dynamische Beschreibung mithilfe der Quantenchromodynamik (QCD), der Eichtheorie der starken Wechselwirkung, ist nicht in naher Zukunft zu erwarten. Weder f\"ur die Kollision im Allgemeinen noch f\"ur den Vorgang der Hadronisierung im Speziellen. Somit sind wir auf phenomenologische Modelle angewiesen.

Die Berechnungen innerhalb der pertubativen QCD (pQCD) mithilfe von gemessenen, parameterisierten Fragmentationsfunktionen sind sehr erfolgreich f\"ur Proton-Proton Kollisionen, aber in Schwerionenkollisionen mit einem dichten Phasenraum ist dieser Ansatz nur f\"ur hohe transversale Impulse anwendbar. Mit den ersten Resultaten von RHIC, die nicht mit pQCD zu erkl\"aren waren, r\"uckte ein phenomenologisches Modell namens Recombination in den Fokus.

Im Gegensatz zu Fragmentation, bei der ein Quark aus einer harten Streuung mit hohem Impuls in viele Hadronen ``zerf\"allt'', die dann nur einen Bruchteil des Impulses tragen, ist die Idee bei der Recombination, dass in einem sehr dichten Medium ein Quark und ein Anti-Quark (drei Quarks) zu einem Meson (Baryon) recombinieren und sich dabei die Impulse addieren.

Eine der vielversprechensten Observablen bei der Suche nach dem QGP ist der sogenannte elliptische Fluss, welcher dem zweiten Fourierkoeffizienten $v_2$ des invarianten transversalen Impulsspectrums entspricht. Er quantifiziert die azimuthale Asymmetry des transversalen Impulses, welche aus der r\"aumlichen Asymmetry in nicht-zentralen Kollisionen entsteht. Da der elliptische Fluss seinem Ursprung entgegenwirkt und damit ein selbstl\"oschendes Verhalten zeigt, ist er damit haupts\"achlich von der fr\"uhen Phase der Kollision gepr\"agt und w\"urde damit direkt vom Fluss der Quarks aus dem QGP abh\"angen.

Der Erfolg der Recombination begr\"undet sich vorallem auf die Vorhersage einer universellen Skalierung des elliptischen Flusses, wonach das Hadron $v_2$ mit der Anzahl der Konstituentenquarks skaliert. Diese Vorhersage ist experimentell gut best\"atigt und die Ergebnisse k\"onnen die Daten von niedrigen bis mittleren transversalen Impulsen sehr exakt beschreiben. W\"ahrend das qualitative Verhalten des elliptischen Fluss dem Skalierungsgesetz der Recombination folgt, h\"angen die quantitativen Ergebnisse von den zwei zus\"atzlichen ``Zutaten'' ab: der Quarkdichteverteilung und der Hadronisierungshyperfl\"ache.

In dieser Arbeit habe ich gezeigt, dass die einflie\ss enden Modelle eine gro\ss e Wirkung in Bezug auf die Flusskoeffizienten haben. F\"ur die Quarkverteilung benutze ich das Blast-Wave Modell, welches, basierend auf hydrodynamischen \"Uberlegungen, ein transversal expandierendes, lokal thermisches System beschreibt. F\"ur die Parameterisierung des Druckgradienten in nicht-zentralen Kollisionen gen\"ugt ein simples geometrisches Argument um nicht nur den elliptischen Fluss $v_2$, sondern auch den n\"achsten Koeffizienten $v_4$ und deren relative St\"arke zu beschreiben. Ein Ergebnis des Blast-Wave Profils ist die beobachtete Massenskalierung von $v_2$ und insbesondere ein negativer Fluss bei niedrigen transversalen Impulsen f\"ur schwere Hadronen und/oder hohe Schwerpunktsenergien. Solche negativen Werte sind jetzt auch in den vorl\"aufigen $J/\psi$ PHENIX-Daten zu sehen und scheinen damit die Anwendung der Blast-Wave Parameterisierung zu st\"utzen. Das ist vorallem im Hinblick auf eine besonders \"uberraschende Vorhersage in diesem Zusammenhang interessant. Denn Recombination in Verbindung mit dem Blast-Wave Modell prognostiziert eines abnehmenden $\langle v_2 \rangle$ f\"ur Schwerpunktsenergien oberhalb von $\sqrt{s}=5\TeV$, welches m\"oglicherweise am LHC beobachtet werden kann.

Die Hadronisierung findet in einem 3-dimensionalen, zeitabh\"angigen Volume statt und f\"allt mit dem kinematischen Ausfieren zusammen, da die hadronische Streuphase nicht ber\"ucksichtigt wird. Eine detaillierte Untersuchung der Hyperfl\"ache w\"urde eine dynamische Behandlung erfordern, aber um grundlegende Einfl\"usse auf die Spektren zu untersuchen ist dieser analytische Ansatz ausreichend, insbesondere weil auch in der Hydrodynamik der Vorgang des Ausfrierens ein absolut nicht-triviales Problem ist.
Wie meinen Ergebnissen zu entnehmen ist, haben eine r\"aumliche Exzentrizit\"at beim Ausfrieren und die radiale Abh\"angigkeit der Ausfrierzeit den gr\"o\ss ten Effekt auf die Flusskoeffizienten und besonders auf deren Verh\"altnis $v_4/(v_2)^2$. W\"ahrend ein einfaches, zirkulares Ausfrieren dieses Verh\"altnis um etwa einen Faktor 2 zu niedrig vorhersagt, erzeugt ein elliptisches Ausfrieren zus\"atzliche Beitr\"age zur Asymmetry beider Koeffizienten und kann damit die experimentellen Daten sehr gut beschreiben. Und die radiale Abh\"angigkeit der Ausfrierzeit beeinflusst dann nochmal zus\"atzlich das Verhalten bei niedrigen transversalen Impulsen.

Die pr\"asentierten Ergebnisse in meiner Arbeit sind ein klarer Hinweis darauf, dass Recombination der dominante Mechanismus zur Hadronisierung bei kleinen mittleren transversalen Impulsen ist, wobei die gute Beschreibung der Flusskoeffizienten und die Konstituentenquark-Skalierung ein starkes Zeichen f\"ur ein QGP in der fr\"uhen Phase der Kollision sind. Trotz der phenomenologischen Natur dieser analytischen Untersuchung zur Hadronisierung in Schwerionenkollisionen durch Recombination, beschreiben die Ergebnisse nicht nur das qualitative Konstituentenquark-Skalierung der experimentellen Daten, sondern auch die quantitative Massenskalierung des elliptischen und hexadekupolen Flusses und das Verh\"altnis der beiden Koeffizienten mit gro\ss em Detail. Aufgrund dieser guten \"Ubereinstimmungen sollten die hervorstechenden Vorhersagen f\"ur den LHC detaillierter in dynamischen Untersuchungen betrachtet werden.

%% Dokument ENDE %%%%%%%%%%%%%%%%%%%%%%%%%%%%%%%%%%%%%%%%%%%%%%%%%%%%%%%%%%
\end{document}